RESEARCH ARTICLE | DECEMBER 08 2023

# Demonstration of a facile and efficient strategy for yield stress determination in large amplitude oscillatory shear: Algebraic stress bifurcation ⊕ ⊘

Pengguang Wang (王鹏广) 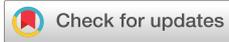 ; Jiatong Xu (徐家通) ⬤ ; Hongbin Zhang (张洪斌) ✉ ⬤

 Check for updates

View
Online

Export
Citation

CrossMark

# Demonstration of a facile and efficient strategy for yield stress determination in large amplitude oscillatory shear: Algebraic stress bifurcation 

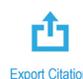

Pengguang Wang (王鹏广), Jiatong Xu (徐家通), and Hongbin Zhang (张洪斌)[a]

**AFFILIATIONS**

Advanced Rheology Institute, Department of Polymer Science and Engineering, School of Chemistry and Chemical Engineering, Shanghai Jiao Tong University, Shanghai 200240, China

[a]Author to whom correspondence should be addressed: hbzhang@sjtu.edu.cn

**ABSTRACT**

The large amplitude oscillatory shear (LAOS) has been extensively studied for understanding the rheological responses of yield stress fluids. However, the employed methodology for determining the yield stress remains uncertain albeit the fact that many classical or plausible methods exist in the literature. Along these lines, herein, based on Fourier transform (FT) rheology, stress decomposition, and stress bifurcation, a new straightforward method termed as algebraic stress bifurcation was developed. More specifically, the main goal was to determine the yield stress and investigate the solid–liquid transition of fluids in LAOS. A simple and efficient mathematical framework was established and verified by the KVHB, Saramito, Giesekus models, and FT rheology. The main strength of this approach is that only the data from the stress/strain sweep are required instead of Lissajous curves. Alternative curves were constructed to demonstrate the non-critical role of both higher harmonics and phenomenological Lissajous curves in determining yield stress. The determined start and end yield points in the solid–liquid transition were compared with the already existing methods. Furthermore, the resulting solid–liquid transition region was analyzed by FT rheology, stress decomposition, and sequence of the physical process to obtain information on nonlinearity and intracycle/intercycle yielding. Our work provides fruitful insights for explaining and reducing the complexities of the stress bifurcation technique by using an easy-to-understand and implement format. Therefore, a concise theoretical framework was introduced for understanding the concept of yield stress, the intercycle yielding process, and the rational choice of yield stress measurement techniques.

Published under an exclusive license by AIP Publishing. https://doi.org/10.1063/5.0174741

## I. INTRODUCTION

Yield stress fluids (YSFs) behave as solids at equilibrium but flow like liquids beyond the application of a critical stress, which is defined as yield stress.[1–3] The yield stress is considered an essential parameter of YSFs and a crucial criterion for the formulation, processing, and functionality of these YSFs in the design and applications of a relatively large number of fluids involving wide fields,[4] e.g., colloidal gels, paints, inks, slurries, cosmetics, toothpaste, foams, wet concrete, oil drilling fluids, and foods, such as butter, mayonnaise, ketchup, and even foods for special medical purposes in dysphagia therapy.

It is also well known that the determined yield stress strongly depends on the test protocols including methods, equipment, and conditions, whereas significant deviations may arise among these yield stress values.[2,5] These deviations are anticipated. For example, the yield stress evaluation during an oscillatory shear flow is related to both the dynamically varying stress and the finite time to complete a solid–liquid transition, where the time of a solid–liquid transition strongly depends on the difference between the applied stress and the yield stress.[6–9] Yielding is experimentally verified as a gradual process, where irreversible rearrangements can be observed below the yield stress.[10–16] An end yield stress may be needed, reflecting the yielding behavior that occurred within a short timescale. For example, for the commercial YSFs applied in many cases, the consumer may not pay attention to the specific yield stress value that triggers the yielding behavior after a long wait. Some yielding processes can occur on a short timescale, e.g., the swallowing behavior of foods for dysphagia therapy.[5] Therefore, an end yield stress under which a significant structural failure occurs in LAOS is needed for the application of YSFs.

As far as the methods of investigating the yielding process and flow properties are concerned, the yield stress determination tests can be divided into the following categories: steady shear, transient shear, and dynamic oscillatory shear tests. Each of these methods has its advantages as well as its deficiencies.

(1) Steady shear. The flow curve to the zero shear rate limit fitted by models (e.g. Herschel–Bulkley (HB) model) produces the yield stress.[17–22] It is known that the yield stress from the sweeping up and down procedures, which is called the static and dynamic yield stress, respectively, may not well correspond for thixotropic YSFs.[23–27] Furthermore, sufficient long times are often required to reach an equilibrium state, and only information on the yield stress can be available, while the yield strain is often inaccessible.

(2) Transient shear. the yield stress value can be obtained from the step shear rate (startup shear), step shear stress (creep), and shear rate/stress ramp, respectively. (i) In the shear startup test, the peak stress is defined as the yield stress.[28–31] However, it should be noted that this method is not always accessible since the stress overshoot may not be present.[31,32] (ii) The creep method is considered one of the most accurate methods for evaluating yield stress. Nevertheless, this method is inefficient due to the long testing time that is required and the prior range requirement of the stress.[1–3,6–9,24,33,34] In addition, similar to the steady shear test, the creep method can also only obtain the yield stress value. In addition, the (iii) transient shear stress ramp tests can also determine the yield strain. Nonetheless, different ramp rates are needed to eliminate the transient effect, and a clear transition may require an extremely slow ramp rate.[7,35,36] On the other hand, the comparative advantage of this method is that not only the yield stress but also the yield strain can be determined.

(3) Dynamic oscillatory shear. A number of determining methods are generated by this test. For instance, (i) the yield stress can be evaluated by the crossover point of $G'$ and $G''$ (i.e., the values generated from the first harmonic of the output signal).[8,14,29,37–39] It should be denoted that the storage modulus $G'$ and loss modulus $G''$ are defined in SAOS, and the concepts of $G'$ and $G''$ are related to the first harmonic as will be discussed in this work. In addition, in some YSFs, the overshoot of $G''$ is used to determine the yield stress.[38,40–42] However, the lack of the crossover point and overshoot sometimes occur and bring limits. (ii) The deviation from the linearly elastic behavior and the viscoelastic regime, which are used to denote the yield point, can also lead to the implementation of several determination methods. For example, the "elastic stress" method adopts the maximum of the product of $G'$ and the strain amplitude $\gamma_0$ versus the stress/strain as the yield stress value.[43] However, it must be noted that the dynamic moduli can be physically explained by taking into account energy storage and dissipation mechanisms, whereas the term elastic stress is just the average energy stored per unit volume per period divided by the strain amplitude.[44] Therefore, although this elastic stress term has been used in the literature to describe elastic stress, it has been roughly recognized as a plausible term and that could be measured to depict yield stress. However, it has been used in stress jump tests that experimentally decompose the total stress into elastic and viscous contributions.[45,46] It should be underlined that this treatment is not always correct for different materials and could lead to inconsistencies among the obtained values of the different methods.[32,43] Another method is based on plotting the shear stress versus shear strain at multiple frequencies. From these curves, a range where the material behavior is strongly non-linear can be derived. Thereby, the onset stress value of the range is considered "apparent yield stress."[47] A series of tests are also required for this approach. Still, another way utilizes the crossover point of two straight lines drawn at low and high strain amplitudes by plotting the stress versus strain (or $G'$ versus strain/strain) as the yield point and the corresponding yield strain point, respectively.[32,42,48–50] In addition, the determination of the yield stress can also be deduced from the plot of the stress/strain amplitude dependence of $I_3/I_1$ (the intensity ratio of the third to the first harmonic) or $I_3/I_1\gamma_0^2$ (i.e., $Q$ value) based on Fourier transform (FT) rheology.[38,51–53] In the plot, the linear relationships appear at a low stress/strain range, and the yield stress can be evaluated from the deviation of this linear behavior. However, this consideration has been pointed out as incapable of giving a consistent criterion to accurately quantify the yield stress value.[48] Based on two-dimensional mechanical correlation spectra using FT rheology, an improving approach was also proposed by introducing the non-linear synchronous self-correlation intensity in which the intensities of the 1st harmonic and higher harmonics were used to identify the yield stress.[48] (iii) The Lissajous figures (stress $\sigma$–strain $\gamma$, and stress $\sigma$–strain rate $\dot{\gamma}$) in LAOS were used to determine the yield stress. For example, a purely theoretical model was developed based on a modified Bingham model.[54] From the Lissajous curve, discontinuities during a whole period can be derived, and the yield stress can be estimated from the positions of these discontinuities. Nevertheless, real YSFs are difficult to be handled with the model. Consequently, Ewoldt et al.[55] suggested the utilization of a perfect plastic dissipation parameter, i.e., the dissipation ratio of the actual dissipated energy to the perfect plastic dissipation, to evaluate the yield stress. It was found that this method may obtain much higher values than those extracted from steady shear.[55] Furthermore, Yu et al.[9] developed a new method, which was named stress bifurcation, to determine the start and the end of the yielding transitions on different YSFs including critical start stress and strain (i.e., the maximum stress and strain to sustain solidlike behavior) as well as end stress and strain rate (i.e., the minimum stress and strain rate above which a complete solid–liquid transition occurs in an oscillatory shear cycle). Unlike the common methods that acquire the knowledge of only one critical yield stress value, the stress bifurcation estimates the yielding onset and end during the solid–liquid transition process. However, raw Lissajous curves and complicated data processing are needed. From the above-mentioned analysis, it is apparent that determining a range of the solid–liquid transition offers new possibilities for deeply understanding the yielding process of the fluids.

Although the concept, judgment, and measurement of the yield stress and yielding process have been studied for several decades, several issues remain elusive. Rational, simple, efficient, and accurate measurement of yield stress has always been fraught with difficulties in rheology. In view of the above-mentioned numerous methods for measuring yield stress and motivated by stress bifurcation, a new algebraic stress bifurcation method was proposed here based on FT rheology, stress decomposition, and stress bifurcation, in terms of the coordinates of some special points in real Lissajous curves that are approximately represented in formulas and equations according to the definition of algebra. This facile and efficient strategy determines the start and end points of the solid–liquid transitions from performing LAOS experiments on typical YSFs, requiring only the stress/strain amplitude sweep data ($G'$, $G''$, and the stress/strain amplitude) instead of using raw data (Lissajous loops) and avoiding cumbersome data processing. On top of that, the elliptical curves based on $I_1$ to approximate the original Lissajous curve ($I_1$-based curves) were constructed via FT rheology to interpret the insignificant contribution of the higher harmonics and the phenomenological Lissajous curves to the determination of the yield stress in algebraic stress bifurcation. Herein, a number of typical yield stress fluids including carbopol gel,[56–58] laponite suspension,[59] polysaccharide solution of xanthan gum[3] and welan gum,[60] gel particle suspension,[61] cellulose nanofiber suspension,[62] body lotion,[63] and ketchup,[64] were used to experimentally verify the rationality and applicability of the method. Furthermore, the yielding evolution in solid–liquid transition regions that were determined by our algebraic stress bifurcation was also thoroughly analyzed to further understand the stress bifurcation and yield process of these YSFs.

## II. THEORY

### A. LAOS

LAOS is considered a powerful tool for investigating the nonlinear yielding process that contains sequentially solidlike and liquidlike behaviors in a single period.[65] The Lissajous curve is used to study the transient behavior of YSFs by plotting elastic (stress vs strain) and viscous (stress vs strain rate) Lissajous curves. Insights have been provided via different LAOS methods,[66] such as FT rheology,[42] Chebyshev polynomials,[67,68] stress decomposition,[68–71] and a fully quantitative sequence of the physical process (SPP) technique.[53,72–74] The interpretations of the LAOS data from such methods intimately depend on their mathematical framework and underlying assumptions.[65,75]

### B. FT rheology

One of the functions of FT is to extract the periodic signals from a complex signal with existing interference signals.[76] In other words, FT uses the sum of a series of trigonometric functions with different frequencies ($\sin\omega t$, $\cos\omega t$, $\sin3\omega t$, $\cos3\omega t$, $\cdots$) to approximate the original signal. FT represents the inherent periodic contributions to a time-dependent signal and displays the corresponding amplitudes and phases (or real and imaginary parts) as a function of the frequency.[42] For a periodic signal $f(x)$ with a period of $2\pi/\omega$, FT series is carried out as follows:[76]

$$f(x) = \sum_{k=0}^{\infty}(a_k\cos k\omega x + b_k\sin k\omega x),$$

where

$$a_k = \frac{1}{\pi}\int_{-\pi/\omega}^{\pi/\omega}f(x)\cos k\omega x dx \quad\text{and}\quad b_k = \frac{1}{\pi}\int_{-\pi/\omega}^{\pi/\omega}f(x)\sin k\omega x dx. \tag{1}$$

For FT rheology, the common stress waveform $\sigma(t)$ for the strain-controlled condition ($\gamma(t) = \gamma_0\sin\omega t$)[42,77] is

$$\sigma(t) = \sum_{n=1}^{\infty}I_n\sin(n\,\omega t + \phi_n)$$
$$= \gamma_0\sum_{n=1}^{\infty}(G_n''\cos n\,\omega t + G_n'\sin n\,\omega t), \tag{2}$$

where $I_n$ is the $n$th harmonic magnitude, $\phi_n$ denotes the phase angle, $\gamma_0$ represents the strain, $\omega$ denotes the angular frequency, $t$ is the time, and $n$ refers to the $n$th harmonic. While both strain- and stress-controlled oscillations have been used, strain-controlled experiments are more common.[65] In addition, the first harmonic moduli of $G_1'$ and $G_1''$ are generally considered $G'$ and $G''$ that commonly obtained from the rheometers in LAOS tests, respectively.[42] Furthermore, only odd higher-order harmonics are expected for typical and idealized experiments. The stress response of viscoelastic materials is typically independent of the shear direction. Thus, the output signals must be an odd function. The demonstrated even harmonics are attributed to the imperfection of the experiment,[78–82] which are relatively small compared with the odd higher harmonics.[42] Therefore, the Fourier series can be described as a function of the odd harmonics,

$$\sigma(t) = \gamma_0\sum_{n=1,odd}^{\infty}(G_n''\cos n\omega t + G_n'\sin n\omega t). \tag{3}$$

Similarly, the strain waveform for the stress-controlled condition with a perfect stress signal, $\sigma(t) = \sigma_{max}\sin\omega t$,[83] can be defined as follows:

$$\gamma(t) = \gamma_0\sum_{n=1}^{\infty}(a_n\cos n\omega t + b_n\sin n\omega t)\text{ or}$$
$$\gamma(t) = \gamma_0\sum_{n=1,odd}^{\infty}(a_n\cos n\omega t + b_n\sin n\omega t). \tag{4}$$

Interpolation can be used to translate the distortion of the strain response with a good stress wave to the stress response with a good strain signal.[67,70] Furthermore, the above-mentioned FT equations can be transformed into a power series ($x = \sin\omega t$ and $y = \cos\omega t$) by using Chebyshev polynomials[68]

$$\sigma(t) = \sum_{n=1}^{\infty}a_n'\cos{}^n\omega t + \sum_{n=0}^{\infty}b_{2n+1}'\sin{}^{(2n+1)}\omega t$$
$$+ \sin\omega t\cos\omega t\sum_{n=0}^{\infty}c_{2n}'\sin{}^{2n}\omega t$$
$$= \sum_{n=1}^{\infty}a_n'y^n + \sum_{n=0}^{\infty}b_{2n+1}'x^{2n+1} + xy\sum_{n=0}^{\infty}c_{2n}'x^{2n}. \tag{5}$$

When even harmonics are excluded, the power series becomes as follows:

$$\sigma(t) = \sum_{n=1,odd}^{\infty}a_n'y^n + \sum_{n=1,odd}^{\infty}b_n'x^n. \tag{6}$$

Furthermore, the coefficients in Eqs. (4) and (6) can be calculated by using Eq. (1) and conducting polynomial fitting, respectively.

## C. Stress decomposition

In the linear region, $G'$ and $G''$ represent the elastic response and the viscous energy dissipation, respectively. Cho et al. developed a useful method to decompose the oscillatory stress response in the nonlinear region into elastic and viscous parts,[69]

$$\sigma(\gamma, \dot{\gamma}/\omega) = \frac{\sigma(\gamma, \dot{\gamma}/\omega) - \sigma(-\gamma, \dot{\gamma}/\omega)}{2}$$
$$+ \frac{\sigma(\gamma, \dot{\gamma}/\omega) - \sigma(\gamma, -\dot{\gamma}/\omega)}{2}$$
$$= \frac{\sigma(x, y) - \sigma(-x, y)}{2} + \frac{\sigma(x, y) - \sigma(x, -y)}{2}. \quad (7)$$

Therefore, the corresponding elastic stress $\sigma'(\gamma, \gamma_0)$ and viscous stress $\sigma''(\dot{\gamma}/\omega, \gamma_0)$ can be derived as follows:

$$\sigma'(x, \gamma_0) = \frac{\sigma(\gamma, \dot{\gamma}/\omega) - \sigma(-\gamma, \dot{\gamma}/\omega)}{2} = \frac{\sigma(x, y) - \sigma(-x, y)}{2}, \quad (8)$$

$$\sigma''(y, \gamma_0) = \frac{\sigma(\gamma, \dot{\gamma}/\omega) - \sigma(\gamma, -\dot{\gamma}/\omega)}{2} = \frac{\sigma(x, y) - \sigma(x, -y)}{2}. \quad (9)$$

This method is suitable for strain-controlled tests because every point $\sigma(x, y)$ has its partners $[\sigma(-x, y)$ and $\sigma(x, -y)]$. However, for the stress-controlled tests, the partner disappears, which can be conveniently solved by using interpolation.[67,70]

Alternatively, Yu et al. provided another form of stress decomposition,[70] which was equal to the method of Cho et al. when even harmonics were ignored,

$$\sigma'(x, \gamma_0) = \frac{\sigma(x, y) + \sigma(x, -y)}{2}, \quad \sigma''(y, \gamma_0) = \frac{\sigma(x, y) + \sigma(-x, y)}{2}. \quad (10)$$

According to the literature,[72,84] stress decomposition suffers from the question of symmetry assumption and fails to accurately capture the physics of yielding. The stress decomposition from Cho et al.[69] can be represented by Fourier series,

$$\sigma'(\gamma, \gamma_0) = \gamma_0 \sum_{n=1}^{\infty} G'_n \sin n\omega t, \quad (11)$$

$$\sigma''(\dot{\gamma}, \gamma_0) = \gamma_0 \left( \sum_{n=1, odd}^{\infty} G''_n \cos n\omega t + \sum_{n=2, even}^{\infty} G'_n \sin n\omega t \right). \quad (12)$$

The above questions are clearly shown via Eqs. (11) and (12). However, when even harmonics are removed, stress decomposition is successful in visually demonstrating the arisen nonlinearity,

$$\sigma'(\gamma, \gamma_0) = \gamma_0 \sum_{n=1, odd}^{\infty} G'_n \sin n\omega t, \quad (13)$$

$$\sigma''(\dot{\gamma}, \gamma_0) = \gamma_0 \sum_{n=1, odd}^{\infty} G''_n \cos n\omega t. \quad (14)$$

Meanwhile, by using Eqs. (13) and (14), there is no need for searching the partner of each point $[\sigma(\gamma, \dot{\gamma})]$ to conquer the issue in which the corresponding partner may not always exist.

## D. Stress bifurcation

Stress bifurcation determines the start and end of the solid–liquid transition to understand the yielding process in LAOS. The general geometric average method[70] has been proposed as the fundament of stress bifurcation. Both this method[70] and the successive stress bifurcation method[3,9,70] require the treatment of raw test data. In detail, the stress response of each point in the oscillatory stress amplitude sweep must be plotted to deduce the elastic [Fig. 1(a)] and the viscous [Fig. 1(b)] Lissajous curves.[3,9,70]

Based on the elastic Lissajous curve, the average stress ($\bar{\sigma}$) can be defined as the arithmetic mean of the total stress at a fixed strain ($\gamma$) to create the $\bar{\sigma}$–$\gamma$ curve, while an $\bar{\sigma}$–$\dot{\gamma}$ curve can be built based on the viscous Lissajous curve. Meanwhile, the mean strain ($\bar{\gamma}$) and the mean strain rate ($\bar{\dot{\gamma}}$) can be further calculated in the same way as the elastic and viscous stresses do. Thereby, six specific points, namely ($\gamma_{max}, \bar{\sigma}_{max}$), ($\bar{\gamma}_{max}, \sigma_{max}$), and ($\gamma_{max}, \sigma_{max}$) coming from the elastic Lissajous curve, ($\dot{\gamma}_{max}, \bar{\sigma}_{max}$), ($\bar{\dot{\gamma}}_{max}, \sigma_{max}$), and ($\dot{\gamma}_{max}, \sigma_{max}$) recognized from the viscous Lissajous curve, can be obtained,[9] where the $\sigma_{max}, \gamma_{max}$, and $\dot{\gamma}_{max}$ correspond to the stress amplitude, strain amplitude, and strain rate amplitude, respectively. The typical treatment procedure is displayed in Fig. 1.

Then, a series of the above-mentioned six points can be obtained by analyzing all the raw data points in the stress sweep test to deduce the stress–strain curves of $\bar{\sigma}_{max}$–$\gamma_{max}$, $\sigma_{max}$–$\bar{\gamma}_{max}$, $\sigma_{max}$–$\gamma_{max}$, and the stress–strain rate curves of $\bar{\sigma}_{max}$–$\dot{\gamma}_{max}$, $\sigma_{max}$–$\bar{\dot{\gamma}}_{max}$, and $\sigma_{max}$–$\dot{\gamma}_{max}$. Accordingly, two bifurcation points can be observed in the stress bifurcation to evaluate the yielding behavior. Thus, the start yield stress $\sigma_{l,s}$ and the start yield strain $\gamma_{l,s}$, as well as the end yield stress $\sigma_{l,e}$ and the end yield strain rate $\dot{\gamma}_{l,e}$ can be defined.

In the linear region, all geometric average curves are straight lines and can be expressed as follows:[9,70]

$$\bar{\sigma} = G_{\bar{\sigma}}\gamma, \quad \sigma = G_{\bar{\gamma}}\bar{\gamma}, \quad \bar{\sigma} = \eta_{\bar{\sigma}}\dot{\gamma}, \quad \sigma = \eta_{\bar{\gamma}}\bar{\dot{\gamma}}, \quad (15)$$

where the slopes are

$$G_{\bar{\sigma}} = G', \quad \eta_{\bar{\sigma}} = G''/\omega, \quad G_{\bar{\gamma}} = |G^*|/\cos\delta, \quad \eta_{\bar{\gamma}} = |G^*|/\omega\sin\delta. \quad (16)$$

The difference between the stress–strain average curves and the stress–shear rate average curves can be evaluated by using the following equation:

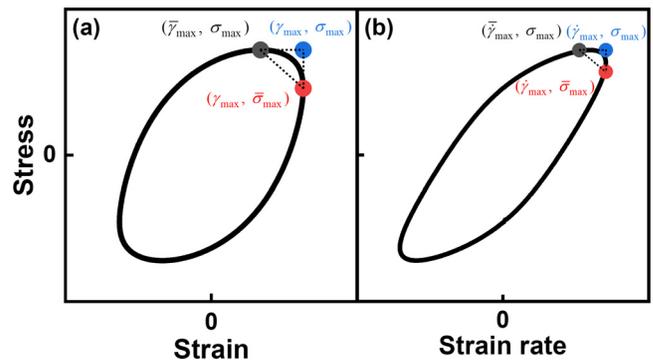

**FIG. 1.** (a) Stress–strain $\sigma$–$\gamma$ Lissajous curve with the points of the maximum stress ($\bar{\gamma}_{max}, \sigma_{max}$), maximum strain ($\sigma_{max}$–$\bar{\gamma}_{max}$), and maximum stress vs maximum strain ($\gamma_{max}, \sigma_{max}$). (b) Stress–strain rate $\sigma$–$\dot{\gamma}$ Lissajous curve with the points of the maximum stress ($\bar{\dot{\gamma}}_{max}, \sigma_{max}$), maximum strain rate ($\dot{\gamma}_{max}, \bar{\sigma}_{max}$), and maximum stress vs maximum strain rate ($\dot{\gamma}_{max}, \sigma_{max}$).

$$G_{\gamma}/G_{\dot{\sigma}} = 1/\cos^2\delta, \quad \eta_{\dot{\gamma}}/\eta_{\sigma} = 1/\sin^2\delta. \tag{17}$$

For solid-like materials, the situation where $\delta$ is close to zero degree endows good superposition of the $\bar{\sigma}_{\max}-\gamma_{\max}$ and $\sigma_{\max}-\bar{\gamma}_{\max}$ curves, while the $\bar{\sigma}_{\max}-\dot{\gamma}_{\max}$ and $\sigma_{\max}-\dot{\bar{\gamma}}_{\max}$ curves show significant differences. Particularly, for liquidlike materials, since the phase angle is close to 90°, a reversed superposition and separation of the above-mentioned curves occurs. Thus, for YSFs, the two $\sigma-\gamma$ curves superpose at low stress levels and bifurcate at high stress levels, while the two $\sigma-\dot{\gamma}$ curves are opposite. Accordingly, two bifurcation points can be observed in the stress bifurcation to evaluate the yielding behavior. Thus, the start yield stress $\sigma_{1,s}$ and the start yield strain $\gamma_{1,s}$ (i.e., the maximum stress and strain to sustain solid-like behavior), as well as the end yield stress $\sigma_{1,e}$ and the end yield strain rate $\dot{\gamma}_{1,e}$ (i.e., the minimum stress and strain rate above which a complete solid–liquid transition occurs in an oscillatory shear cycle) can be defined.

In brief, the previously proposed stress bifurcation method is essentially based on the geometric average of elastic and viscous Lissajous curves, requiring the treatment of the raw data. Furthermore, the physical interpretations of bifurcation during the solid–liquid transition can be rationalized, which depends on the phase angle $\delta$.

### E. SPP

The SPP formalism that has been used in recent years to great success to accurately reflect the sequence of processes occurring in the response of YSFs to LAOS, is considered a comprehensive theoretical approach for understanding the time-resolved behavior of yield stress fluids.[65] Rogers and Lettinga[72] suggested that SPP is efficient for treating non-linear stress responses, which is an innovative technique based on the physical approach rather than linear algebra assumptions to analyze LAOS responses. SPP handles stress waveforms as representing a sequence of physical processes, offering viscous and elastic contributions to treat the infinite harmonic series provided by FT rheology. The methods of SPP applied in this work include:

(1) SPP on a yield stress fluid[53] (exhibiting repeating elastic extension-yielding-flow-reformation) determines the rate-dependent static and dynamic yield stresses, post-yield regime by apparent cage modulus, and the connection between steady-shear and oscillatory test as follows:

  (i) SPP observes Lissajous curves by plotting stress versus strain and stress versus strain rate;

  (ii) The apparent cage modulus in the post-yield regime ($G_{\mathrm{cage}}$) can be defined as follows:

$$G_{\mathrm{cage}} = d\sigma/d\gamma|_{\sigma=0}. \tag{18}$$

It is apparent that

$$\lim_{\delta,\gamma_0\to0} G_{\mathrm{cage}} = \lim_{\delta,\gamma_0\to0} d\sigma/d\gamma|_{\sigma=0} = \lim_{\gamma_0\to0} G', \tag{19}$$

where $\delta$ is the phase angle.

  (iii) A strain can be acquired at the point of maximum stress ($\gamma$ at $\sigma_{\max}$). Then, the flow curves from LAOS can be obtained by plotting stress versus strain rate at the post-yielding regions of the stress signals. Therefore, the flow data from LAOS can be compared with those of the steady-shear.

(2) Kamani *et al.*[65] and Park and Rogers[85] further defined two transient moduli, $G_t'(t)$ and $G_t''(t)$. The transient moduli consider the stress response of a three-dimensional space $[(x, y, z)]$, and the position of each point [the position vector $\vec{P}(t)$] can be expressed as follows:

$$\vec{P}(t) = (\gamma(t), \quad \dot{\gamma}(t)/\omega, \quad \sigma(t)) = (x, y, z). \tag{20}$$

Then, three vectors were defined, including the unit tangent ($\vec{T}$), normal ($\vec{N}$), and binormal ($\vec{B}$) vectors,

$$\vec{T}(t) = \vec{P}'(t)/\|\vec{P}'(t)\|, \quad \vec{N}(t) = \vec{T}'(t)/\|\vec{T}'(t)\|, \\ \vec{B}(t) = \vec{T}(t) \times \vec{N}(t), \tag{21}$$

where $\vec{T}$ is the normalized time derivative of $\vec{P}(t)$ and the direction of velocity, $\vec{N}$ denotes tangent to $\vec{T}$ and the direction of acceleration, and $\vec{B}$ represents the cross product of $\vec{T}$ and $\vec{N}$. Then, $\vec{B}(B_\gamma, B_{\dot{\gamma}/\omega}, B_\sigma)$ is obtained, and $G_t'(t)$ as well as $G_t''(t)$ can be defined as follows:

$$G_t'(t) = -B_\gamma/B_\sigma(t), \quad G_t''(t) = -B_{\dot{\gamma}/\omega}(t)/B_\sigma(t). \tag{22}$$

Here, $G_t'(t)$ and $G_t''(t)$ were calculated as partial derivatives of the stress vs strain and strain rate, which demonstrate the transient influence of both the strain and strain rate on the stress separately.

According to Kamani *et al.*[65] and Park and Rogers,[85] the definition of the transient moduli is one of the main differences between SPP and other methods that use one or two-dimensional space introducing FT or Lissajous curves. These works were based on secant values ($\sigma/\gamma$ and $\sigma/\dot{\gamma}$) or total derivatives of the stress response ($d\sigma/d\gamma$ and $d\sigma/d\dot{\gamma}$). For these methods, the total derivatives can be given by the following equations:

$$\frac{d\sigma}{d\gamma} = \frac{\partial\sigma}{\partial\gamma} + \left(\frac{\partial\sigma}{\partial\dot{\gamma}}\right)\left(\frac{d\dot{\gamma}}{d\gamma}\right), \quad \frac{d\sigma}{d\dot{\gamma}} = \frac{\partial\sigma}{\partial\dot{\gamma}} + \left(\frac{\partial\sigma}{\partial\gamma}\right)\left(\frac{d\gamma}{d\dot{\gamma}}\right). \tag{23}$$

Therefore, the additional strain rate $\left(\frac{\partial\sigma}{\partial\dot{\gamma}}\right)\left(\frac{d\dot{\gamma}}{d\gamma}\right)$ and strain contributions $\left(\frac{\partial\sigma}{\partial\gamma}\right)\left(\frac{d\gamma}{d\dot{\gamma}}\right)$ were considered. By contrast, the transient moduli represent the single effect of the strain and strain rate.

By using $G_t'$ and $G_t''$, Rogers *et al.* proposed the phase angle $\delta_t = \arctan(G_t''/G_t')$ and the velocity of phase angle $d\delta_t/dt$.[86–88] The $\delta_t$ close to zero denotes a nearly perfectly elastic behavior. The phase angle near $\pi/2$ represents a liquidlike response. Meanwhile, the peak of $d\delta_t/dt$ is close to the position at which the stress changes rapidly with the strain, which is regarded as yielding and the transitions between the elastic and viscous responses.

## III. MATERIALS AND METHODS

### A. Materials

Carbopol 980 powder was purchased from Lubrizol/Noveon Consumer Specialties, Shanghai, China, and laponite powder (grade LAPONITE RD) was bought from BYK Additives & Instruments Co., Ltd, Shanghai, China. Xanthan gum was provided by CP Kelco, Shanghai, China, whereas Curdlan was obtained from Mitsubishi Corporation Life Sciences, Tokyo, Japan. $\kappa$-Carrageenan ($\kappa$-car) was obtained from Danisco (China) Co., Ltd., Kunshan, China, and CNF was purchased from Tianjin Woodelf Biotechnology Co. Ltd., Tianjin, China. Welan gum was obtained from Zhejiang DSM Zhongken Biotechnology Co., Ltd., Zhejiang, China. The body lotion was bought from Shanghai Wenfeng Hairdressing Beauty Treatment Co., Ltd., Shanghai, China. Ketchup was purchased from Mccormick (Guangzhou) Food Co., Ltd., Guangzhou, China. Sodium hydroxide (NaOH), hydrochloric acid (HCl), liquid paraffin, Span 80, and cyclohexane were acquired from Titan, Shanghai, China. The water used in all experiments was pretreated with the Milli-Q System (Millipore Corporation, Bedford, MA, USA). All samples and chemicals were used without further purification.

### B. Sample preparation

All samples were prepared at room temperature. Carbopol gels and laponite suspensions at different concentrations were prepared with small modifications according to the previously reported works in the literature.[5] Carbopol gels were obtained by mixing carbopol powder and water, which were stirred (1000 rpm, 15 min). Then, the pH of the prepared solutions was adjusted to seven using NaOH. The laponite suspensions were prepared by dispersing laponite powder in water (1000 rpm, 10 min). After that, laponite suspensions were adjusted to pH = 7 by using HCl. The aqueous solutions of xanthan gum were obtained according to a previously reported procedure.[5] Xanthan gum powder was dispersed in water under violent stirring for 12 h. Then, xanthan gum solutions were centrifuged until bubbles were removed to obtain homogeneous solutions. Welan gum solution was prepared the same as xanthan gum. 0.65 wt. % CNF suspension was prepared by diluting the commercial 1.3 wt. % CNF aqueous suspension.

Curdlan/κ-car double network hydrogel particles were prepared according to a previously reported work.[8,9] 2 wt. % curdlan mixture was prepared by adding curdlan powder to water and stirring at 1000 rpm and 50 °C for 12 h. After that, 2 wt. % κ-car powder was dissolved for further stirring (1000 rpm, 12 h). Then, liquid paraffin and span 80 were added to prepare an emulsion (1000 rpm, 4 h). The volume ratio was 25:50:1 (Curdlan/κ-car double hydrogel:liquid paraffin:span 80). Next, the emulsion was heated to 95 °C (cross-linking the curdlan network) and then cooled to room temperature (cross-linking the κ-car network). Subsequently, the emulsion was washed with petroleum until liquid paraffin and span 80 were removed. The prepared curdlan/κ-car hydrogel particles were freeze-dried. Finally, the dried curdlan/κ-car particles were rehydrated for 12 h to prepare suspensions (particle size: ∼8 μm) with different concentrations for rheology tests, which was proved to be sufficient to make the rheological properties of the suspensions reproducible.

### C. Rheological measurement

All tests were executed on a stress-controlled rotational rheometer (HaakeMars III, Thermo Fisher, Germany) using parallel plates (diameter: 60 mm, gap: 0.5 mm) sealed with low-viscosity silicone oil (50 mPa s) at 25 °C. Waterproof sandpaper (3 M, particle size: ∼20 μm) was also used to mitigate the impact of wall slip. Oscillatory stress amplitude sweeps were carried out with different stress ranges and frequencies. Twenty cycles were applied for major experiments, and the raw torque and displacement were recorded as 512 points per cycle. As far as the tests at different frequencies are concerned, the rest time depends on the frequency to ensure an equivalent test time. The pre-oscillatory-shear was set with a strain of 5 for 5 min and then rested. In addition, the rest time was set enough for the sample to reach equilibrium (the change of $G'$ was within ±5% in 5 h).[9]

Thixotropy experiments were executed by setting a shear rate sweep from 0.01 to 1000 s⁻¹, followed by a sweep from 1000 to 0.01 s⁻¹ at a fixed duration of 50 s. The rest time for each sample was equal to that of the corresponding oscillatory stress amplitude sweep.

## IV. RESULTS AND DISCUSSION

### A. Algebraic stress bifurcation (ASB)

The concept and measurement of yield stress is still debatable.[2,3] It has also been denoted that the yield stress evaluation varies with the test protocols, whereas deviations arise among the obtained yield stress values.[2] The stress bifurcation method[7] is based on the general geometric average method[70] to provide the start (i.e., the maximum point to sustain solidlike behavior) and end (i.e., the minimum point to trigger a complete solid–liquid transition in a cycle) yield points. Therefore, stress bifurcation takes into account the timescale, which is necessary because many processes happen within a short timescale (e.g., the swallowing process[5]).

Unlike the above-mentioned stress bifurcation method, in the introduced algebraic stress bifurcation (ASB), the raw data are not required to be treated. Figure 2(a) shows the underlying idea of the proposed ASB strategy.

The establishment of the developed ASB is based on FT rheology (Sec. II B), stress decomposition (Sec. II C), and stress bifurcation (Sec. II D). For a truly elastic material, the maximal strain $\gamma_{0max}$ obtained at the maximal stress $\sigma_{max}$ is equal to $\gamma_0$. On the contrary, for a truly viscous material, $\epsilon_{max}$ is acquired at $\gamma = 0$.[53] This is nicely visualized in Fig. 2(a) with two idealized material responses under an oscillatory shear, one perfectly elastic response [i.e., $\sigma'(t) = G'\gamma_0 \sin \omega t$ and $\gamma(t) = \gamma_0 \sin \omega t$] and one perfectly viscous response [i.e., $\sigma''(t) = G''\gamma_0 \sin(\omega t + 0.5\pi) = G''\gamma_0 \cos \omega t$ and $\gamma(t) = \gamma_0 \sin \omega t$].[42,53] Then, by superposing the pure elastic and pure viscous curves in Fig. 2(a), alternative Lissajous curves can be represented using the following Eqs. (24) and (25), respectively:

$$\sigma(t) = \gamma_0(G' \sin \omega t + G'' \cos \omega t), \qquad (24)$$

$$\gamma(t) = \gamma_0 \sin \omega t, \quad \dot{\gamma}(t) = \omega \gamma_0 \cos \omega t. \qquad (25)$$

The algebraic elastic and viscous Lissajous curves were plotted in Figs. 2(b) and 2(c), respectively. The six specific points denoted in Figs. 2(b) and 2(c) can also be determined in the following way.

First, the algebraic elastic and viscous Lissajous curves, $\sigma$–$\gamma$ and $\sigma$–$\dot{\gamma}/\omega$, can be represented using the following equations, respectively:

$$(G'^2 + G''^2)\gamma^2 - 2\sigma G'\gamma + \sigma^2 - G''^2\gamma_0^2 = 0, \qquad (26)$$

$$(G'^2 + G''^2)\dot{\gamma}^2 - 2\sigma G''\dot{\gamma}\omega + \sigma^2\omega^2 - G'^2\gamma_0^2\omega^2 = 0. \qquad (27)$$

After that, the coordinates of the maximum stresses $(\gamma_{\sigma max}, \sigma_{max})$ in Fig. 2(b) and $(\dot{\gamma}_{\sigma max}, \sigma_{max})$ in Fig. 2(c), maximum strain $(\gamma_0, \sigma_{\gamma 0})$ in Fig. 2(b), and maximum strain rate $(\dot{\gamma}_0, \sigma_{\dot\gamma 0})$ in Fig. 2(c) in the algebraic elastic and viscous Lissajous curves can be deduced $(\sigma_{max} = \gamma_0\sqrt{G'^2 + G''^2})$, respectively, as follows:

$$\gamma_{\sigma max} = \frac{\gamma_0}{\sqrt{\left(\frac{G''}{G'}\right)^2 + 1}}, \quad \dot{\gamma}_{\sigma max} = \frac{\omega\gamma_0}{\sqrt{\left(\frac{G'}{G''}\right)^2 + 1}}; \qquad (28)$$

$$\sigma_{\gamma 0} = \sigma(\gamma_0) = G'\gamma_0, \quad \sigma_{\dot\gamma 0} = \sigma(\dot{\gamma}_0/\omega) = G''\dot{\gamma}_0/\omega = G''\gamma_0. \qquad (29)$$

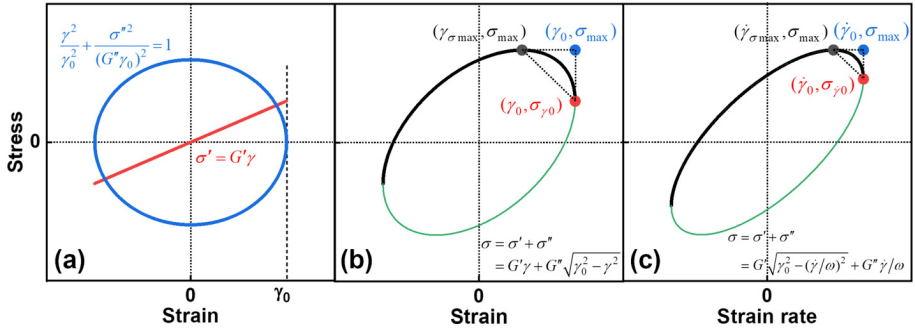

FIG. 2. Schematic illustration of the algebraic stress bifurcation (ASB) principle: (a) rheological responses of a perfectly elastic material (red curve) and a purely viscous material (blue curve); algebraic (b) elastic and (c) viscous Lissajous curves.

As far as the coordinates of $(\gamma_0, \sigma_{\gamma 0})$ and $(\dot{\gamma}_0, \sigma_{\dot{\gamma} 0})$ are concerned, they can be represented as $(\gamma_0, G'\gamma_0)$ and $(\dot{\gamma}_0, G''\gamma_0)$ according to Eq. (29).

As a result, based on the proposed ASB approach, the three points $(\bar{\gamma}_{\sigma max}, \sigma_{max}), (\gamma_{\sigma max}, \bar{\sigma}_{max}), (\gamma_{\sigma max}, \sigma_{max})$ in Fig. 1(a) and the three points $(\bar{\dot{\gamma}}_{\sigma max}, \sigma_{max}), (\dot{\gamma}_{\sigma max}, \bar{\sigma}_{max}), (\dot{\gamma}_{\sigma max}, \sigma_{max})$ in Fig. 1(b), which were obtained by the geometric average method using raw data,[20] can be alternatively calculated from the stress/strain sweep data $(G', G'')$ as follows:

$$\left( \gamma_{\sigma max} \left( = \frac{\gamma_0}{\sqrt{\left(\frac{G''}{G'}\right)^2 + 1}} \right), \sigma_{max} \right),$$
$$(\gamma_0, \sigma_{\gamma 0}(= G'\gamma_0)), \quad (\gamma_0, \sigma_{max}), \qquad (30)$$

$$\left( \dot{\gamma}_{\sigma max} \left( = \frac{\omega\gamma_0}{\sqrt{\left(\frac{G'}{G''}\right)^2 + 1}} \right), \sigma_{max} \right),$$
$$(\omega\gamma_0, \sigma_{\dot{\gamma} 0}(= G''\gamma_0)), \quad (\omega\gamma_0, \sigma_{max}). \qquad (31)$$

Figures 2(a-ii) and 2(a-iii) describe the three points of $(\gamma_{\sigma max}, \sigma_{max}), (\gamma_0, \sigma_{\gamma 0})$, and $(\gamma_0, \sigma_{max})$, as well as the three points of $(\dot{\gamma}_{\sigma max}, \sigma_{max}), (\dot{\gamma}_0, \sigma_{\dot{\gamma} 0})$, and $(\dot{\gamma}_0, \sigma_{max})$, respectively.

Briefly, the first step of ASB is to establish the elastic and viscous $I_1$-based curves [Figs. 2(b) and 2(c)] by using sweep data of $G'$ and $G''$ [Eqs. (30) and (31)]. Afterward, the above-mentioned six points [Figs. 2(b) and 2(c), Eqs. (30) and (31)] can be directly obtained from the functions of the two Lissajous curves by carrying out mathematical calculations. Therefore, it is clear that the essential difference between ASB and stress bifurcation is that the approximate coordinates of the above-mentioned six specific points can be deduced by ASB from the dynamic stress/strain sweep (i.e., $G'$ and $G''$ vs stress/strain amplitude), whereas the stress bifurcation requires the processing of raw data (i.e., Lissajous curves shown in Fig. 1).

The reason for stress bifurcation based on the general geometric average method has been described by Yu *et al.*[9] as those statements in Sec. II D. For the proposed ASB approach, the physical reasons for the stress bifurcation can also be rationalized from Eqs. (32) and (33). Moreover, Fig. 3 shows the reason for bifurcation.

For the $\sigma-\gamma$ curve [Fig. 3(a)], the horizontal ($D_h$) and vertical ($D_v$) differences between the points of $(\gamma_{\sigma max}, \sigma_{max})$ and $(\gamma_0, \sigma_{\gamma 0})$ can be expressed as follows:

$$D_h = \left( \gamma_0 \Big/ \sqrt{(G''/G')^2 + 1} \right) \Big/ \gamma_0 = \frac{1}{\sqrt{\left(\frac{G''}{G'}\right)^2 + 1}},$$
$$D_v = G'\gamma_0/\sigma_{max} = \frac{1}{\sqrt{\left(\frac{G''}{G'}\right)^2 + 1}}. \qquad (32)$$

It is apparent that the $D_h$ and $D_v$ parameters between $(\gamma_{\sigma max}, \sigma_{max})$ and $(\gamma_0, \sigma_{\gamma 0})$ are close to the value of one for solidlike materials and zero for liquidlike materials, representing that the points of $(\gamma_{\sigma max}, \sigma_{max})$ and $(\gamma_0, \sigma_{\gamma 0})$ show good superposition and significant differences, respectively.

Similarly, for the $\sigma-\dot{\gamma}$ curve [Fig. 3(b)], the $D_h$ and $D_v$ between the points of $(\dot{\gamma}_{\sigma max}, \sigma_{max})$ and $(\dot{\gamma}_0, \sigma_{\dot{\gamma} 0})$ can be estimated as follows:

$$D_h = D_v = \frac{1}{\sqrt{\left(\frac{G'}{G''}\right)^2 + 1}}. \qquad (33)$$

In the same way, the $D_h$ and $D_v$ parameters between $(\dot{\gamma}_{\sigma max}, \sigma_{max})$ and $(\dot{\gamma}_0, \sigma_{\dot{\gamma} 0})$ are close to the value of zero for solidlike materials and one for liquidlike materials, indicating that the points of $(\dot{\gamma}_{\sigma max}, \sigma_{max})$ and $(\dot{\gamma}_0, \sigma_{\dot{\gamma} 0})$ show significant differences and good superposition, respectively.

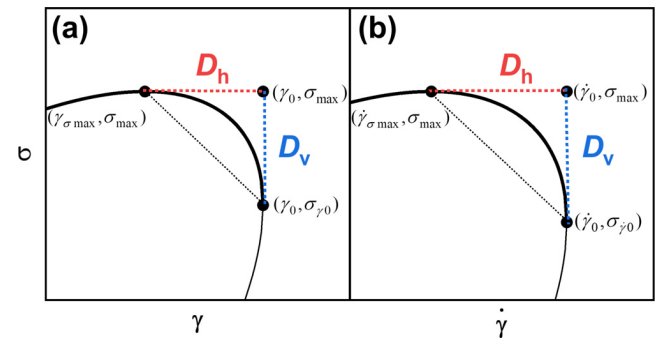

FIG. 3. Depiction of the horizontal ($D_h$) and vertical ($D_v$) differences between the points of $(\gamma_{\sigma max}, \sigma_{max})$ and $(\gamma_0, \sigma_{\gamma 0})$ in the algebraic elastic Lissajous curve (a) and the points of $(\dot{\gamma}_{\sigma max}, \sigma_{max})$ and $(\dot{\gamma}_0, \sigma_{\dot{\gamma} 0})$ in the algebraic viscous Lissajous curve (b).

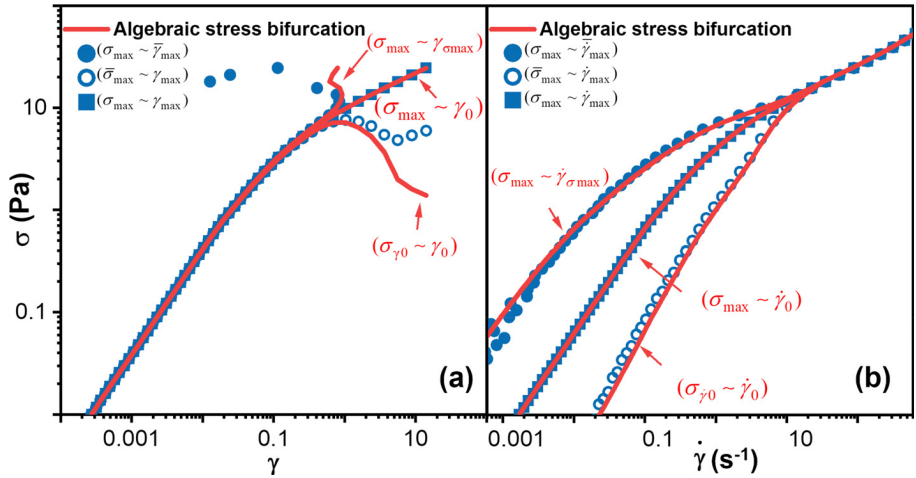

It should be also noted that in practical applications of ASB, the construction of $I_1$-based curves [Figs. 2(b) and 2(c)] is needless because the relationship between the data of a sweep plot ($G'$, $G''$) and the six points of ASB has been deduced from the $I_1$-based curves based on Eqs. (30) and (31). Thus, ASB gives a more facile and efficient approach to determining the crucial six points.

Accordingly, three $\sigma-\gamma$ curves ($\sigma_{max}-\gamma_{\sigma max}$, $\sigma_{\gamma 0}-\gamma_0$, and $\sigma_{max}-\gamma_0$) and three $\sigma-\dot{\gamma}$ curves ($\sigma_{max}-\dot{\gamma}_{\sigma max}$, $\sigma_{\gamma 0}-\dot{\gamma}_0$, and $\sigma_{max}-\dot{\gamma}_0$) can be obtained based on the analyses of the whole sweep points. Thus, the two bifurcation points can be readily recognized and determined from these curves, as can be ascertained from Figs. 4(a) and 4(b) (red lines). As can be seen, good agreement was achieved compared to those obtained by stress bifurcation.[9] On top of that, as shown in Fig. 4, except for the data in large deformation in the $\sigma-\gamma$ curves [Fig. 4(a)], a perfect overlap was recorded for the $\sigma \sim \gamma$ [Fig. 4(a)] and $\sigma \sim \dot{\gamma}$ [Fig. 4(b)] curves. The proposed ASB evaluates the proximity between $G'$ and $G''$. In other words, a measure to judge the relative change of $G'$ and $G''$ is offered. Accordingly, the start and end of the solid–liquid transition can be achieved when $G'$ is close to and away from $G''$ during the stress sweep, respectively. In more detail, the ASB method can appraise the start and end points of the solid–liquid transition by observing two bifurcation points (thereby solid–liquid transition region is determined). As seen in Fig. 4, the start and the end of the solid–liquid transition determined by the proposed ASB method are almost the same as those by the stress bifurcation.[9] A detailed discussion about the features of $\sigma \sim \gamma$ ($\sigma_{max}-\gamma_{\sigma max}$, $\sigma_{\gamma 0}-\gamma_0$, and $\sigma_{max}-\gamma_0$) and the $\sigma-\dot{\gamma}$ curves ($\sigma_{max}-\dot{\gamma}_{\sigma max}$, $\sigma_{\gamma 0}-\dot{\gamma}_0$, and $\sigma_{max}-\dot{\gamma}_0$) will be further presented in Sec. IV B. The excellent consistency of the extracted results from ASB with stress bifurcation implies the rationality and feasibility of this approach. The detailed reasons will be discussed in Sec. IV C.

The proposed ASB method allows the judgment of start and end yield points based only on the knowledge of $G'$, $G''$, and stress amplitude, which significantly minimizes the complexity of the stress bifurcation technique without requiring the processing of raw data. According to Eqs. (30) and (31), this method can significantly simplify data processing, allowing a facile and efficient application of less specialized software packages, such as Origin® and Excel, to treat rheological data, as compared with the professional software of MATLAB.

However, the ASB method lacks the information of higher harmonics, which limits itself to be applied for LAOS analysis. Therefore, the ASB method is typically designed as a kind of yield stress determination method (i.e., simply obtaining the start and end yield points from LAOS experiments).

## B. Verification from typical YSFs

In this section, ASB was verified by performing experiments on four kinds of representative samples, including carbopol gel, laponite suspension, xanthan gum solution, and hydrogel particle suspension. The thixotropy properties of the four samples were first investigated (Appendix A provides further detailed information), and the values of $\sigma_y$ are presented in Table I. Then, stress-controlled LAOS experiments were carried out on the samples of carbopol gels, laponite suspensions, xanthan gum solutions, and hydrogel particle suspensions at different concentrations and frequencies (The results were shown and discussed in Appendix B). These obtained LAOStress results correspond well with those in the literature.[56–60,90] After that, these results were analyzed by stress bifurcation and ASB methods, which were plotted and investigated in Appendix C. Based on Appendixes A–C, the high similarity between the results of ASB and stress bifurcation can be observed, and the validity of ASB method is thus further verified.

**TABLE I.** Comparison of the yield stress values for 0.1 wt. % carbopol gels, 1.25 wt. % laponite suspensions, 1.25 wt. % xanthan gum solutions, 0.65 wt. % CNF suspensions, 0.4 wt. % welan gum solutions, 100 wt. % body lotion emulsions, and 100 wt. % ketchup obtained by different measurement methods.

| | | Yield stress (Pa) | | | | |
|---|---|---|---|---|---|---|
| | HB | Creep | Oscillatory shear | | | LAOS |
| Sample | $\sigma_{0,H}$ | $\sigma_c$ | $\sigma_{d,e}$ | $\sigma_{d,p}$ | $\sigma_{d,c}$ | $\sigma_{l,s}$ | $\sigma_{l,e}$ |
| Carbopol gel | 6.3 | 7.8–9.2 | 6.6 | 8.3 | 9.2 | 6.7 | 12.5 |
| Laponite | 0.5 | 3.8–4.5 | 4.3 | 4.4 | 4.8 | 4.5 | 5.2 |
| Xanthan gum | 7.2 | 12.4–14.4 | 21.7 | 25.4 | 27.1 | 23.1 | 31.5 |
| CNF | 1.6 | 5.4–6.3 | 3.7 | 6.0 | 4.6 | 2.1 | 9.2 |
| Welan gum | 3.8 | 4.6–5.4 | 8.2 | 8.4 | 9.8 | 7.9 | 10.6 |
| Body lotion | 3.1 | 34.0–39.0 | 26.6 | 50.0 | 37.8 | 24.5 | 53.8 |
| Ketchup | 40.9 | 76–82 | 68.0 | 73.4 | 90.3 | 65.0 | 104.5 |

### C. Interpretation of the good overlaps between results of two methods based on FT rheology

ASB was verified by FT rheology as shown in Figs. 5–7. The results of LAOS tests on 0.1 wt. % carbopol gel ($G'$, $G''$, stress amplitude, and Lissajous curves) were selected as the representative data to analyze the rheology behavior. Dynamic moduli vs stress are displayed in Fig. 5(a). The solid–liquid transition region of 0.1 wt. % carbopol gel determined by ASB was in the strain range of 0.35–2.4.

The $x$–$y$ projection FT fingerprint was also used to show the nonlinearities for each higher harmonic by plotting stress vs $n$th harmonics [Fig. 5(b)], which demonstrated the normalized harmonic intensities ($I_n/I_1$) ($I_1$ is the first harmonic magnitude) and provided a visual "snapshot" of the onset and growth of harmonics as stress increases.[91] In addition, other representative FT rheology results are shown in Fig. 5, including the intensities of $I_3/I_1$, $I_5/I_1$, and $I_7/I_1$ changed with stress [Fig. 5(c)] and the results of the FT rheology from the frequency domain at the strain amplitudes of 0.35, 2.4, and 5.6 [Fig. 5(d)]. In Appendix D, more discussion of the FT results was provided.

The main mathematical difference between the stress bifurcation and ASB is that the former considers the whole Fourier series ($I_1$, $I_2$, $I_3$, $\cdots$, equal to the raw data of Lissajous curves), while the latter only introduces a part of the Fourier series ($I_1$, $G'$ and $G''$). Hence, the good overlaps between the results from the stress bifurcation and the ASB methods can be explained; in other words, the reason why higher harmonics can be not considered. The raw data (Lissajous curves from rheometer) and the corresponding $I_1$-based curves from ASB are plotted together in Figs. 6(a) and 6(b), including the strain amplitudes of 0.15, 0.35, 0.69, 1.6, 2.4, and 5.6. Figure 6(a-i) illustrates the elastic Lissajous curves consisting of the raw data at 0.15 strain and the corresponding algebraic curve [$\sigma(\gamma) = \sigma'(\gamma) + \sigma''(\gamma)$] by combining a purely elastic response ($\sigma'(\gamma) = G'\gamma$) and a purely viscous response ($\sigma''(\gamma) = \pm G''(\gamma_0^2 - \gamma^2)^{0.5}$), where $G' = 25$ Pa, $G'' = 6.3$ Pa, and $\gamma_0 = \sigma_{max}/\sqrt{G'^2 + G''^2} = 0.15$ were obtained from the rheometer software without any additional treating. Similarly, the situations of other strain amplitudes are displayed in Figs. 6(a-ii)–6(a-vi) with the used parameters.

Figure 6(a) shows that the raw data and $I_1$-based curves correspond well at relatively low stress amplitudes, where deviations arise with the increase in stress. However, although deviations occur in the second and fourth quadrants, the curves show good overlaps in the first and third quadrants that are the quadrants of concern for stress bifurcation and ASB. Therefore, in Fig. 6(a), the small differences between raw data and algebraic elastic Lissajous curves from ASB were visually demonstrated. Furthermore, the comparisons were also made by plotting viscous Lissajous curves in Fig. 6(b), which yielded the same conclusions as Fig. 6(a).

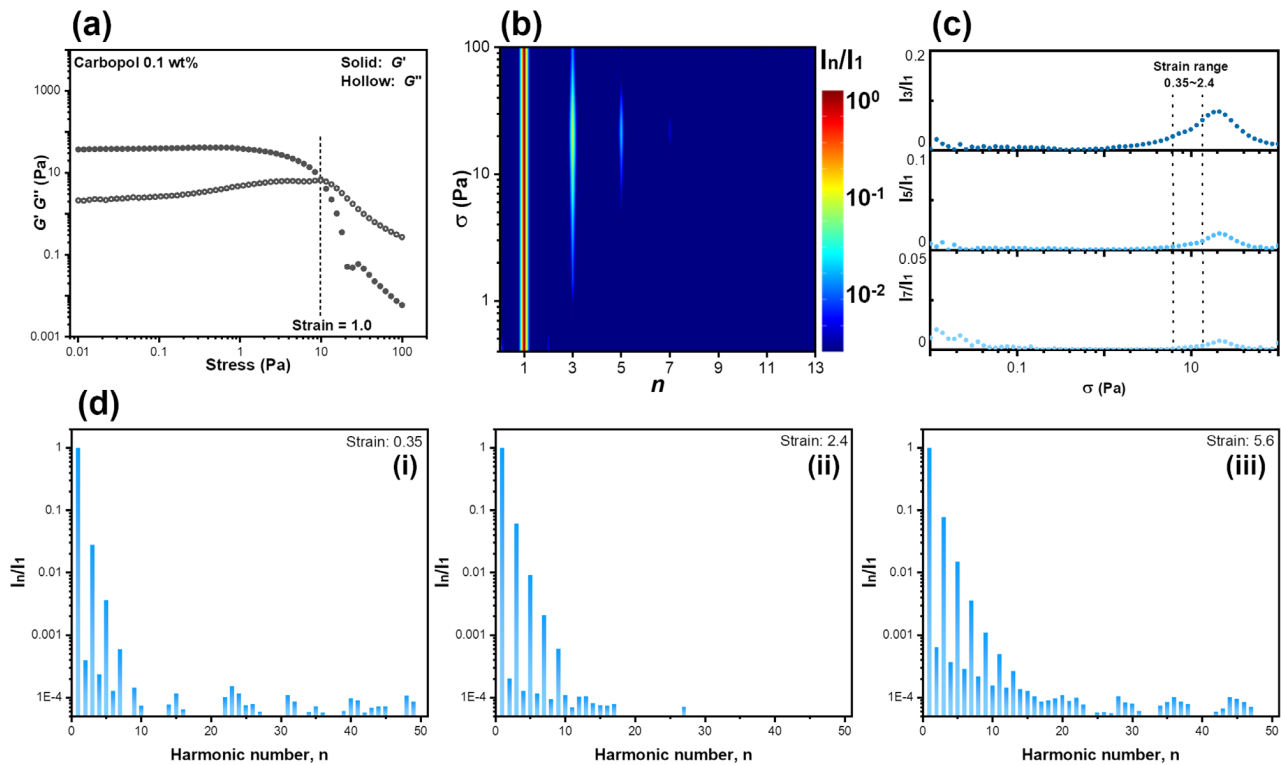

**FIG. 5.** FT analysis for the stress sweep test results of 0.1 wt. % carbopol gel at 1 Hz (the first step of the mechanistic explanation of ASB based on the FT analysis): (a) dynamic moduli of the amplitude sweep test; (b) stress-frequency projection ($\sigma$ vs ($n$-$\omega$) plot); (c) normalized $I_3/I_1$, $I_5/I_1$, and $I_7/I_1$ as a function of the stress amplitude; (d) the FT spectra of strain amplitude at (i) 0.35, (ii) 2.4, and (iii) 5.6 showing $I_n/I_1$ as a function of $n$th harmonic. Solid–liquid transition region for 0.1 wt. % carbopol gel at 1 Hz in the strain of 0.35–2.4.

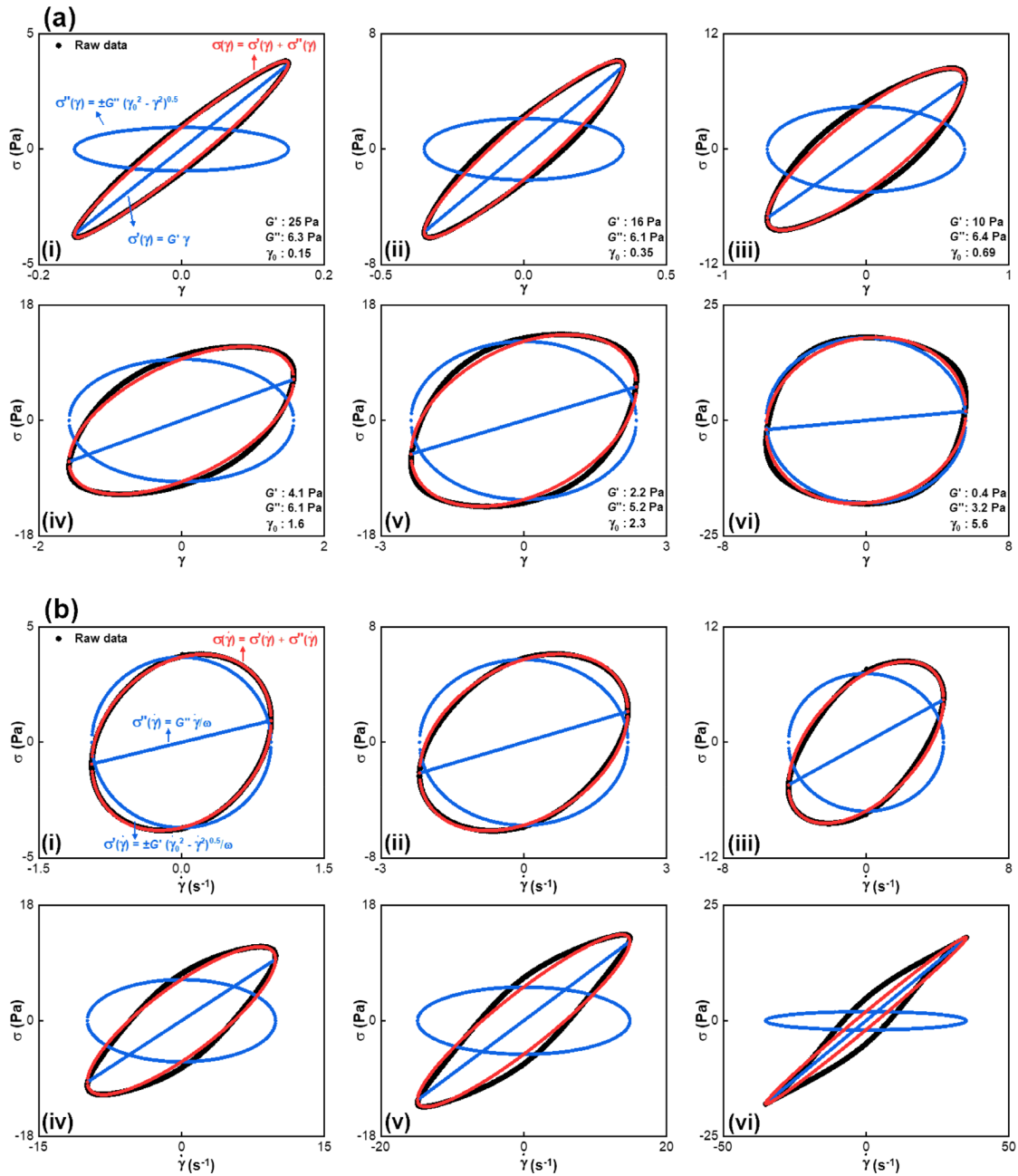

**FIG. 6.** The contributions of the 1st harmonic to the base of ASB: $I_1$-based curves (the second step of the mechanistic explanation of ASB based on FT analysis, data: from the results displayed in Fig. 5): (a) the comparisons between raw data and the corresponding algebraic elastic Lissajous curves at different strain amplitudes: (i) 0.15, (ii) 0.35, (iii) 0.69, (iv) 1.6, (v) 2.4, and (vi) 5.6 and (b) the comparisons between raw data and the corresponding $I_1$-based curves.

The highest harmonic of higher harmonics, $I_3$, was also introduced to interpret the good overlaps between stress bifurcation and ASB. The situations by considering $I_3$ are as follows:

$$\sigma(t) = \gamma_0(G' \sin \omega t + G'' \cos \omega t + G'_3 \sin 3\omega t + G''_3 \cos 3\omega t), \quad (34)$$

$$\gamma(t) = \gamma_0 \sin \omega t, \quad \dot{\gamma}(t) = \gamma_0 \omega \cos \omega t, \quad (35)$$

$$\sigma_{\gamma 0} = \sigma(t)|_{t=\frac{\pi}{2\omega}} = \gamma_0(G' - G'_3), \quad (36)$$

$$\sigma_{\dot{\gamma} 0} = \sigma(t)|_{t=0} = \gamma_0(G'' + G''_3). \quad (37)$$

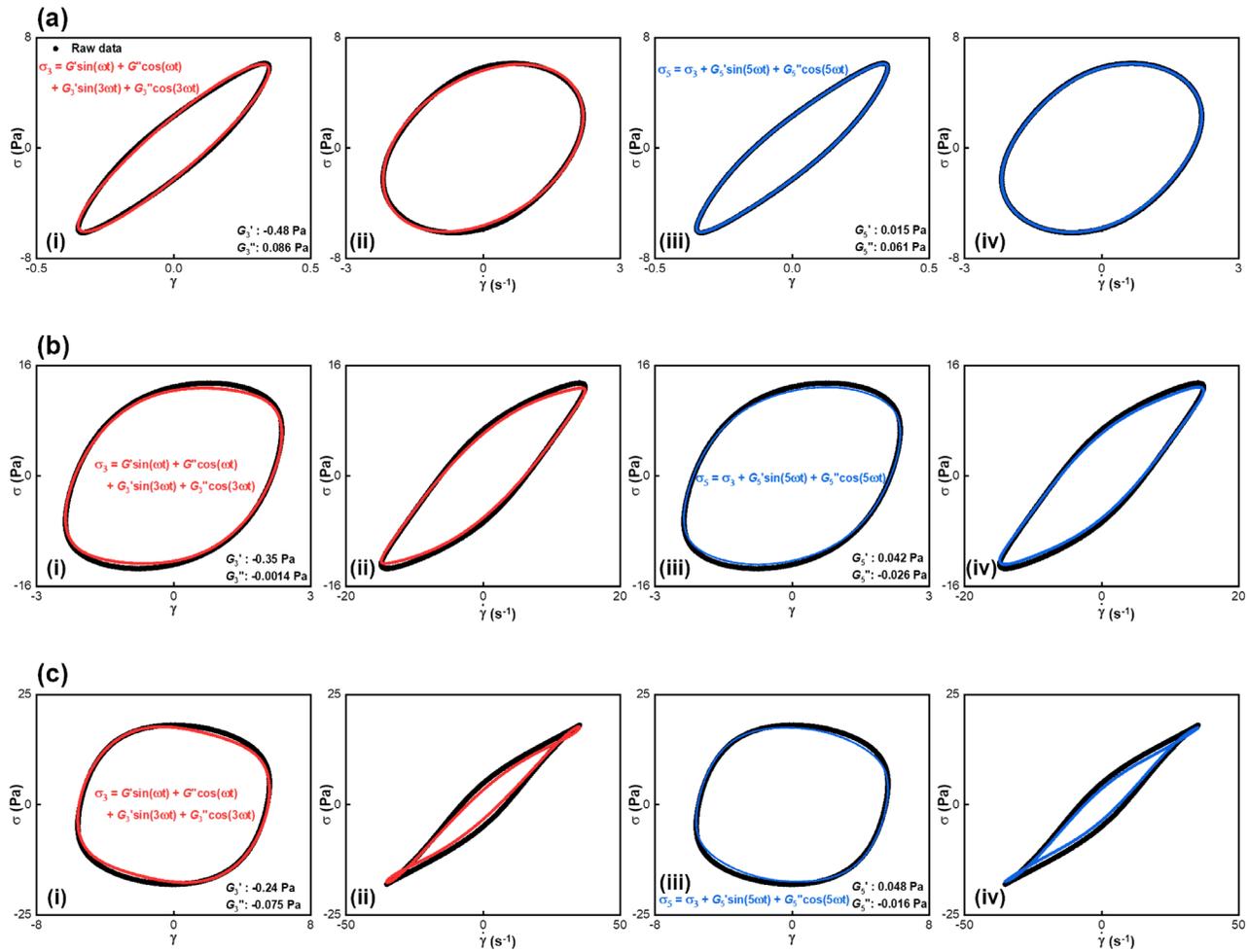

**FIG. 7.** Impact of third and fifth harmonics on $I_1$-based curves (the last step of the mechanistic explanation of ASB based on FT analysis, data: from the results displayed in Fig. 5): (a) the $I_1$-based curves at the strain amplitude of 0.35 considering $I_3$ and $I_5$: the $I_1$-based (i) elastic and (ii) viscous curves by introducing third harmonic; the $I_1$-based (iii) elastic and (iv) viscous curves by introducing third and fifth harmonics; the $I_1$-based elastic and viscous curves at the strain amplitude of (b) 2.4 and (c) 5.6.

In addition, $\gamma_{\sigma\text{max}}$ can be obtained by solving the following equation:

$$\frac{1}{\gamma_0\omega}\frac{d\sigma(x)}{dx} = (G'\sqrt{1-x^2} - G''x + 3G'_3\sqrt{1-(3x-4x^3)^2} - 3G''_3(3x-4x^3) = 0, \quad (38)$$

where $x = \sin\omega t, x \in [0,1]$, $\gamma_{\sigma\text{max}} = \gamma_0 x$, and $\dot{\gamma}_{\sigma\text{max}} = \omega\sqrt{\gamma_0 - \gamma_{\sigma\text{max}}^2}$. From Eq. (38), finding the analytical solution seems difficult and thus the numerical solution can be involved. In addition, by using interpolation, this question can be solved to obtain the analytical solution of $\gamma_{\sigma\text{max}}$ by the FT of the distorted strain,

$$\gamma(t) = \gamma_0(b_1\sin\omega t + a_1\cos\omega t + b_3\sin 3\omega t + a_3\cos 3\omega t), \quad (39)$$

$$\sigma(t) = \sigma_{\text{max}}\sin\omega t, \quad (40)$$

$$\gamma_{\sigma\text{max}} = \gamma_0(b_1 - b_3), \quad (41)$$

whereas in Sec. II B, detailed information was provided.

For the strain amplitude of 0.35 (the start yield point), considering $I_1$ and $I_3$ [$G' = 16$, $G'' = 6.1$, $G'_3 = -0.48$, $G''_3 = 0.086$ Pa, $\gamma_0 = 1$ (assumed for simplicity)], the values of $\sigma_{\gamma 0}$, $\gamma_{\sigma\text{max}}$, $\sigma_{\dot{\gamma}0}$, and $\dot{\gamma}_{\sigma\text{max}}$ were 16.48 Pa, 0.9534, 6.186 Pa, and 1.896. If only $I_1$ was considered, these values were 16, 0.9344, 6, and 2.238 with corresponding errors of 3%, 2%, 3%, and 15%, respectively. Note that the magnitude of the error in $\dot{\gamma}_{\sigma\text{max}}$ does not influence the judgment of the start yield point. As far as the strain amplitude of 2.4 is concerned (the end yield point, $G' = 2.2$ Pa, $G'' = 5.2$, $G'_3 = -0.35$, $G''_3 = -0.0014$ Pa, $\gamma_0 = 1$), the values of $\sigma_{\gamma 0}$, $\gamma_{\sigma\text{max}}$, $\sigma_{\dot{\gamma}0}$, and $\dot{\gamma}_{\sigma\text{max}}$ were 2.55 Pa, 0.4079, 5.1986 Pa, and 5.737. These were 2.2 Pa, 0.3896, 5.2 Pa, and 5.787 by only considering $I_1$, where the corresponding errors were 16% ($\sigma_{\gamma 0}$ does not influence the judgment of the end yield point), 5%, 0.03%, and −9%. Although the introduction of third harmonics had no obvious impact on the results of ASB, the higher harmonics are still important [Figs. 7(a)–7(c)] since they can help to revise the point coordinates and adjust $I_1$-based curves closer to raw data.

Therefore, it is not surprising that the results of the stress bifurcation (considering raw Lissajous curves) and ASB (introducing $G'$, $G''$, and stress amplitude) are in good agreement. In other words, since the raw data consists of infinite Fourier series, only $I_1$ is needed to be considered for stress bifurcation, which is the ASB. More importantly, although some errors in the results from stress bifurcation and ASB were observed, the same conclusion can be reached. After all, in most cases, an error in point coordinates has negligible influence on determining whether two points overlap or separate, while the error was only ~2% by introducing $I_3$ in a real situation. In short, small deviations were recorded between the results of ASB and the values obtained by involving third harmonics than possesses the highest intensity within the higher harmonics. The good overlaps between the results of ASB and stress bifurcation were, thus, interpreted. In other words, for ASB, the application of only $G'$, $G''$, and the stress/strain amplitude instead of raw Lissajous loops is feasible and rational, where higher harmonics and phenomenological Lissajous curves are unnecessary in the determination of the yield stress for this method.

### D. Verification from models

All YSFs present stress bifurcation in LAOS experiments. To better understand such phenomena, the verification of ASB is further carried out by introducing models, in other words, how the ASB works on models with clear physical interpretations. The verification of the ASB method from the perspective of the rheological constitutive model[9,92] was further carried out in this section. Before the verification, representative results were generated in Figs. 8(a) and 8(b), which correspond well with those in the literature by setting the same parameters.[92]

#### 1. KVHB model

Based on the original Bingham model, the modified Bingham model[68,70,71] describes a fluid with yield stress by combining Hookean solid and steady shear flow. Then, the Hookean model was replaced by the Kelvin–Voigt model, and the power law model under steady shear was adopted to present the Kelvin–Voigt–Herschel–Bulkley (KVHB) model.[9] By combining the Kelvin–Voigt model and the Herschel–Bulkley model, the KVHB model is expressed as

$$\dot{\gamma}(t) = \frac{1}{G}\frac{d\sigma(t)}{dt} - \lambda\ddot{\gamma}(t) + U(\sigma(t)-\sigma_y)\left(\frac{\sigma(t)-\sigma_y}{K}\right)^{1/n}$$
$$- U(-\sigma(t)-\sigma_y)\left(\frac{\sigma(t)+\sigma_y}{K}\right)^{1/n}, \quad (42)$$

where $G$ denotes modulus, $\lambda$ is the relaxation time, $\sigma_y$ refers to yield stress, $U(x) = 0$ and 1 when $x$ is lower and higher than 0, respectively, $K$ is the consistency index, and $n$ is the flow index. For $|\sigma| \leq \sigma_y$, $\sigma(t) = G\dot{\gamma}(t) + \eta\dot{\gamma}(t)$, i.e., the Kelvin–Voigt model, while for $|\sigma| > \sigma_y$, the KVHB model becomes the Herschel–Bulkley model ($\sigma(t) = \sigma_y + K\dot{\gamma}^n$).

#### 2. Saramito model

The Saramito model[92] is a three-dimensional model for elastoviscoplastic fluid flows, which combines the Bingham viscoplastic model and the Oldroyd viscoelastic model.[93] The Saramito model can deduce fundamental flows: uniaxial elongation, simple shear flow, and large amplitude oscillatory shear, showing the elastoviscoplastic behavior of a viscoelastic solid at low stress and a fluid at stresses above yield stress. The Saramito model is given by

$$\frac{\eta_m}{G}\frac{d\sigma}{dt} + \max\left(0, \frac{|\sigma|-\sigma_y}{|\sigma|}\right)\sigma = \eta_m\dot{\gamma}, \quad (43)$$

where $\eta_m$ is microscopic viscosity, $\sigma$ represents the stress tensor, and $\dot{\gamma}$ denotes the deformation rate tensor. When $|\sigma| \leq \sigma_y$, the Kelvin–Voigt model is obtained while for $|\sigma| > \sigma_y$ the model reduces to the Oldroyd model. Then, by further introducing $D = \dot{\gamma}/2 = (\nabla v + \nabla v^{\mathsf{T}})/2$, $W = (\nabla v - \nabla v^{\mathsf{T}})/2$ (the vorticity tensor), $We = \lambda U/L$ (dimensionless Weissenberg number, where $U$ and $L$ are some characteristic velocity and length of the flow, respectively), $Bi = \tau_0 L/\eta_0 U$ (dimensionless Bingham number, where $\eta_0$ is the total viscosity), $\sigma_d = \sigma - \frac{1}{3}\text{tr}(\sigma)I$, and $\xi \geq 0$ (a material parameter), Eq. (43) can be written as

$$We\frac{d\sigma}{dt} + (1 + \xi\text{tr}\sigma)\max\left(0, \frac{|\sigma_d|-Bi}{|\sigma_d|}\right)\sigma = 2\frac{\eta_m}{\eta_0}D, \quad (44)$$

where the Gordon–Schowalter derivative $\frac{d\sigma}{dt} = \frac{\partial\sigma}{\partial t} + v \cdot \nabla\sigma + \sigma \cdot W - W \cdot \sigma - a(\sigma \cdot D + D \cdot \sigma)$ ($a=0$: the Jaumann derivative of tensor; $a=1$: the upper-convected derivative; and $a=-1$: the lower-convected derivative). The total stress is given by $\sigma$ plus the term $-pI + 2\eta D$ (i.e., the contribution from the viscous body).

#### 3. Giesekus model

The Giesekus model[94] describes the total stress signal of a fluid as the summation of the solvent stress contribution ($\eta_s\dot{\gamma}$) and the polymeric stress contribution ($\sigma_p$).[74,95] The dimensionless nonlinearity parameter $\alpha$ is in the range of $0 \sim 1$, where $\alpha = 0$ and 1 indicate no and maximal flow orientation coupling, respectively.[96] The Giesekus model is given by ($a=1$),

$$\sigma = \eta_s\dot{\gamma} + \sigma_p, \quad (45)$$

$$\sigma_p + \lambda\frac{d\sigma_p}{dt} + \alpha\frac{\lambda}{\eta_p}\sigma_p \cdot \sigma_p = 2GD, \quad (46)$$

where $\sigma_p$ is the polymer stress tensor, $\eta_s$ is the solvent viscosity, $G$ is the polymer shear modulus, and $\alpha$ is the mobility factor.

#### 4. Raw and $I_1$-based curves from the above-mentioned models and FT rheology

As shown in Figs. 8(c)–8(e), the KVHB model, Saramito model, and Giesekus model were used to generate representative sequential Lissajous curves (raw Lissajous curves, blue lines). After that, these data were treated by FT to give the values of $G'$ and $G''$. Finally, $I_1$-based curves (red lines) were constructed based on $G'$ and $G''$.

Meanwhile, in Fig. 8(f), FT rheology was applied to deduce supernon-linear Lissajous curves. The raw Lissajous curves from the 0.1 wt. % carbopol gel were processed by FT to obtain Fourier coefficients. Then, the values of $G'$ and $G''$ were not changed, and the intensities of all higher harmonics were amplified three times ($A=3$). Therefore, the nonlinearity of the 0.1 wt. % carbopol gel was greatly promoted to give supernon-linear Lissajous curves (max($I_3/I_1$) ≈ 0.3). As a result, in Fig. 8, different kinds of Lissajous curves were successfully constructed to provide visual representations for the differences between the raw and $I_1$-based curves.

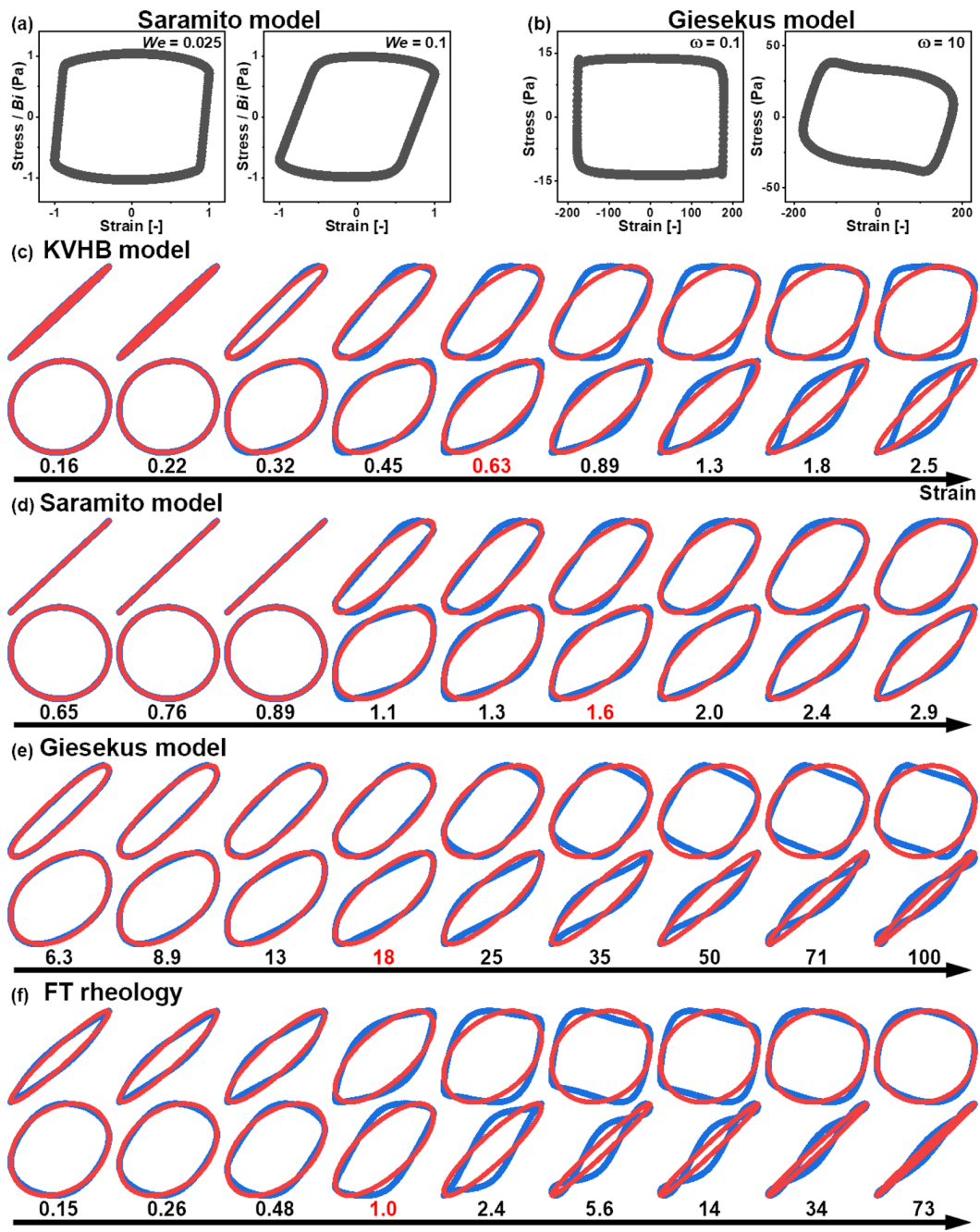

FIG. 8. Models to give simulated Lissajous curves for the verification of the ASB method. (a) Representative results from the Saramito model ($a = 1$, $Bi = 3$, $\alpha = 1$, and $\xi = 0$). Left panel: $We = 0.025$. Right panel: $We = 0.1$. (b) Representative results from the Giesekus model ($a = 1$, $\lambda = 1$ s, $\eta_m / \eta_0 = 0.3$, $\eta_s = 0.01$ Pa·s, $\eta_p = 10$ Pa·s, $G = \eta_p / \lambda$ $= 10$ Pa, and $\gamma_0 = 178$). Left panel: $\omega = 0.1$ rad·s$^{-1}$. Right panel: $\omega = 10$ rad·s$^{-1}$. Comparisons between the normalized Lissajous curves from models (blue lines) and the normalized $I_1$-based curves (red lines) at different strain amplitudes (in each panel, the upper and lower rows are the elastic and viscous Lissajous curves), including the results from (c) KVHB model (f = 0.01 Hz, $G = 100$ Pa, $n = 0.5$, $\sigma_y = 20$ Pa); (d) Saramito model; (e) Giesekus model; (f) FT rheology (A = 3. Sample: 0.1 wt. % carbopol gel).

As shown in Figs. 8(c)–8(f), it is clear that these raw Lissajous curves (blue lines) become more non-linear and viscous as the strain amplitude is increased. Meanwhile, the three points of $(\dot{\gamma}_{\max}, \sigma_{\max})$, $(\dot{\gamma}_{\max}, \dot{\sigma}_{\max})$, and $(\gamma_{\max}, \sigma_{\max})$ in each raw elastic Lissajous curve and the three points of $(\bar{\dot{\gamma}}_{max}, \sigma_{max})$, $(\dot{\gamma}_{max}, \bar{\sigma}_{max})$, and $(\dot{\gamma}_{max}, \sigma_{max})$ in each raw viscous Lissajous curve were separated from each other and gathered together with the increase in the strain amplitude, respectively. Thus, the phenomenon of bifurcation happened.

Furthermore, it is also shown that significant differences arise between the raw (blue lines) and algebraic (red lines) Lissajous curves. However, it should be highlighted that the points of $(\bar{\dot{\gamma}}_{max}, \sigma_{max})$, $(\dot{\gamma}_{max}, \bar{\sigma}_{max})$, $(\dot{\gamma}_{max}, \sigma_{max})$, $(\bar{\dot{\gamma}}_{max}, \sigma_{max})$, $(\dot{\gamma}_{max}, \bar{\sigma}_{max})$, and $(\dot{\gamma}_{max}, \sigma_{max})$ in these $I_1$-based curves reflected those separations and gatherings for the six points for the raw Lissajous curves to some extent. Thus, although these raw and corresponding $I_1$-based curves have huge differences, the judgment of separation and gathering for the six points $[(\bar{\dot{\gamma}}_{max}, \sigma_{max})$, $(\dot{\gamma}_{max}, \bar{\sigma}_{max})$, $(\dot{\gamma}_{max}, \sigma_{max})$, $(\bar{\dot{\gamma}}_{max}, \sigma_{max})$, $(\dot{\gamma}_{max}, \bar{\sigma}_{max})$, and $(\dot{\gamma}_{max}, \sigma_{max})]$ will not be significantly affected.

For example, when the scale of the $x$-axis is 0–1, the two points of $x = 0.2$ and 0.5 can be regarded as separate, where the two points of $x = 0.2$ and 0.7 are also separated. Meanwhile, compared with the above situation, the two points of $x = 0.20$ and 0.21 can be considered close to each other, which does not affect the judgment that the two points of $x = 0.20$ and 0.22 are also close to each other, although the difference between 0.20 and 0.22 is two times that between 0.20 and 0.21.

Moreover, the simulated results of stress bifurcation and ASB by treating the obtained Lissajous curves from the three models and FT rheology were plotted in Figs. 9 and 10. Different kinds of sequential raw Lissajous curves were calculated by changing parameters that were denoted in each panel. In brief, as shown in Figs. 9 and 10, the phenomena of bifurcation were shown in all panels. The start yield points provided by the stress bifurcation and ASB in every stress–strain plot correspond well with each other. Furthermore, except for the results from the Saramito model [Fig. 9(b)], the two kinds of end yield points in every stress–strain rate plot have good agreements. Although huge deviations between the results from the stress bifurcation and ASB in the stress–strain rate plots arose for the Saramito model [Fig. 9(b)], acceptable end yield stress values were provided.

Furthermore, by using these models, the physical meanings of the start and end yield points can be interpreted. For example, the ASB curves obtained by treating the results using the KVHB model in Fig. 9(a) clearly show that, although the yield stress is set as 20 Pa, the resulting start and end yield stress values are positively related to the magnitude of viscosity (the consistency index $K$) of fluid. This phenomenon shows the complexity of the concept and determination of the yield stress.[2,3,6–9] The demonstrated start and end yield points can be interpreted based on the KVHB model consisting of a Kelvin–Voigt model and a HB model, where the viscosity of the fluid in the HB model will generate shorter movement with the increase in the $K$ value at the same stress level. In other words, at the oscillatory shear condition, the contribution of the HB model to the total strain rate decreases with the increase in the $K$ value while other parameters are constants. More specifically, if the $K$ value is increased, a larger stress/longer timescale is needed to reach a similar state as the situation with a low $K$ value. Therefore, within a limited timescale (an oscillatory cycle), the demonstrated start and end yield stress rises with the increase in the $K$ value when the yield stress is set as a constant. This also demonstrates that the yield stress determination depends on the test protocols, and that the timescale plays an important role in such determination processes.[2,3,6–9]

Figure 9(a) also shows that the start yield point is not significantly dependent on the relaxation time of the Kelvin–Voigt model. Meanwhile, the end yield stress/strain rate is negatively associated with the relaxation time because higher stress is needed to transform the system to a nearly pure viscous fluid when the Kelvin–Voigt model can relax in a shorter timescale. Then, the end yield strain rate is also increased.

Figure 9(b) also shows the start yield point of 20 Pa in all panels because the low strain rate causes a weak viscous stress provided by the viscous body, and the major external stress is loaded on the spring. The end yield stress is decreased with the increase in the rigidity of the spring since more external stress is transferred to the friction element. In other words, a complete solid–liquid transition within an oscillatory cycle (the extensive structural destruction) occurs at a lower stress level. The end yield stress is also positively related to the frequency, where the start and end yield stresses are almost the same at a low frequency (0.001 Hz). This also denotes that the degree of structural damage can be associated with the timescale, where a shorter timescale corresponds to a larger end yield stress.

As shown in Fig. 10, the Giesekus model is a versatile differential constitutive relation for the non-linear viscoelasticity (shear and extension).[97] In the linear region, the Giesekus model is reduced to the Jeffreys model containing Maxwell fluid and Newtonian elements, which can predict the nonlinear shear-thinning viscosity and extensional thickening and describe the rheological behaviors of polymer solutions. This model was also applied as "test equation" to generate nonlinear responses[67,97,98] under imposed high oscillatory deformations, providing comparisons with other models.[80,99]

The deviations in the ASB region between the results of stress bifurcation and ASB in the stress–strain rate panels of Fig. 10(a) are attributed to the stress overshoots of the responses given by the Giesekus model. In other words, the phenomena of secondary loops arise in the viscous Lissajous curves (or the maximum stress values are achieved in the second and fourth quadrants). By contrast, the corresponding $I_1$-based curves in ASB are close to those of a pure viscous fluid. Then, the deviations between the two kinds of bifurcation curves occur. However, such deviations do not affect the judgment of two bifurcation points as shown in Fig. 10(a). This denotes that secondary loops do not affect the functionality of ASB. Figure 10 shows that the results of stress bifurcation and ASB method correspond well by processing the LAOS responses from not only the models defined possessing the yield stress but also the Giesekus constitutive equation for polymer solutions.

As shown in Fig. 10(b), the nonlinearities (i.e., the intensities of higher harmonics) of the experimental data of carbopol gels were amplified 1.5–3 times (A = 1.5–3) to investigate the role of higher harmonics. Figure 10(b) implies that the stress bifurcation and ASB method can provide almost the same bifurcation points even if the intensities of higher harmonics are set 3 times the original nonlinearity. Therefore, this phenomenon shows the insignificant role of nonlinearities/higher harmonics in the ASB method.

## 5. The connection between the bifurcation method and QL-LAOS

Last but not least, the quasilinear LAOS (QL-LAOS)[100,101] has been proposed based on the phenomenon that a fluid can present only the first harmonic in the MAOS/LAOS region when the frequency is set high enough because the structural transition may not occur during an oscillatory shear cycle. Therefore, QL-LAOS is similar to SAOS, to some extent, composing a class of constant-microscopic-state motions. One of the differences between QL-LAOS and SAOS is that different stress amplitudes correspond to different specific microscopic states,[101] while analyzing SAOS and QL-LAOS are considered the same. In other words, in QL-LAOS, the material parameter is the function of the frequency and stress amplitude, which differs from the SAOS situation. Here, we envisaged using stress bifurcation and ASB in the QL-LAOS. First, different Lissajous curves based on the Saramito model were plotted in Fig. 11(a). As can be seen, significant differences are shown between the blue and red curves at the frequency of 2 Hz.

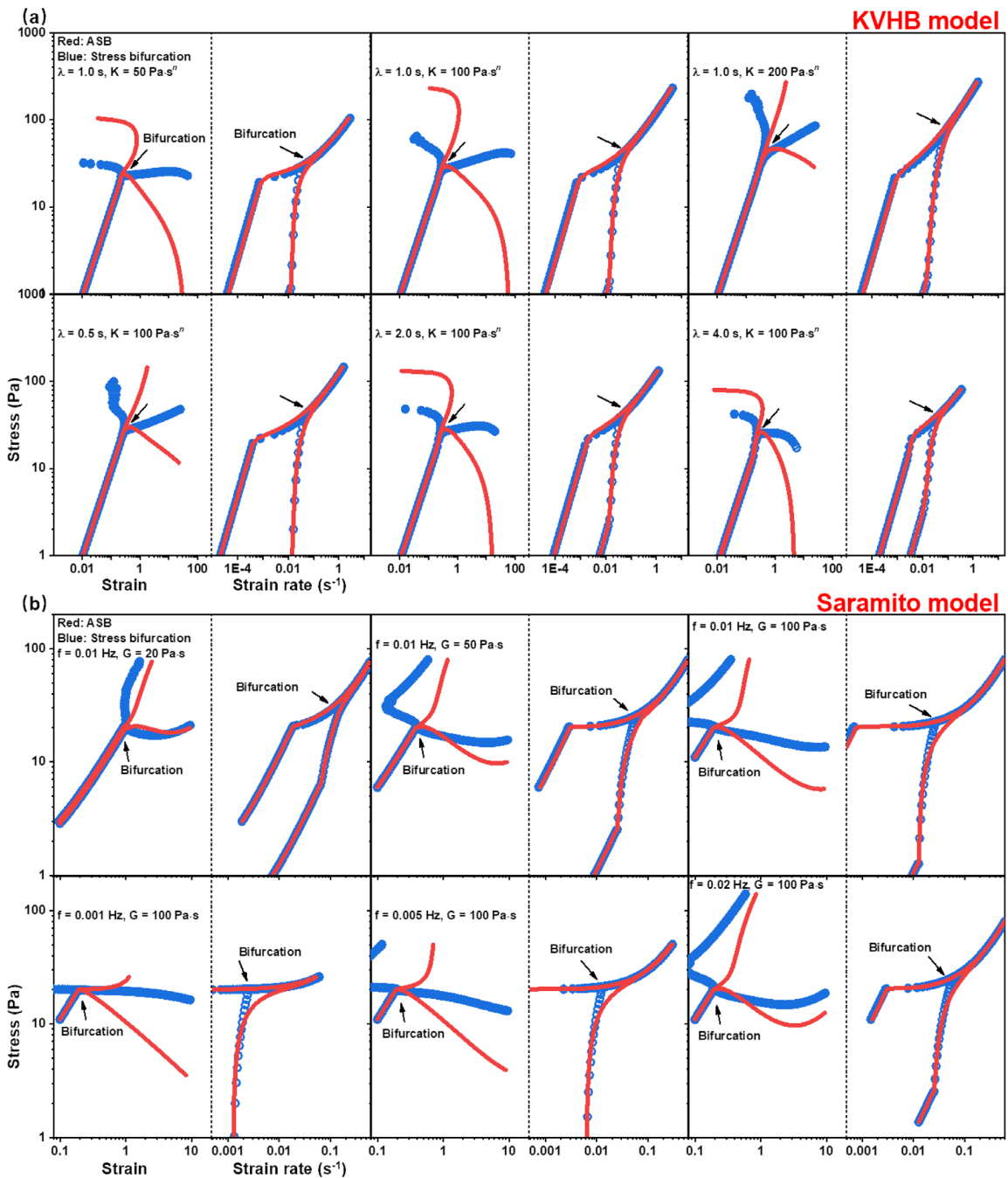

FIG. 9. Comparisons between the results of stress bifurcation (blue lines) and ASB (red lines) by treating sequential Lissajous curves generated from models, including (a) KVHB model ($f = 0.01$ Hz, $G = 100$ Pa, $n = 0.5$, and $\sigma_y = 20$ Pa) and (b) Saramito model ($a = 1$, $\eta_p = 100$ Pa·s, $\eta_m/\eta_0 = 1$, and $\sigma_y = 20$ Pa).

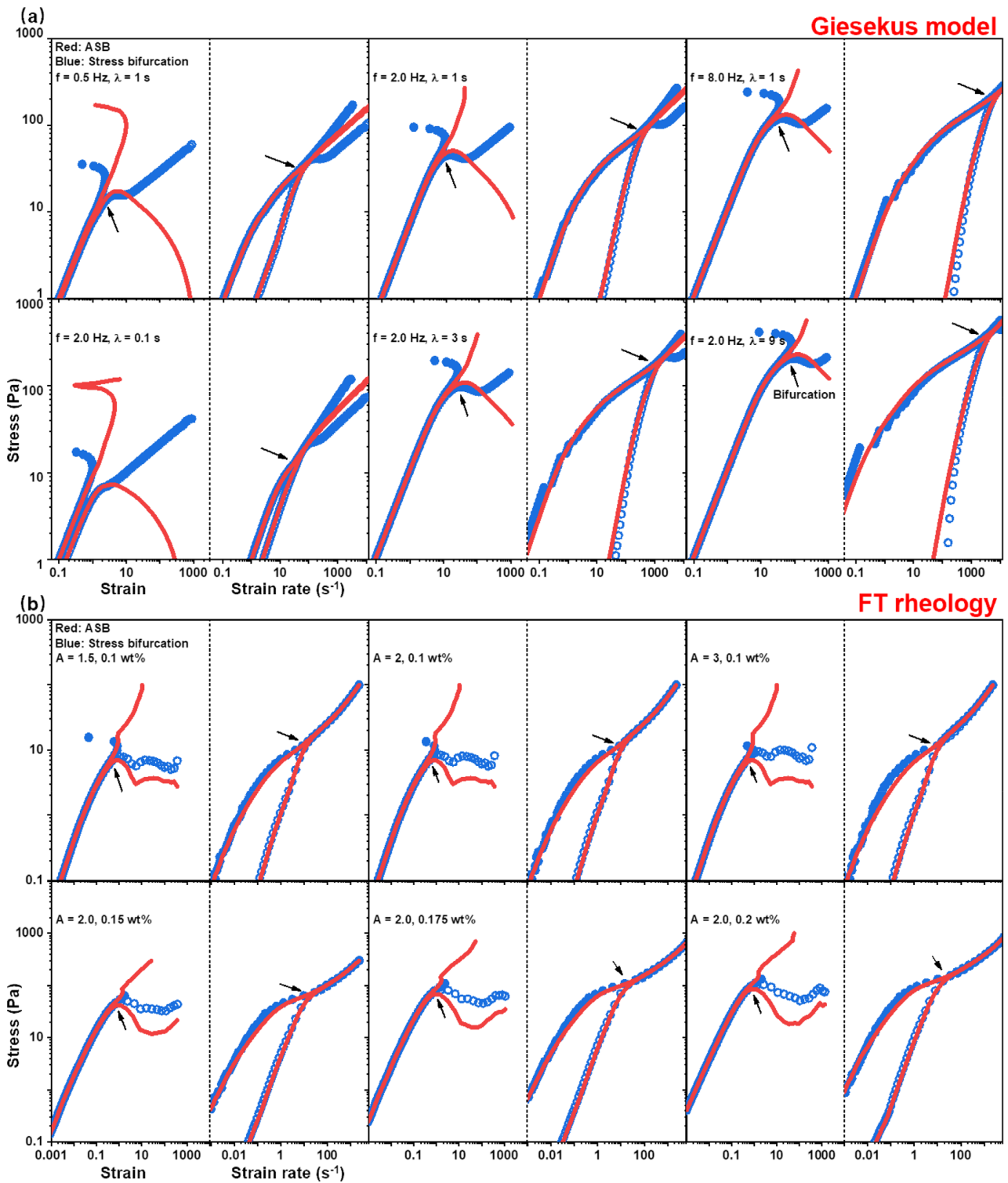

**FIG. 10.** Comparisons between the results of stress bifurcation (blue lines) and ASB (red lines) by treating sequential Lissajous curves generated from a model and the FT rheology method, including (a) Giesekus model ($a = 1$, $\alpha = 0.3$, $\eta_s = 0.01$ Pa s, $\eta_p = 10$ Pa s, and $G = \eta_p / \lambda = 10$ Pa) and (b) FT rheology (the concentration of each applied carbopol gel is correspondingly denoted).

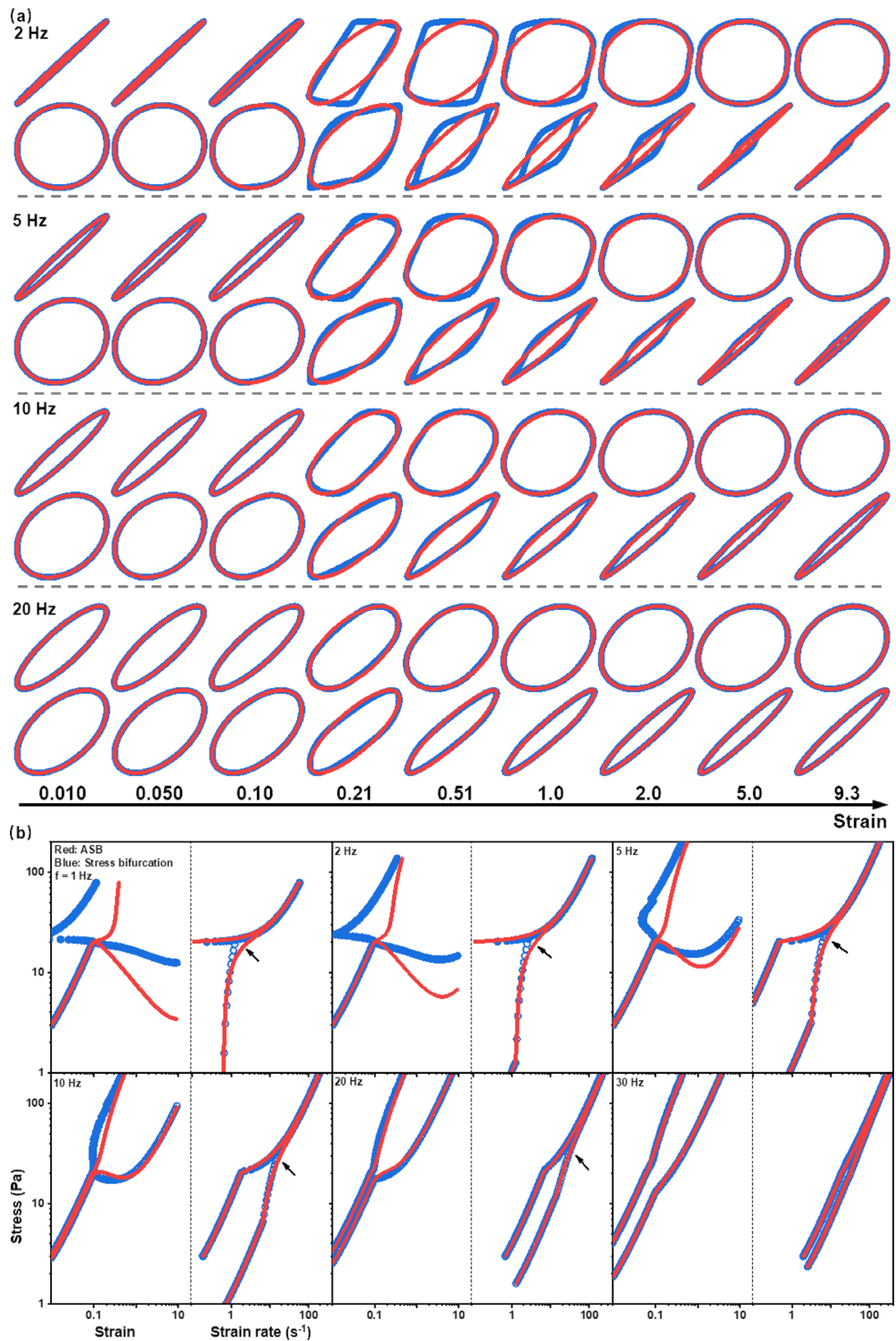

**FIG. 11.** The comparison of stress bifurcation and ASB in QL-LAOS. (a) Comparison between the normalized Lissajous curves from the Saramito model (blue lines) and the normalized $I_1$-based curves (red lines) at different frequencies (in each panel, the upper and lower rows are the elastic and viscous Lissajous curves), including the frequencies of 2, 5, 10, and 20 Hz ($a = 1$, $\eta_p = 1\,\text{Pa s}$, $\eta_m/\eta_0 = 1$, $\sigma_y = 20\,\text{Pa}$, and $G = 200\,\text{Pa}$). (b) Comparison between the results of stress bifurcation (blue lines) and ASB (red lines) at different frequencies ($a = 1$, $\eta_p = 100\,\text{Pa s}$, $\eta_m/\eta_0 = 1$, and $\sigma_y = 20\,\text{Pa}$, $G = 200\,\text{Pa}$).

However, the deviations gradually decrease with the increase in frequency. At 20 Hz, the two kinds of curves well overlap, reflecting that the original LAOS responses are close to sinusoidal waveforms, which corresponds well with the conclusion given by de Souza Mendes *et al.*[100,101] Then, the differences between the quasilinear LAOS responses and sinusoidal signals can be represented by the deviations between the curves provided by the stress bifurcation and ASB, as denoted in Fig. 11(b). As a result, plotting a series of Lissajous curves on the Pipkin space can be represented by the panels in Fig. 11(b).

In addition, it is noticed that the elastic Lissajous curves in the SAOS region in Fig. 11(a) are gradually transformed from nearly pure elastic behavior to viscoelastic behavior with the increase in frequency, which is different from the common phenomenon. This originates from the inherent characteristic of the Saramito model that contains a viscous body in parallel with other elements. When the frequency is high enough, the contribution of this viscous body becomes significant because of the high strain rate amplitude.

Furthermore, the QL-LAOS has indicated that there exist relationships between LAOS methods (stress decomposition as well as the $S$ and $T$ ratios) and LAOS models (i.e., Kelvin–Voigt model possessing the shear modulus $G_{KV}$ and viscosity $\eta_{KV}$ as well as the Maxwell model containing $G_{MW}$ and $\eta_{MW}$).[102] For example, at the strain-controlled condition, the QL-LAOS on the stress decomposition and the Kelvin-Voigt model gives the following equations:

$$G_{KV}(\omega, \gamma_0, t) = \sigma' / \gamma = \sum_{n=1,odd}^{\infty} G_n'(\omega, \gamma_0) \sin n\omega t / \sin \omega t, \quad (47)$$

$$\eta_{KV}(\omega, \gamma_0, t) = \sigma'' / \dot{\gamma} = \sum_{n=1,odd}^{\infty} G_n''(\omega, \gamma_0) \cos n\omega t / \sin \omega t. \quad (48)$$

At the stress-controlled condition of $\sigma(t) = \sigma_{max} \sin \omega t$, the relationship between this decomposition and the Maxwell model deduces the equations,

$$G_{MW}(\omega, \sigma_{max}, t) = \sigma_{max} \cos \omega t / \gamma_0 \sum_{n=1,odd}^{\infty} nb_n(\omega, \sigma_{max}) \cos n\omega t, \quad (49)$$

$$\eta_{MW}(\omega, \sigma_{max}, t) = \sigma_{max} \sin \omega t / \gamma_0 \sum_{n=1,odd}^{\infty} -n\omega a_n(\omega, \sigma_{max}) \sin n\omega t. \quad (50)$$

Since the stress decomposition is considered in the stress bifurcation and ASB, the connection between stress bifurcation and QL-LAOS can be found involving L Hospital Theory (e.g., $\lim_{\theta \to -0.5\pi} \cos \theta / \cos n\theta = 1/n$),

$$\sigma_{\gamma 0} = \gamma_0 \sum_{n=1,odd}^{\infty} (-1)^{(\frac{n-1}{2})} G_n'(\omega, \gamma_0) = G_{KV}\left(\omega, \gamma_0, \frac{\pi}{2\omega} \text{ or } \frac{3\pi}{2\omega}\right), \quad (51)$$

$$\sigma_{\dot{\gamma} 0} = \gamma_0 \sum_{n=1,odd}^{\infty} (-1)^{(\frac{n-1}{2})} G_n'(\omega, \gamma_0) = \eta_{KV}\left(\omega, \gamma_0, \frac{\pi}{2\omega} \text{ or } \frac{3\pi}{2\omega}\right), \quad (52)$$

$$\gamma_{\sigma max} = \gamma_0 \sum_{n=1,odd}^{\infty} (-1)^{(\frac{n-1}{2})} b_n(\omega, \sigma_{max})$$
$$= \sigma_{max} / nG_{MW}\left(\omega, \sigma_{max}, \frac{\pi}{2\omega} \text{ or } \frac{3\pi}{2\omega}\right), \quad (53)$$

$$\dot{\gamma}_{\sigma max} = \omega\gamma_0 \sum_{n=1,odd}^{\infty} (-1)^{(\frac{n+1}{2})} na_n(\omega, \sigma_{max})$$
$$= \sigma_{max} / \eta_{MW}\left(\omega, \sigma_{max}, \frac{\pi}{2\omega} \text{ or } \frac{3\pi}{2\omega}\right). \quad (54)$$

As a result, the curves provided by the stress bifurcation/ASB can represent the material functions according to Thompson *et al.*[102] Based on the above four equations [Eqs. (51)–(54)], it is clear that the information from both strain-controlled and stress-controlled conditions (i.e., carrying out the FT of distorted stress signals and distorted strain signals to obtain Fourier coefficients) is needed, where interpolation can be used.[67,70] Therefore, although this work proposes a method named algebraic "stress" bifurcation, it can also be regarded as a "strain" bifurcation.

To sum up, the ASB method is verified from the perspective of the model, where the KVHB model, Saramito model, Giesekus model, and FT rheology were applied. Therefore, the ASB method is not only highly suitable for treating real samples to provide precise start and end yield points, but also applicable in ideal situations by introducing models to obtain acceptable results. Most importantly, the results from models also reflect one of the key points in this work: the determinations of separation and gathering for the six spots (($\bar{\dot{\gamma}}_{max}, \sigma_{max}$), ($\dot{\gamma}_{max}, \bar{\sigma}_{max}$), ($\gamma_{max}, \sigma_{max}$), ($\bar{\dot{\gamma}}_{max}, \sigma_{max}$), ($\dot{\gamma}_{max}, \bar{\sigma}_{max}$), and ($\dot{\gamma}_{max}, \sigma_{max}$)) are not seriously dependent on the specific coordinates. In other words, the error of the ASB method relative to the stress bifurcation method will not significantly influence the determinations of the start and end yield points. The ASB method shows the correct physics for the above cases, and the interpretation of the experimental data are then reliable. Finally, the use of stress bifurcation and ASB in the QL-LAOS is conceived and the connection between the two bifurcation methods and QL-LAOS is demonstrated.

### E. Verification of the determined start and end yield stresses

The theory, verification, and FT interpretation of ASB have been introduced and thoroughly analyzed in the above-mentioned descriptions. Then, the rationality of the results from ASB was compared with those obtained using different yield stress determination methods, as was evaluated using several typical YSFs. The results are shown and compared in Figs. 12 and 21–23 as well as Table I. It is found that the start and end yield stresses from these typical YSFs are rational and correspond well with those from other yield stress determination methods. More information of these yield stress values can be found in Appendix E.

In addition, the $\sigma-\gamma$ curves of ASB inherently provided the comparison between the start yield point and the maximum value of the elastic stress [Fig. 12(b)], as described by Eq. (30). To sum up, ASB should be a promising method for the yield stress determination.

### F. Significance of the solid–liquid transition region

The rationalization of ASB has been discussed. Then, the rheological behaviors of typical YSFs around the determined solid–liquid transition regions were systematically investigated in the following to provide deeper insights into the yield processes of samples. The fine yielding evolution between the start yield point and end yield point can be demonstrated by taking into account different approaches, including changing cycle and point numbers (Fig. 24), stress decomposition (Figs. 13, 25, and 26), and several strategies of SPP (Figs. 14, 27, 28, 29, and 30).

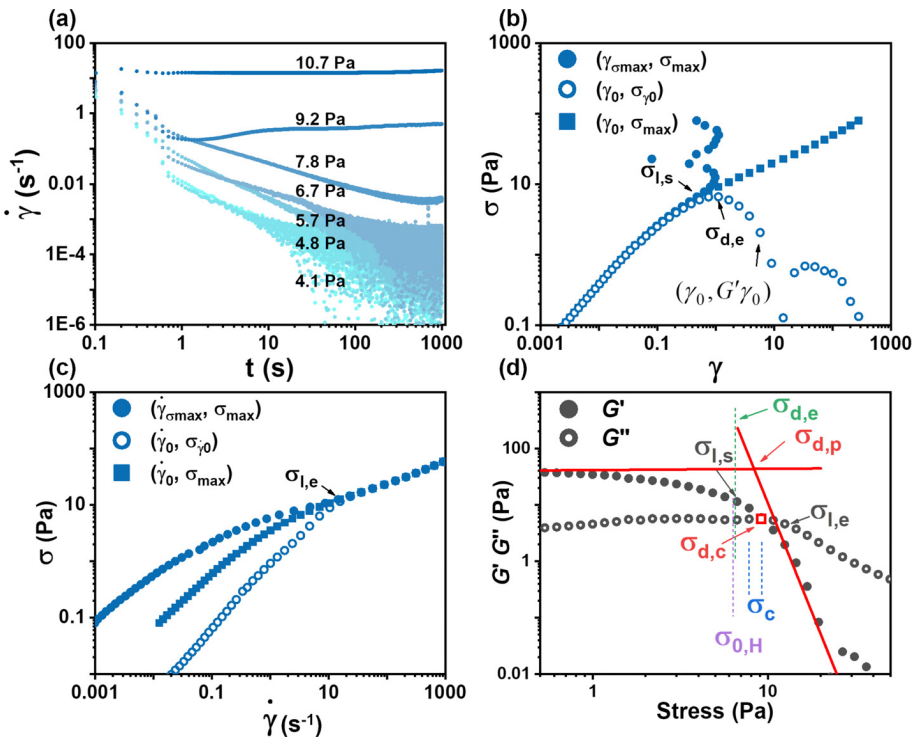

**FIG. 12.** Comparison of yield stresses obtained by using different methods (sample: 0.1 wt. % carbopol gels): (a) creep ($\sigma_c$), (b) elastic stress method ($\sigma_{d,e}$) and stress vs strain from ASB ($\sigma_{l,s}$), (c) stress vs strain rate from ASB ($\sigma_{l,e}$), (d) the intersection of two power-law extrapolations at low- and high-stresses in $G'$ vs the stress amplitude ($\sigma_{d,p}$), the characteristic modulus where $G' = G''$ ($\sigma_{d,c}$), and the comparison between these above-determined yield stress. Frequency: 1 Hz.

The solid–liquid transition region determined by ASB was further investigated by changing the numbers of the sampling points and the oscillation cycles required for each sampling point. It is found that the start and end yield stresses had insignificant relationships with the number of both the sampling points and repeated cycles. This result reflects that ASB can give reliable values and conclusions, which is important for the application of this method. Further detailed information was provided in Appendix F.

Stress decomposition and SPP were also involved in providing more insights into the yield processes of these samples. The mathematical frameworks of stress decomposition and SPP can refer to Secs. II C and II E, respectively, which are applied to generate the following results in this section. In addition, it should be highlighted that the blue curves in Fig. 6 represent the responses of pure elastic materials and perfect viscous samples by using the values of $G'$ and $G''$ generated from the LAOS experiment. Thus, by correspondingly adding the blue curves, the obtained red elliptic curves represent the linear viscoelastic behaviors. Then, in Fig. 7, different kinds of Lissajous curves were constructed by introducing different numbers of higher harmonics. By contrast, the results from the stress decomposition plotted in Fig. 13 were generated by using the raw data/Lissajous curves that contain all harmonics. Therefore, the blue curves in Fig. 6 are different from the curves in Fig. 13.

The stress decomposition allows the separation of the elastic part and viscous part from LAOS results.[69,70] The data of several samples were treated by stress decomposition, as described in Sec. II C including 0.1 wt. % carbopol gels (Fig. 13), 3 wt. % hydrogel particle suspensions (Fig. 25, Appendix G), 1.25 wt. % laponite suspensions [Fig. 26(a), Appendix G], and 1.25 wt. % xanthan gum solutions [Fig. 26(b), Appendix G]. The solid–liquid transition regions of the four samples are denoted in the corresponding panels. Figure 13(a) depicts the region after the start yield point, while Fig. 13(b) shows the region around the solid–liquid transition. As can be observed, the elastic and viscous parts were odd functions of strain and strain rate, respectively. At low stress amplitudes, the elastic [Fig. 13(b-i)] and viscous [Fig. 13(b-ii)] parts were linear. After the start yield point (strain = 0.35), the slope of the elastic curve decreased and the intensity of the viscous part sharply increased. Around the end yield point (strain = 2.4), the elastic curve bent and the slope of the viscous curve gradually decreased. In the post-yield region, the intensity of the elastic part varied little, and the viscous part gradually increased. The elastic parts were higher and smaller than the viscous parts at low and high stress levels, respectively, indicating the manifestation of a solid–liquid transition during the stress sweep. The yield behavior of 3 wt. % hydrogel particle suspension evaluated by stress decomposition was similar to the carbopol gel. For the strong thixotropic fluids {laponite suspension [Fig. 26(a)] and xanthan gum solution [Fig. 26(b)]}, the instantaneous collapse and strengthening of the elastic and viscous parts can be also observed during the solid–liquid transition, respectively.

As the stress amplitude increased, nonlinearities raised (e.g., Figs. 13(a-i) and 13(a-ii)]. Although the viscous parts vs angle were close to sinusoidal functions [e.g., Figs. 13(a-iv) and 13(b-iv)], the elastic parts were gradually distorted as the stress amplitude increased [e.g., Figs. 13(a-iii) and 13(b-iii)]. The nonlinearity of the elastic part was considerably larger than the viscous part and stress response. This effect clearly denotes that the stress decomposition was more sensitive to nonlinearity than FT rheology that only considers the nonlinearity of the total stress.[69] If even harmonics are removed from the raw Lissajous curves, the elastic part and viscous part, separated by using the stress decomposition, represent the pure contributions of the elastic moduli $G'_n$ and the viscous moduli $G''_n$ in all odd harmonics, respectively. Compared with the viscous parts, the more distorted elastic parts denote that the $G'_n$ values of all odd higher harmonics have more significant effects on the elastic parts than the cases for the viscous parts.

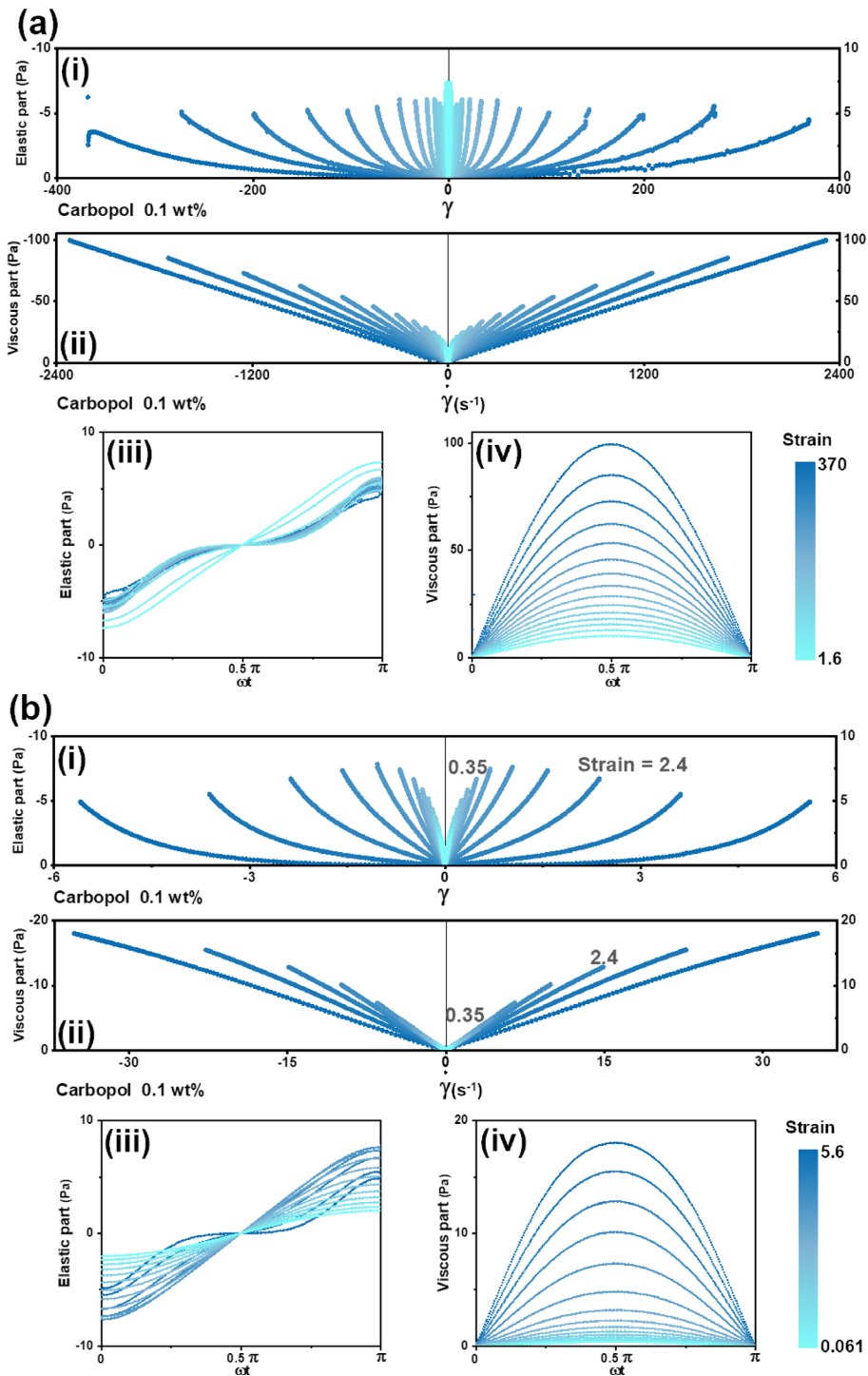

**FIG. 13.** Stress decomposition to divide Lissajous curves into elastic and viscous parts (sample: 0.1 wt. % carbopol gel). (a) The results in the strain amplitude range of 1.6 ∼370: (i) elastic stress vs strain; (ii) viscous stress vs strain rate; (iii) elastic and (iv) viscous stress vs angle. (b) The results in the strain amplitude range of 0.061–5.6.

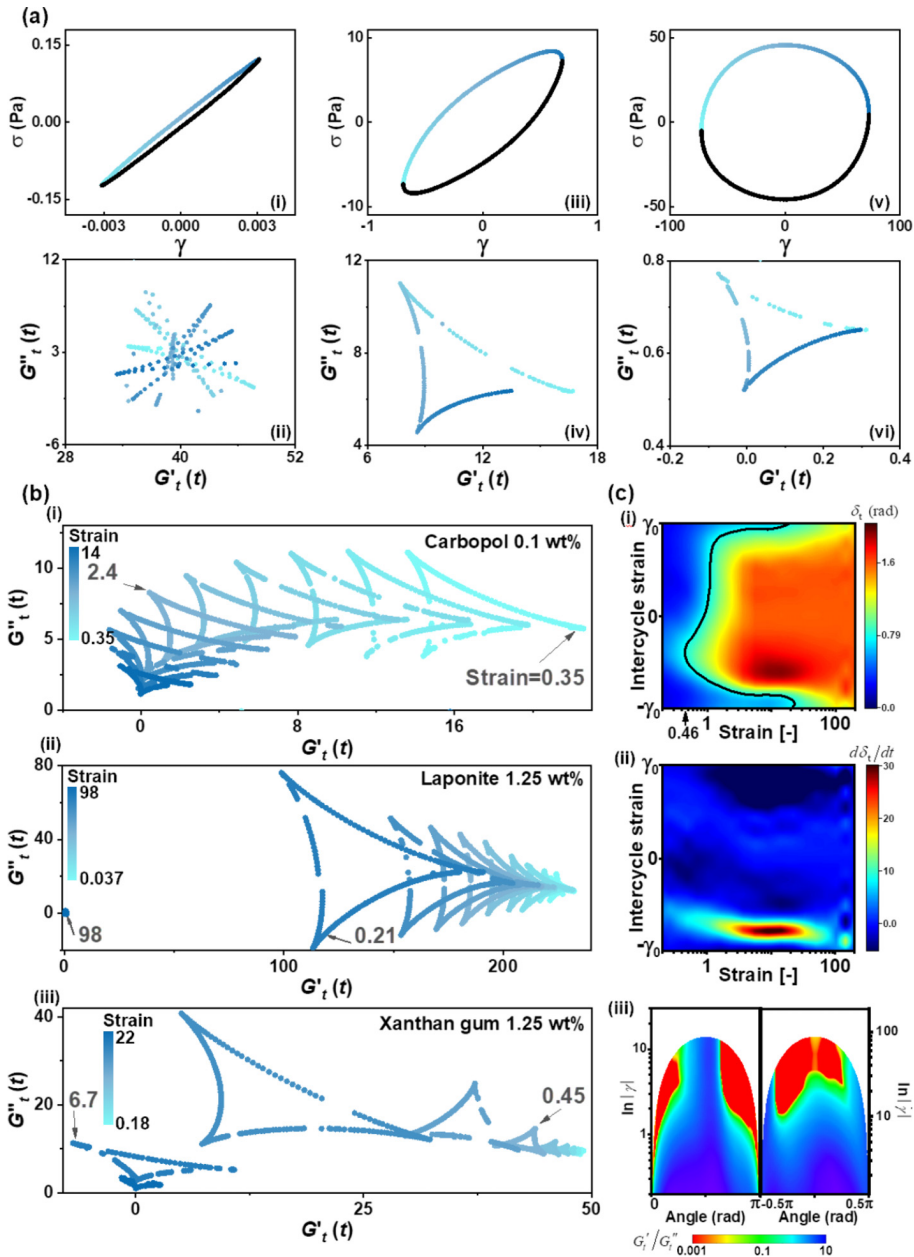

FIG. 14. Transient modulus and transient-modulus-related methods in SPP. (a) Transient modulus: the Lissajous curves at the strain amplitudes of (i) 0.0031, (iii) 0.69, and (v) 73; the Cole–Cole plots at the strain amplitudes of (ii) 0.0031, (iv) 0.69, and (vi) 73. (b) The Cole–Cole plots at different points: (i) 0.1 wt. % carbopol gel, (ii) 1.25 wt. % laponite suspension, and (iii) 1.25 wt. % xanthan gum. (c) Transient-modulus-related methods in SPP to treat 0.1 wt. % carbopol gel: (i) contour plot of phase angle $\delta_t$; (ii) contour plot of the phase angle velocity $d\delta_t/dt$. The denoted strain values represent the first appearance of $\delta_t = \pi/4$ ($G_t' = G_t''$); (iii) $\ln|\dot{\gamma}|$ (intracycle strain rate) and $\ln|\dot{\gamma}|$ (intracycle strain rate) vs angle [$\theta$, $\gamma(t) = \gamma_0 \sin\theta$] to show the $G_t'/G_t''$ ratio with the color mapping. Frequency: 1 Hz.

Several methods of SPP were further introduced to study the solid–liquid transition regions of 0.1 wt. % carbopol gel, 1.25 wt. % laponite suspension, and 1.25 wt. % xanthan gum solution (Figs. 14 and 27–30). In Appendixes H–J, the elastic Lissajous curve, $G_{\text{cage}}$, the amount of the strain acquired in a Lissajous loop ($\gamma$ at $\sigma_{\max}$), and the LAOS-based flow curve were investigated. In SPP, $G_t'(t)$ and $G_t''(t)$ are called the transient moduli to describe the elastic and viscous transition inside a Lissajous loop.[103,104] More specifically, $G_t'(t)$ and $G_t''(t)$ amplify the non-linear behavior of the stress/strain response inside a Lissajous loop.[105] The rheological behavior is presented by the moving $G_t'(t)$ and $G_t''(t)$ in a transient Cole–Cole plot as illustrated in Fig. 14(a). The structural transition inside an elastic Lissajous curve [Fig. 14(a-iii)] is directly associated with the Cole–Cole plot [Fig. 14(a-iv)]. The increased and decreased $G_t'(t)$ indicate the manifestation of stiffening and softening, respectively. Similarly, thickening and thinning can be determined by the upward and downward movements of $G_t''(t)$, respectively.

Intra-cycle transitions at different stresses are shown in Fig. 14(b). The Cole–Cole plots exhibit changes in the size and position of the deltoid with the increase in stress, where the central coordinates of the deltoid are ($G'$, $G''$).[105] Therefore, the position shift represents the changed dynamic moduli. The size of the deltoid was positive to the range of the intra-cycle rheological transition. At small stress levels, the increasing stress promoted the intra-cycle structural transition. Then, a large stress amplitude led to a short time for structural reformation. Therefore, the intra-cycle structural transitions and deltoid sizes became smaller. After that, the deltoid evolved to a single point under extremely large stress, which denoted the rheological equilibrium state without intra-cycle rheological transition. For carbopol gel (non-thixotropic fluid), the change in the deltoid size gradually occurred within the solid–liquid transition region. By contrast, tremendous changes took place for laponite suspension and xanthan gum solution (thixotropic fluids) between $\sigma_{L,s}$ and $\sigma_{L,e}$, indicating the violent intra-cycle structural transitions. Moreover, negative values of $G_t'(t)$ arise at high stress levels in Figs. 14(b-i) and 14(b-iii), indicating that the increased strain leads to a decrease in stress, which originated from the strain-induced structural breakdown. $G_t'(t) = 0$ was attributed to plastic deformation.[85]

For 0.1 wt. % carbopol gel, the instantaneous $\delta_t$ and dimensionless $d\delta_t/dt$ are displayed in Figs. 14(c.i) and 14(c.ii). The positions of $\arctan(G_t''/G_t') = 0$ $(G_t'(t) \gg G_t''(t))$, $= \pi/4$ $(G_t' = G_t'', \gamma_0 = 0.46)$, and $= \pi/2$ $(G_t' = 0)$ can be distinguished from the blue, cyan, and red regions, respectively. Meanwhile, the peaks of $\delta_t$ and $d\delta_t/dt$ were closely related to the intracycle yielding position, which visually showed that the intracycle yielding position became closer to the minimum strain with the increase in strain amplitude. In addition, it was found that the critical strain of the first appearance of $\arctan(G_t''/G_t') = \pi/4$ was 0.46, which corresponded to the start yield strain of 0.35 provided by the ASB. As shown in Fig. 30, the critical strain from SPP and the start yield strain from the ASB were 0.82 and 0.45, respectively. By viewing the raw data, although the two kinds of strain values were different, the critical strain from SPP of the 0.1 wt. % CARBOpol gel was between the sampling strain point of 0.35 and the adjacent point of 0.48. The same goes for the 1.25 wt. % xanthan gum

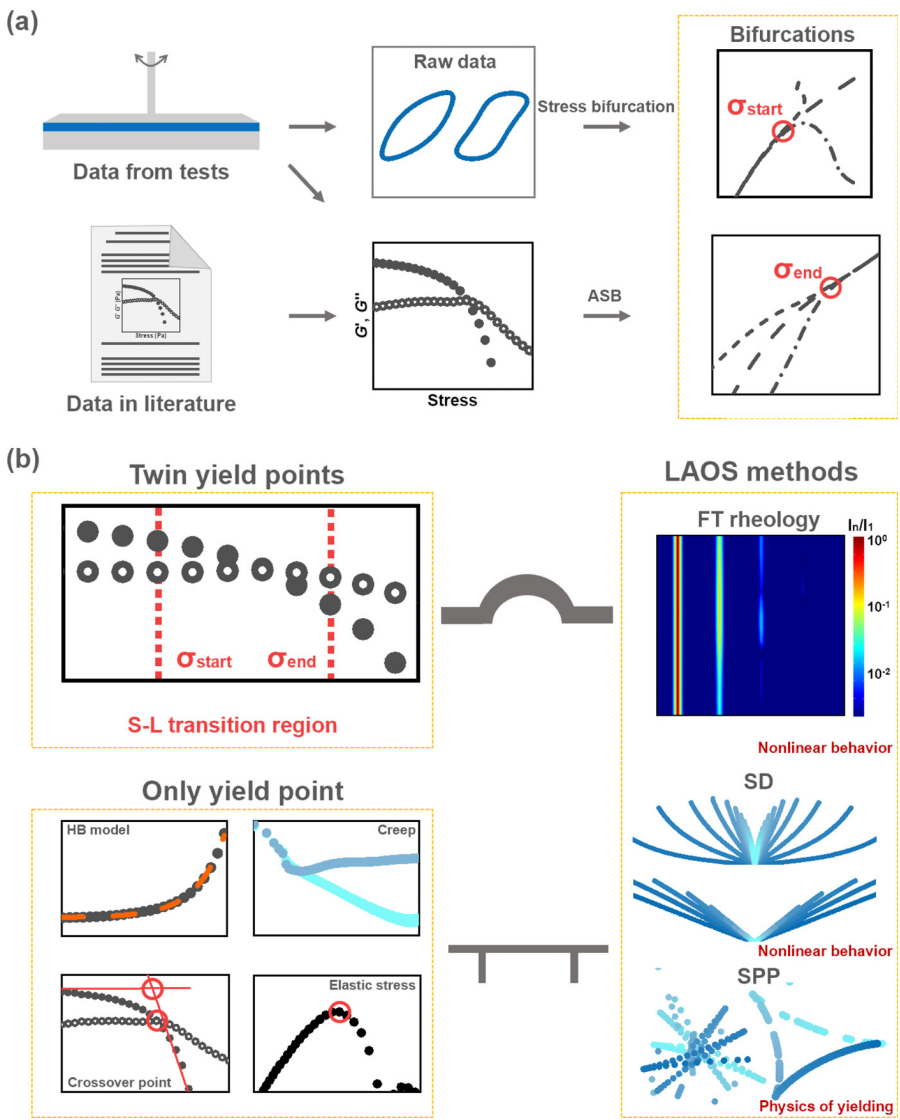

**FIG. 15.** Schematic illustration of the functionality of algebraic stress bifurcation (ASB): (a) ASB simplifies the data processing, requiring only the data of $G'$, $G''$ in sweep tests without raw data (i.e., Lissajous curves); (b) ASB is accessible to define a solid–liquid transition region with a start yield point including information of start yield stress and strain, and an end yield point including end yield stress and strain rate, unlike the common approaches that acquire only one critical yield stress value. Furthermore, ASB determines a quantitative solid–liquid transition region, providing a new bridge connecting with other classic approaches, such as FT rheology (studying the arisen nonlinearity during a LAOS stress/strain sweep), stress decomposition (investigating the intracycle and intercycle nonlinearity during a LAOS stress/strain sweep), and SPP (revealing the intracycle and intercycle physics of yielding), for a panoramic analysis of the yield behavior in YSFs.

solution. This phenomenon also shows that the start yield point from the ASB is rational.

Figure 14(c-iii) describes sequential intracycle solid–liquid transitions through the $G_t'/G_t''$ values at each angle step. When $G_t'/G_t''$ is close to zero, the sample is more liquidlike than solidlike (the red regions). The blue regions indicate that $G_t'/G_t''$ are much larger than zero, reflecting the solidlike response. In Fig. 14(c-iii), the blue and red regions were shrunk and expanded as the strain amplitude increased, respectively. This result also shows the intracycle and intercycle time-dependent rheological behaviors.

Briefly, in this section, it was demonstrated that the solid–liquid transition region determined by ASB can be analyzed by the LAOS methods to provide further understanding of the nonlinearity as well as the intracycle and intercycle yielding. According to Sec. IV C and this section, ASB had the ability to link the yield stress determination methods and the yield behavior investigation tools.

To sum up, Fig. 15 schematically illustrates the functionality of ASB that was established based on FT rheology, stress decomposition, and stress bifurcation. The requirement of only $G'$, $G''$, and the stress/strain amplitude endowed this method with the ability to treat LAOS results without raw Lissajous loops, e.g., raw data being unavailable from the literature [Fig. 15(a)]. Furthermore, ASB can establish a bridge between the yield stress determination and the yield process demonstration by defining a solid–liquid transition region with a start yield point and an end yield point [Fig. 15(b)] instead of only using one yield stress value available from other methods. Furthermore, the determined solid–liquid transition region can be integrated into other powerful rheological tools, such as FT rheology, stress decomposition, and SPP, to analyze the start and end of the macroscopic yield process in the YSFs.

## V. CONCLUSIONS

In this work, based on FT rheology, stress decomposition, and stress bifurcation, an easy-to-implement algebraic stress bifurcation method was proposed to determined the start and end yield points as well as the solid–liquid transition in LAOS. The most salient feature of this method is that the high harmonics and phenomenological Lissajous curves have been confirmed to be unnecessary in the determination of the start and end yield points, which was examined and explained by the KVHB model, Saramito model, Giesekus model, and FT rheology. Meanwhile, the use of stress bifurcation and algebraic stress bifurcation in the QL-LAOS is conceived. The rationality of the determined start and end yield points in the solid–liquid transition was proved as compared with those obtained by other determination methods for a number of typical YSFs. The resulting two yield stress values are the same as those from the existing stress bifurcation method. Furthermore, the algebraic stress bifurcation determines a quantitative solid–liquid transition region, providing thus a new bridge connecting with other methods for the yield stress determination.

The algebraic stress bifurcation significantly reduces the complexities of data processing, providing hence a new approach for determining the start and end points of the solid–liquid transition that requires only $G'$, $G''$, and the stress/strain amplitude instead of raw Lissajous loops. When raw data are unavailable in the literature, only based on the available stress/strain amplitude sweep data, this method can also provide the start and end yield stress values. Furthermore, this formalism of an easy and efficient mathematical framework allows the application of less specialized data processing software packages, such as

Origin® and Excel, instead of professional MATLAB, facilitating a more expansive application of LAOS.

## ACKNOWLEDGMENTS

This work was supported by the National Natural Science Foundation of China (Grant Nos. 22073062 and 21174075) and the Fundamental Research Funds for the Central Universities. Professor W. Yu is acknowledged for constructive opinions and discussions.

## AUTHOR DECLARATIONS

### Conflict of Interest

The authors have no conflicts to disclose.

### Author Contributions

**Pengguang Wang:** Conceptualization (equal); Data curation (equal); Formal analysis (equal); Investigation (equal); Methodology (equal); Resources (equal); Validation (equal); Visualization (equal); Writing – original draft (equal). **Jiatong Xu:** Formal analysis (supporting); Writing – original draft (supporting). **Hongbin Zhang:** Conceptualization (equal); Formal analysis (equal); Funding acquisition (lead); Investigation (equal); Methodology (equal); Project administration (lead); Supervision (lead); Validation (equal); Visualization (equal); Writing – original draft (equal); Writing – review & editing (lead).

## DATA AVAILABILITY

The data that support the finding of this study and the software for the determination of yield stress using the algebraic stress bifurcation method are available from the corresponding author upon reasonable request.

## NOMENCLATURE

### Parameter Interpretation

| | |
|---|---|
| $a_n$ | The Fourier coefficient of $n$th harmonic |
| $a_n'$ | The Chebyshev coefficient of power series |
| $b_n$ | The Fourier coefficient of $n$th harmonic |
| $b_n'$ | The Chebyshev coefficient of power series |
| $c_n'$ | The Chebyshev coefficient of power series |
| $D_h$ | The horizonal difference between the maximum stress and maximum strain |
| $D_v$ | The vertical difference between the maximum stress and maximum strain |
| $G_{cage}$ | The modulus of YSFs at zero stress, Pa |
| $G'$ | Elastic modulus, Pa |
| $G''$ | Loss modulus, Pa |
| $G_n'$ | $n$th harmonic elastic modulus, Pa |
| $G_n''$ | $n$th harmonic loss modulus, Pa |
| $G_t'(t)$ | Transient elastic modulus, Pa |
| $G_t''(t)$ | Transient viscous modulus, Pa |
| $G_{\bar{\gamma}}$ | The slope of stress-mean strain curve, Pa |
| $G_{\bar{\sigma}}$ | The slope of mean stress–strain curve, Pa |
| $I_n$ | The intensity of $n$th harmonic, Pa |
| $K$ | Consistency coefficient, Pa s$^n$ |

| | | | |
|---|---|---|---|
| $n$ | Non-Newtonian index | $\sigma_y$ | Yield stress, Pa |
| $t$ | Time, s | $\sigma_{max}$ | Stress amplitude, Pa |
| $\gamma$ | Strain | $\sigma_{l,s}$ | Start yield stress, Pa |
| $\bar{\gamma}$ | Mean strain | $\sigma_{l,e}$ | End yield stress, Pa |
| $\dot{\gamma}$ | Strain rate, s$^{-1}$ | $\sigma_{max}\text{-}\gamma_{\sigma max}$ | Stress amplitude–strain at maximum stress curve of ASB |
| $\dot{\bar{\gamma}}$ | Mean strain rate, s$^{-1}$ | | |
| $\gamma_0$ | Strain amplitude of ASB | $\sigma_{\gamma 0}\text{-}\gamma_0$ | Stress at maximum strain–strain amplitude curve of ASB |
| $\gamma_{max}$ | Strain amplitude of stress bifurcation | | |
| $\dot{\gamma}_{max}$ | Strain rate amplitude, s$^{-1}$ | $\sigma_{max}\text{-}\dot{\gamma}_{\sigma max}$ | Stress amplitude–strain at maximum stress rate curve of ASB |
| $\dot{\gamma}_{l,s}$ | Start yield strain | | |
| $\dot{\gamma}_{l,e}$ | End yield strain rate, s$^{-1}$ | $\sigma_{\dot{\gamma} 0}\text{-}\dot{\gamma}_0$ | Stress at maximum strain rate–strain rate amplitude curve of ASB |
| $(\gamma_0, \sigma_{\gamma 0})$ | The coordinates of the maximum strain in algebraic elastic Lissajous curve | | |
| | | $\sigma_{max}\text{-}\bar{\gamma}_{max}$ | Stress amplitude–maximum of mean strain curve of stress bifurcation |
| $(\gamma_{\sigma max}, \sigma_{max})$ | The coordinates of the maximum stress in algebraic elastic Lissajous curve | | |
| | | $\bar{\sigma}_{max}\text{-}\gamma_{max}$ | Maximum of mean stress–strain amplitude curve of stress bifurcation |
| $(\dot{\gamma}_0, \sigma_{\dot{\gamma} 0})$ | The coordinates of the maximum strain rate in algebraic viscous Lissajous curve | | |
| | | $\sigma_{max}\text{-}\dot{\bar{\gamma}}_{max}$ | Stress amplitude–maximum of mean strain rate curve of stress bifurcation |
| $(\dot{\gamma}_{\sigma max}, \sigma_{max})$ | The coordinates of the maximum stress in algebraic viscous Lissajous curve | | |
| | | $\bar{\sigma}_{max}\text{-}\dot{\gamma}_{max}$ | Maximum of mean stress–strain rate amplitude curve of stress bifurcation |
| $(\gamma_{max}, \bar{\sigma}_{max})$ | End point of the mean stress–strain curve of the elastic Lissajous plots | | |
| | | $\sigma'$ | Elastic shear stress, Pa |
| $(\bar{\gamma}_{max}, \sigma_{max})$ | End point of the stress-mean strain curve of the elastic Lissajous plots | $\sigma''$ | Viscous shear stress, Pa |
| | | $\phi_n$ | The phase angle of $n$th harmonic, $^{\circ}$ |
| $(\dot{\bar{\gamma}}_{max}, \sigma_{max})$ | End point of the stress-mean strain rate curve of the viscous Lissajous plots | $\omega$ | Angular frequency, rad/s |
| $(\dot{\gamma}_{max}, \bar{\sigma}_{max})$ | End point of the mean stress–strain rate curve of the viscous Lissajous plots | | |

## APPENDIX A: THIXOTROPY OF DIFFERENT SAMPLES

| | |
|---|---|
| $\eta_{\dot{\gamma}}$ | The slope of stress-mean strain rate curve, Pa s |
| $\eta_{\bar{\sigma}}$ | The slope of mean stress–strain rate curve, Pa s |
| $\sigma$ | Stress, Pa |
| $\sigma_{0,H}$ | Yield stress from the HB model of steady shear, Pa |
| $\sigma_c$ | Yield stress from creep method, Pa |
| $\sigma_{d,c}$ | Yield stress from the cross over point of $G'$ and $G''$, Pa |
| $\sigma_{d,e}$ | Yield stress from elastic stress method, Pa |
| $\sigma_{d,p}$ | Yield stress from the intersection of two power-law extrapolations at low and high stress ranges in $G'$ vs. stress amplitude, Pa |

Thixotropy of four samples are characterized by down and up shear rate sweep (Fig. 16) including 1 wt. % carbopol gel, 1.25 wt. % laponite suspension, 1.25 wt. % xanthan gum, and 3 wt. % hydrogel particle suspension. 1 wt. % carbopol gel and 3 wt. % hydrogel particle suspension exhibit weak thixotropy while 1.25 wt. % laponite suspension and 1.25 wt. % xanthan gum solution present evident hysteresis.

The yield stresses ($\sigma_y$) in steady-state shear were explored by using the HB model,[106,107]

$$\sigma = \sigma_y + K\dot{\gamma}^n, \quad \sigma > \sigma_y, \tag{A1}$$

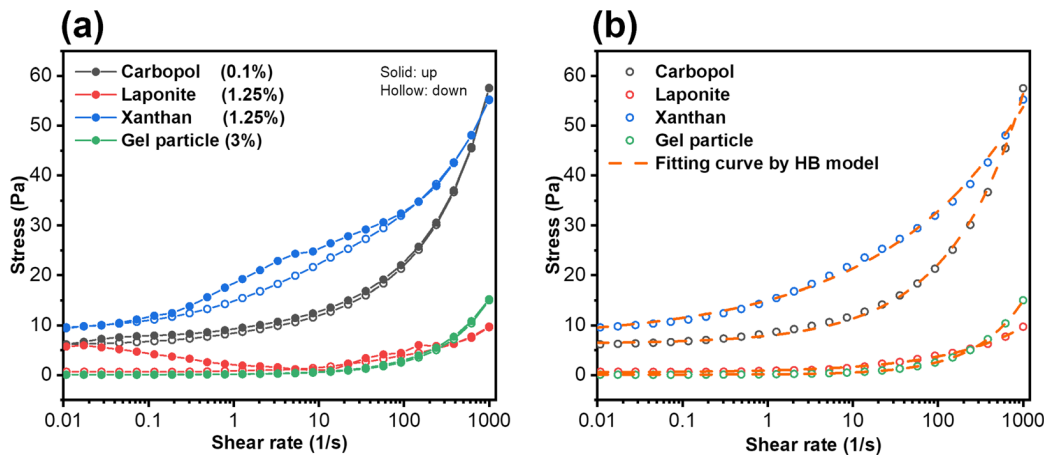

FIG. 16. The thixotropic properties of four samples: (a) the flow curves of 0.1 wt. % carbopol gel, 1.25 wt. % laponite suspension, 1.25 wt. % xanthan gum, and 3 wt. % hydrogel particle suspension; (b) yield stress measurements by steady-state shear (dotted lines refer to the fitting data of the HB model).

**TABLE II.** Comparison of the coordinates of the start ($\sigma_{l,s}$ and $\gamma_{l,s}$) and end ($\sigma_{l,e}$ and $\dot{\gamma}_{l,e}$) of the solid–liquid transition for carbopol gels, laponite suspensions, xanthan gum solutions, and gel particle suspensions at different concentrations (C) and frequencies (f).

| Sample | C (wt. %) | f (Hz) | Algebraic stress bifurcation | | | | Stress bifurcation | | | |
|--------|-----------|--------|-----------|-----------|-----------|-----------|-----------|-----------|-----------|-----------|
| | | | $\sigma_{l,s}$ | $\gamma_{l,s}$ | $\sigma_{l,e}$ | $\dot{\gamma}_{l,e}$ | $\sigma_{l,s}$ | $\gamma_{l,s}$ | $\sigma_{l,e}$ | $\gamma_{l,e}$ |
| Carbopol | 0.2 | 1 | 82.3 | 0.49 | 154 | 35.6 | 82.3 | 0.49 | 154 | 40.9 |
| | 0.175 | 1 | 57.7 | 0.40 | 126 | 42.4 | 57.7 | 0.40 | 126 | 48.2 |
| | 0.15 | 1 | 33.8 | 0.35 | 74.0 | 29.1 | 33.8 | 0.35 | 74.0 | 33.4 |
| | 0.125 | 1 | 16.5 | 0.31 | 36.3 | 20.4 | 16.5 | 0.31 | 36.3 | 24.1 |
| | 0.1 | 1 | 6.2 | 0.35 | 15.6 | 22.7 | 6.2 | 0.35 | 15.6 | 26.7 |
| | 0.1 | 0.5 | 6.0 | 0.37 | 13.2 | 24.4 | 6.0 | 0.37 | 13.2 | 27.2 |
| | 0.1 | 0.2 | 6.0 | 0.38 | 11.3 | 26.0 | 6.0 | 0.38 | 11.3 | 29.0 |
| | 0.1 | 0.1 | 6.0 | 0.38 | 11.2 | 38.8 | 6.0 | 0.38 | 11.2 | 41.7 |
| Laponite | 2 | 1 | 44.1 | 0.16 | 46.4 | 1356 | 44.1 | 0.16 | 46.4 | 1392 |
| | 1.75 | 1 | 32.1 | 0.20 | 42.5 | 422 | 32.1 | 0.20 | 42.5 | 455 |
| | 1.5 | 1 | 19.7 | 0.20 | 25.7 | 62.6 | 19.7 | 0.20 | 25.7 | 64.4 |
| | 1.25 | 1 | 4.3 | 0.27 | 5.0 | 542 | 4.3 | 0.27 | 5.0 | 561 |
| | 1.25 | 0.5 | 3.8 | 0.35 | 4.4 | 1869 | 3.8 | 0.35 | 4.4 | 1928 |
| | 1.25 | 0.2 | 3.8 | 0.26 | 4.4 | 3263 | 3.8 | 0.26 | 4.4 | 3327 |
| | 1.25 | 0.1 | 3.8 | 0.24 | 4.9 | 13430 | 3.8 | 0.24 | 4.9 | 13455 |
| Xanthan | 2 | 1 | 33.8 | 0.34 | 54.2 | 20.7 | 33.8 | 0.34 | 20.7 | 30.8 |
| | 1.75 | 1 | 28.8 | 0.39 | 46.1 | 35.1 | 28.8 | 0.39 | 35.1 | 46.0 |
| | 1.5 | 1 | 23.1 | 0.42 | 36.7 | 56.1 | 23.1 | 0.42 | 56.1 | 71.5 |
| | 1.25 | 1 | 19.7 | 0.45 | 31.5 | 42.0 | 19.7 | 0.45 | 42.0 | 60.1 |
| | 1 | 1 | 16.2 | 0.57 | 25.6 | 75.9 | 16.2 | 0.57 | 75.9 | 95.3 |
| | 1.25 | 0.5 | 19.6 | 0.48 | 26.9 | 27.1 | 19.6 | 0.48 | 26.9 | 35.8 |
| | 1.25 | 0.2 | 16.8 | 0.43 | 22.9 | 19.0 | 16.8 | 0.43 | 22.9 | 23.4 |
| | 1.25 | 0.1 | 12.3 | 0.30 | 19.6 | 17.5 | 12.3 | 0.30 | 19.6 | 20.3 |
| Gel particle | 5 | 1 | 73.3 | 0.76 | 102.6 | 448 | 73.3 | 0.76 | 102.6 | 474 |
| | 4.5 | 1 | 12.1 | 0.24 | 31.0 | 29.3 | 12.1 | 0.24 | 31.0 | 34.2 |
| | 4 | 1 | 4.2 | 0.20 | 16.9 | 23.4 | 4.2 | 0.20 | 16.9 | 27.1 |
| | 3.5 | 1 | 0.99 | 0.09 | 6.5 | 19.1 | 0.99 | 0.09 | 6.5 | 21.7 |
| | 3 | 1 | 0.40 | 0.10 | 2.2 | 15.4 | 0.47 | 0.13 | 2.2 | 17.4 |
| | 3 | 0.5 | 0.35 | 0.09 | 2.35 | 21.5 | 0.35 | 0.08 | 2.35 | 24.3 |
| | 3 | 0.2 | 0.30 | 0.06 | 1.68 | 14.4 | 0.19 | 0.04 | 1.68 | 16.8 |
| | 3 | 0.1 | 0.26 | 0.06 | 1.68 | 15.8 | 0.22 | 0.05 | 1.68 | 17.4 |

where $K$ is the consistency coefficient and $n$ denotes the non-Newtonian index. The values of $\sigma_y$ are presented in Table II. The fitting correlation coefficients $R^2$ are in the range of 0.997–0.999 within the whole shear rate range.

## APPENDIX B: LAOS RESULTS OF DIFFERENT SAMPLES

The reasons for the yield behavior of these YSFs are quite different. More specifically, carbopol gel consists of microgels that are disorderly arranged to resist external stress at high concentrations.[41] The platelets of laponite suspension can form aggregations from the interaction between the net positive charge on the edges and the net negative charge on the faces.[108] These aggregations can further establish a network to present a solid–liquid transition. For the xanthan gum solution, hydrogen bonds contribute to weak gel-like structures.[109] The suspension of hydrogel particles leads to the formation of a network when the volume ratio is above a critical percolation value.[110] Figure 16 demonstrate the results of $G'$ and $G''$ vs stress for the four kinds of samples. The changes in $G'$ and $G''$ were small, and $G'$ were higher than $G''$ at the low range of stress, indicating the existence of linear viscoelastic behavior and elastic domination. Then, $G'$ decreased with the increased applied stress, and $G''$ and the viscous contribution dominated the stress response, indicating the network collapsed and a liquid-like behavior was induced. Similar rheological results have been also reported for carbopol gel,[56–58] laponite,[59] and xanthan gum.[60,90]

It should be noted that the shear banding effect of the laponite suspensions[111] is also shown in Fig. 17(b). The shear banding effect of the laponite suspensions indicates that a jump in shear rate occurs during the upward steady shear test, which reflects an unattainable region in the flow curve is observed. This can be attributed to the fact that laponite suspensions are close to the ideal thixotropic yield stress fluid (thixotropic response: the viscosity decreases with time in the shear test, leading to the structural avalanche at the imposed stress[1–3]). Therefore, the structure of a laponite suspension can be violently destroyed above the yield stress. Then, an unattainable region in the flow curve arises. Figure 16(b) also shows the shear banding effect. More discussions were provided in Appendix C.

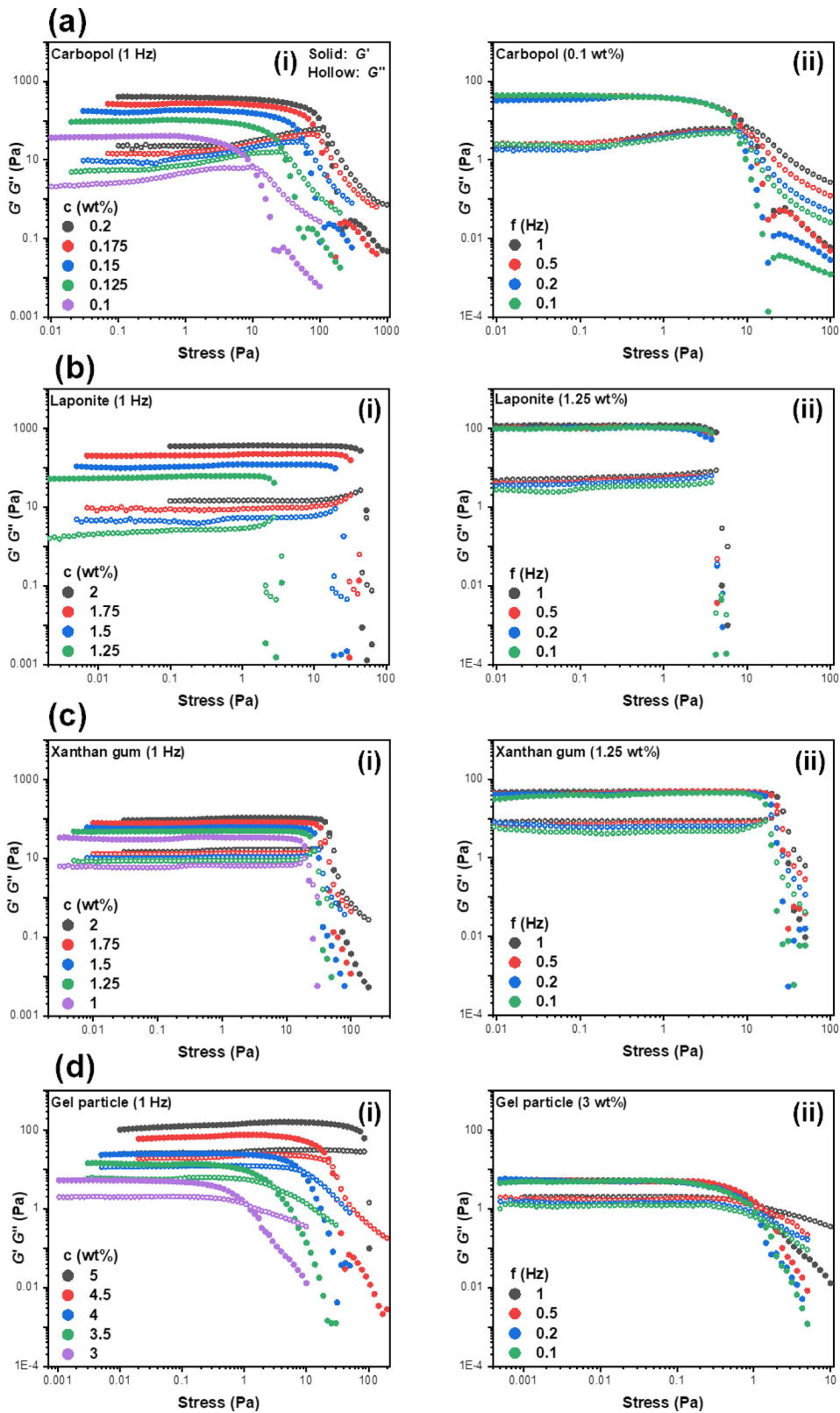

**FIG. 17.** Dynamic moduli as a function of stress amplitudes in amplitude sweep tests for various YSFs: (a) carbopol gels, (b) laponite suspensions, (c) xanthan gum solutions, and (d) hydrogel particle suspensions at different (i) concentrations and (ii) frequencies.

## APPENDIX C: RESULTS FROM STRESS BIFURCATION AND ASB

The acquired results of stress bifurcation and ASB on carbopol gels and laponite suspensions at different concentrations and frequencies are demonstrated in Fig. 18. By using the $\sigma$–$\gamma$ curves of stress bifurcation as an example in Fig. 18(a.i), the $\sigma_{max}$–$\gamma_{\sigma max}$ and $\sigma_{\gamma 0}$–$\gamma_0$ curves superposed at a low stress level, where a linear dependence of stress on strain can be found, indicating an ideal linear solid-like behavior. Deviation appears with the increase in the stress and the superposition remains, which shows the nonlinear solid-like behavior before the solid–liquid transition. Bifurcation happens when the stress continues to increase. Therefore, $\sigma_{L,s}$ and $\gamma_{L,s}$ can be obtained, which is regarded as the critical point to sustaining solid-like behavior. For the $\sigma-\dot\gamma$ curves in Fig. 18(a-ii) ($\sigma_{max}-\dot\gamma_{\sigma max}$ and $\sigma_{\dot\gamma 0}-\dot\gamma_0$ curves), the change in the curves was opposite of that of the $\sigma-\gamma$ curves. The obtained $\sigma_{L,e}$ and $\dot\gamma_{L,e}$ denote also the critical point above which complete solid–liquid transition happens.

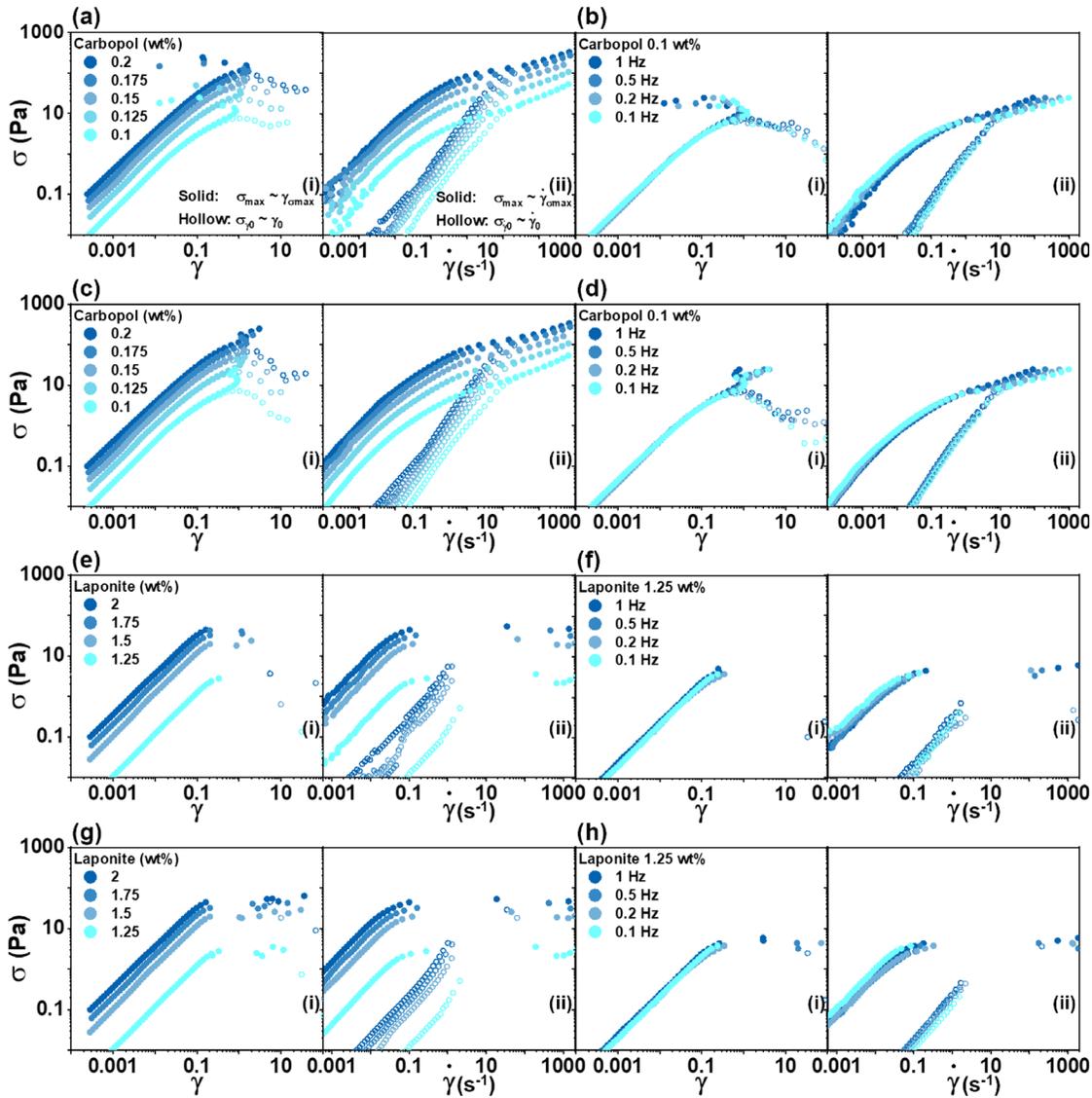

FIG. 18. Comparison of the acquired results from stress bifurcations and ASBs for carbopol gels and laponite suspensions at different concentrations and frequencies: (a) stress bifurcations between $\sigma_{max}$–$\gamma_{\sigma max}$ curves and $\sigma_{\gamma 0}$–$\gamma_0$ curves (i), $\sigma_{max}$–$\dot\gamma_{\sigma max}$ and $\sigma_{\gamma 0}$ ~ $\dot\gamma_0$ curves (ii) for carbopol gels at different concentrations from LAOStress; (b) stress bifurcations for 0.1 wt. % carbopol gels at different frequencies; (c) ASBs for carbopol gels at different concentrations (i) results from Eq. (30) and (ii) results from Eq. (31); (d) ASBs for 0.1 wt. % carbopol gels at different frequencies; (e) stress bifurcations for laponite suspensions at different concentrations; (f) stress bifurcations for 1.25 wt. % laponite suspensions at different frequencies; (g) ASBs for laponite suspensions at different concentrations; and (h) ASBs for 1.25 wt. % laponite suspensions at different frequencies.

The curves, $\sigma_{L,s}$ and $\sigma_{L,e}$, become higher with the increase in the concentration. The curves of the stress bifurcation for carbopol gels at different frequencies are displayed in Fig. 18(b). The $\sigma-\gamma$ curves at the low stress level were independent of the frequency, which corresponds to the outcomes of Yu et al.[9] However, the $\sigma-\dot\gamma$ curves exhibited evident frequency dependency. In addition, the influence of frequency can refer to the work by Yu et al.[9] The results of ASB for carbopol gels are displayed in Figs. 18(c) and 18(d) by using the data only from Fig. 17(a), which yields the same conclusions as stress bifurcation does. The above-mentioned disciplines are applicable for the results of laponite suspensions [Figs. 18(e)–18(h)], xanthan gum solutions [Figs. 19(a)–19(d)], and hydrogel particle suspensions [Figs. 19(e)–19(h)].

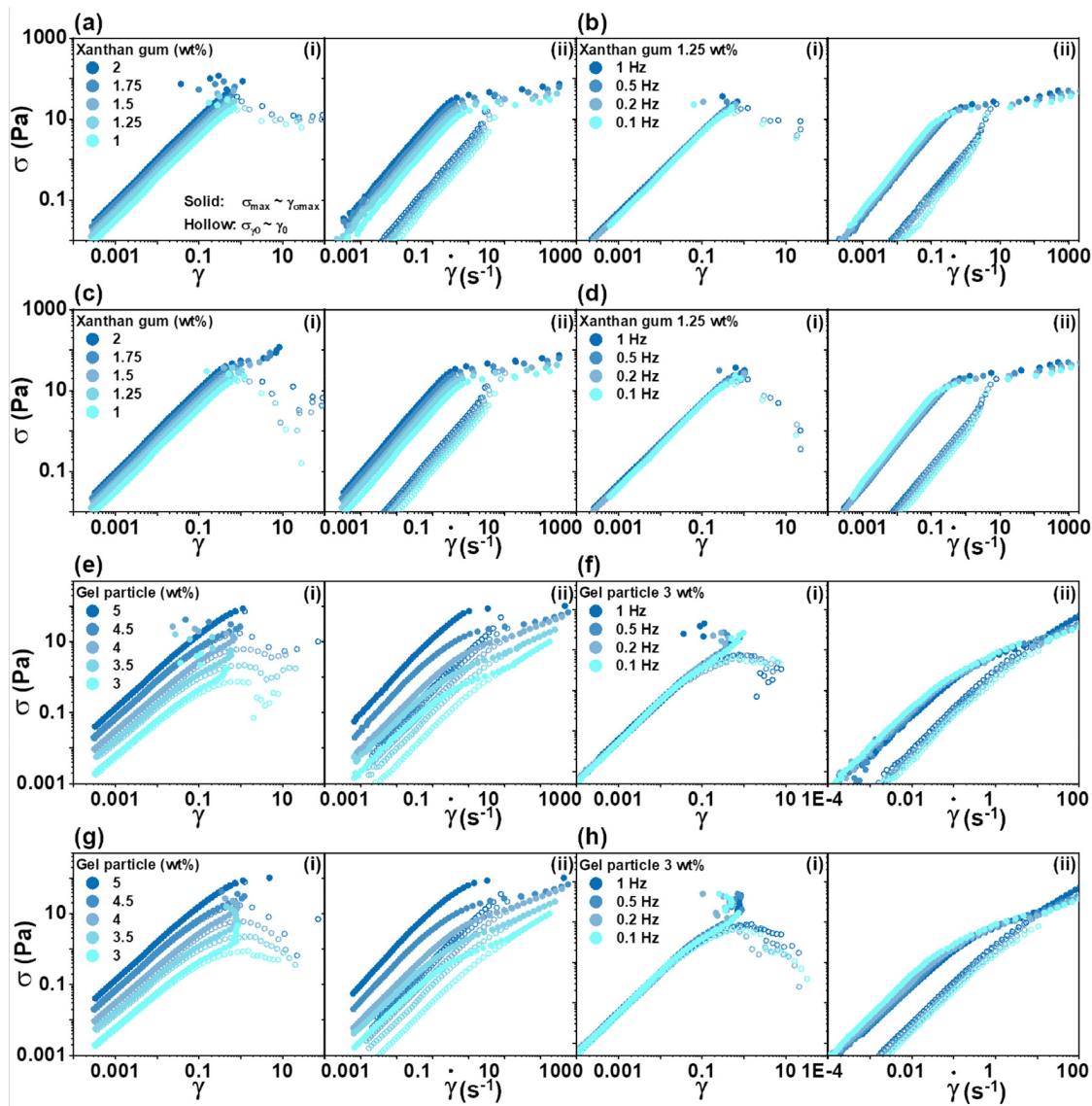

FIG. 19. Comparison of the acquired results from stress bifurcations and ASBs of xanthan gum solutions and hydrogel particle suspensions at different concentrations and frequencies: stress bifurcations for xanthan gum solutions at different (a) concentrations and (b) frequencies (1.25 wt. %); ASBs for xanthan gum solutions at different (c) concentrations and (d) frequencies; stress bifurcations for hydrogel particle suspensions at different (e) concentrations and (f) frequencies (3 wt. %); ASBs for hydrogel particle suspensions at different (g) concentrations and (h) frequencies.

Finally, the results of stress bifurcation from Lissajous curves and ASB were correspondingly plotted together in Fig. 20, including carbopol gels [Figs. 20(a) and 20(b)], laponite suspensions [Figs. 20(c) and 20(d)], xanthan gum solutions [Figs. 20(e) and 20(f)], and hydrogel particle suspensions [Figs. 20(g) and 20(h)]. Figure 20 indicates impressive excellent overlaps between the stress bifurcation and ASB. Furthermore, the coordinates of the start ($\sigma_{l,s}$ and $\gamma_{l,s}$ from the $\sigma-\gamma$ curves) and end ($\sigma_{l,e}$ and $\dot{\gamma}_{l,e}$ from the $\sigma-\dot{\gamma}$ curves) yield points determined by ASB and stress bifurcation are presented in Table I. It was found that the values of $\sigma_{l,s}$, $\gamma_{l,s}$, and $\sigma_{l,e}$ from ASB and stress bifurcation correspond well.

Small deviations between the results of $\sigma_{l,s}$ and $\gamma_{l,s}$ from the two methods may arise when the samples (gel particle suspensions) present relatively weak viscoelasticity. However, this effect does not significantly affect the judgment of the solid–liquid transition region. In addition, the value of $\dot{\gamma}_{l,e}$ from stress bifurcation was larger than that of ASB because the strain rate $\dot{\gamma}(t)$ is the derivative of strain $\gamma(t)$ with time where the transient strain rate can exceed $\omega\gamma_0$, which can be attributed to the distortion of the output strain signal under stress-controlled conditions. To sum up, Fig. 20 and Table II present the high similarity between the results of ASB and stress bifurcation.

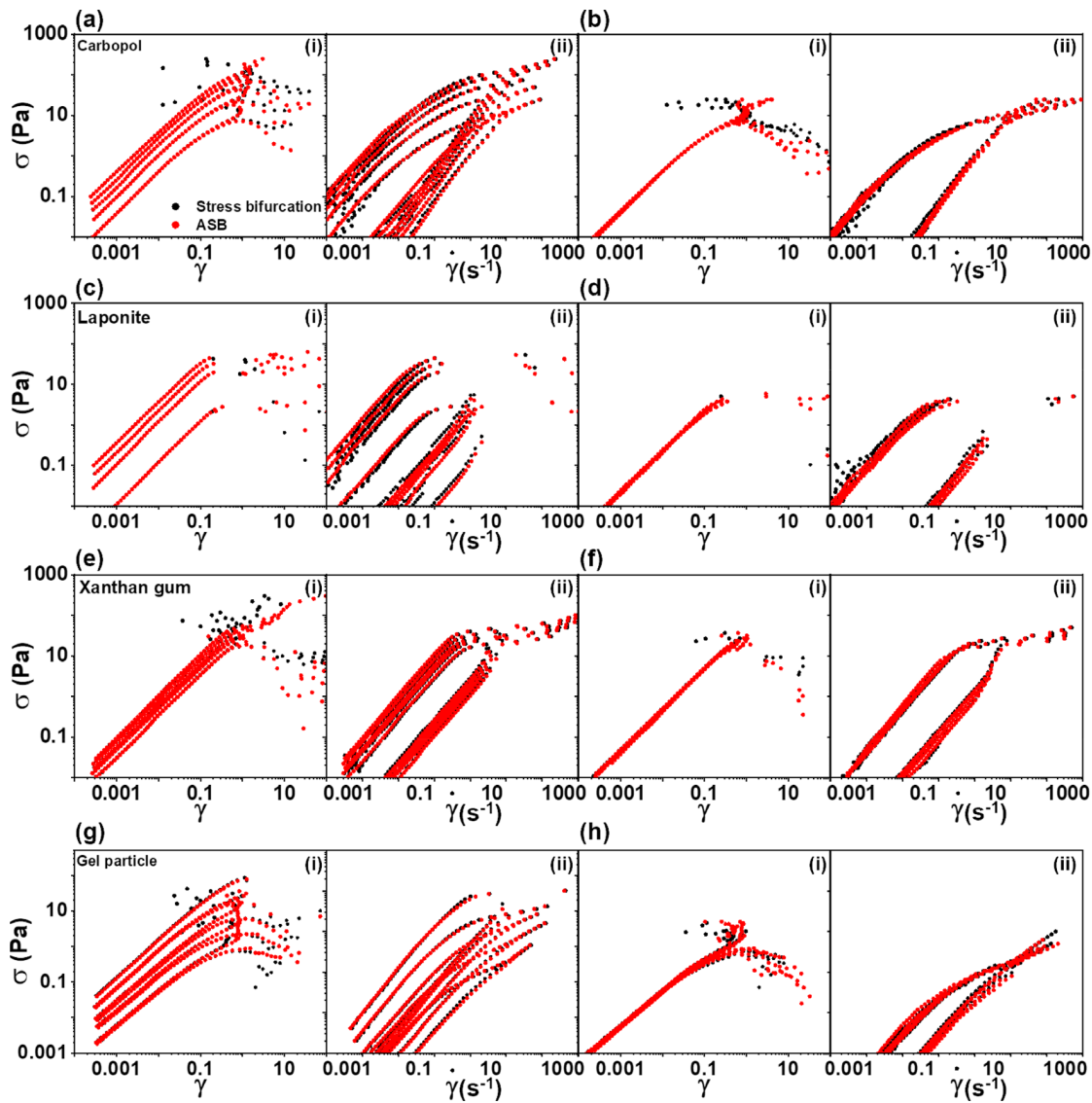

**FIG. 20.** The comparison between stress bifurcations (black points) and ASBs (red points) for carbopol gels, laponite suspensions, xanthan gum solutions, and hydrogel particle suspensions at different concentrations and frequencies: carbopol gels at different (a) concentrations and (b) frequencies; laponite suspensions at different (c) concentrations and (d) frequencies; xanthan gum solutions at different (e) concentrations and (f) frequencies; and hydrogel particle suspensions at different (g) concentrations and (h) frequencies.

It should be highlighted that when the applied stress exceeds the yield stress of a laponite suspension, this sample can no longer generate a stress exceeding the yield stress, which is clearly shown in Figs. 18(e)–18(h). This phenomenon is attributed to the shear banding effect of the laponite suspensions,[111] which is also implied by the creep test demonstrated in Appendix E.

## APPENDIX D: SUPPLEMENTARY DISCUSSION FROM FT RHEOLOGY

The intensities of $I_3/I_1$, $I_5/I_1$, and $I_7/I_1$ changed with stress are depicted in Fig. 5(c), where the start and end yield points are denoted. The intensities of $I_3/I_1$, $I_5/I_1$, and $I_7/I_1$ were close to zero at low stress levels, whereas the nonlinearities at very low stress amplitudes can be attributed to the sensitivity and resolution of the rheometer. After that, these intensities increased with the stress, which is directly related to the structural transformation[112,113] (more specifically, the yield process within a single Lissajous loop leads to significantly higher harmonics[53]). Then, at the high stress region, the carbopol gel gradually becomes a purely viscous fluid with the increase in the stress along with the decreased intensities of $I_3/I_1$, $I_5/I_1$, and $I_7/I_1$. It can also be found that the obvious increase and peak of $I_3/I_1$ occurred before the start yield point and after the end yield point, respectively. In addition, the significantly increased point of $I_7/I_1$ was closer to the start yield point than that of $I_3/I_1$, whereas, $I_3/I_1$, $I_5/I_1$, and $I_7/I_1$ achieved maximum values at the same point.

Figure 5(d) displays the results of the FT rheology from the frequency domain at the strain amplitudes of 0.35, 2.4, and 5.6. Therefore, the harmonics of the three points near the solid–liquid transition can be clearly demonstrated. The signal at 5.6 strain exhibited the most harmonics that possess relatively high intensities ($I_3$–$I_{13}$) than 2.4 strain ($I_3$–$I_9$), and then 0.35 strain ($I_3$–$I_7$), which shows the increased nonlinearity as the stress increased. Furthermore, even harmonics, which can be attributed to the imperfection of the experiment,[78–82] were smaller than odd harmonics, which is one of the typical properties of FT rheology.

## APPENDIX E: SUPPLEMENTARY VERIFICATION OF ASB RESULTS

The determination of the yield stress was carried out by using different methods for carbopol gels (Fig. 12), laponite suspensions [Fig. 21(a)], xanthan gum solutions [Fig. 21(b)], CNF suspensions [Fig. 22(a)], welan gum solutions [Fig. 22(b)], body lotion emulsions [Fig. 23(a)], and ketchup [Fig. 23(b)] including steady shear ($\sigma_{0,H}$, HB model, Fig. 16), creep [$\sigma_c$, e.g., Fig. 12(a)], ASB [$\sigma_{l,s}$ and $\sigma_{l,e}$, e.g., Figs. 12(b) and 12(c)], elastic stress [$\sigma_{d,e}$, e.g., Fig. 12(c)], the intersection of two power-law extrapolations at both low and high stress ranges in $G'$ vs stress [$\sigma_{d,p}$, e.g., Fig. 12(d)], and crossover point of $G'$ and $G''$ [$\sigma_{d,c}$, e.g., Fig. 12(d)]. For the creep method, when the shear stresses below the yield stress, the shear rate decreases with time because of the solidlike behavior of YSFs. If the applied stresses are higher than the yield stress, the flow of a YSF begins and the shear rate can finally remain a constant. The critical stress value at which the flow begins is regarded as the yield stress.

The yield stresses of these samples from different methods were found to approximatively follow the order: $\sigma_{0,H} < \sigma_{d,e} \approx \sigma_{l,s} \leq \sigma_{d,p}$ and $\sigma_{d,c} < \sigma_{l,e}$ and $\sigma_{0,H} < \sigma_c < \sigma_{d,c}$. The derived stress sweep results of CNF suspensions, welan gum solutions, and ketchup correspond well with those reported in the literature.[60,62,64] The obtained $\sigma_{0,H}$ had the lowest value because the microstructure was destructed at a high shear rate during the downward steady shear. The $\sigma_{l,e}$ possessed the highest value, indicating an occurrence of a complete solid–liquid transition in the sample. The relative size between $\sigma_c$ and $\sigma_{l,s}$ varied with the conditions. On the one hand, steady shear under constant stress $\sigma_{max}$ damaged the structure more significantly than oscillatory shear that applied a stress signal changed within the range of $-\sigma_{max} \sim \sigma_{max}$ [$\sigma_c < \sigma_{l,s}$, Figs. 12(a), 22(a), 23(a), and 23(b)]. On the other hand, the start yield point may appear previously along with the partial failure of the structure, leading hence to a decreased $G'$, where the flow does not happen, considering the complex microstructure of some fluids [$\sigma_c > \sigma_{l,s}$, Figs. 21(a) and 22(b)]. For the thixotropic laponite suspension, $\sigma_c \approx \sigma_{l,s}$ was achieved. Meanwhile, compared with the samples presenting $\sigma_c > \sigma_{l,s}$, the creep results within the timescale of 0–1 s of the YSFs behaving $\sigma_c < \sigma_{l,s}$ showed that the shear rate values were comparatively low and increased relatively slowly with the increase in the applied shear stress, where the time of 1 s corresponded the time completing a whole oscillation cycle (f = 1 Hz). In other words, when the stress amplitude of $\sigma_c$ was applied in the oscillatory shear, the samples possessing $\sigma_c < \sigma_{l,s}$ will remain solidlike behavior because there is no enough time during an oscillation cycle to complete the solid–liquid transition as that happened in the creep experiment. The creep tests on the laponite suspension show the shear banding effect of the laponite suspension, reflecting the thixotropic property and the yield stress above which apparent flow occurs immediately.

The creep method is considered one of the most accurate methods for evaluating yield stress.[1–3,6–9,24,33,34] Although the yield stress determination from the point of $G' = G''$ has been widely applied,[32,37] the obtained yield stress may be higher than the real yield stress because the YSF has yielded. Furthermore, the yield stress determination from the intersection may be uncertain, which is dependent on the chosen stress/strain amplitude range and affected by the quality of data. In addition, it should be underlined that the elastic stress method is not always correct for different materials and could lead to inconsistencies among the obtained values of the different methods.[32,43]

## APPENDIX F: EFFECTS OF SAMPLING POINT AND OSCILLATION CYCLE ON ASB

The solid–liquid transition region determined by ASB was investigated on non-thixotropic [Figs. 24(a) and 24(b)] and thixotropic [Figs. 24(c) and 24(d)] YSFs by changing the numbers of the sampling points [Figs. 24(a) and 24(c)] and the oscillation cycles required for each sampling point [Figs. 24(b) and 24(d)]. As can be seen in Fig. 4, ASB can present the movements of ($\dot{\gamma}_0, \sigma_{\gamma 0}$), ($\dot{\gamma}_0, \sigma_{\gamma 0}$), ($\gamma_{\sigma max}, \sigma_{max}$), and ($\dot{\gamma}_{\sigma max}, \sigma_{max}$) that are close to those points of the raw data (i.e., Lissajous curves) according to the descriptions of Sec. IV C. Furthermore, from the positions of the highest and rightmost points in the elastic and viscous Lissajous plots, the shape evolution of the Lissajous curve during the solid–liquid transition can be qualitatively expressed. It was found that the solid–liquid transition of carbopol gel is a continuous process that is not practically related to the number of sampling points and oscillation cycles [Figs. 24(a)–24(f)]. Meanwhile, the viability of ASB was further verified. For laponite suspension, the increased number of points shortened the stress gradient between the two sampling points, resulting thus in more precise results [Figs. 24(g-i) and 24(g-ii)]. Meanwhile, the increased number of cycles slightly decreased the start yield stress and significantly increased the end yield strain rate [Figs. 24(g-iii) and 24(g-iv)], indicating that shear induced structural changes.

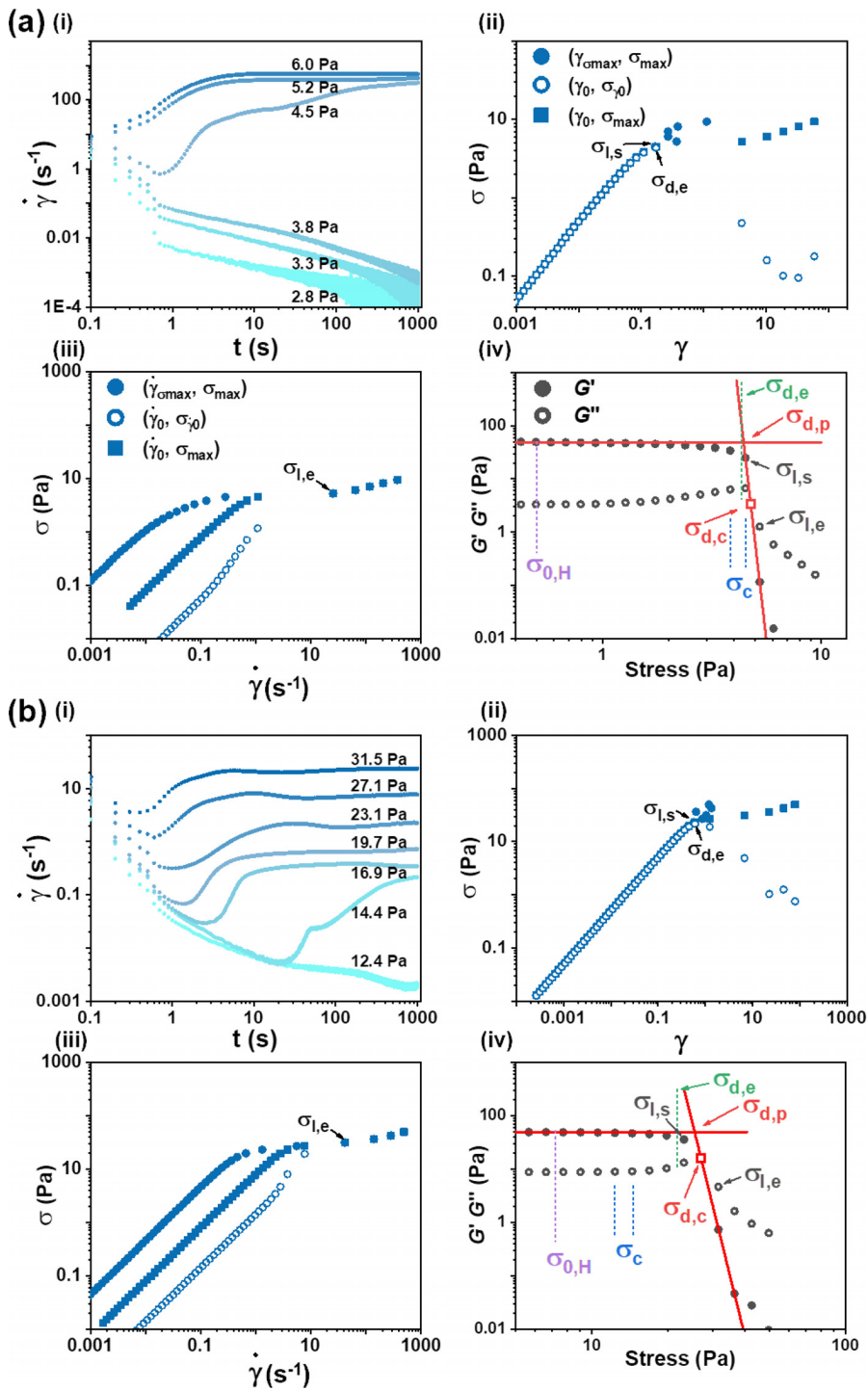

FIG. 21. Comparison of yield stresses obtained by using different methods: (a) 1.25 wt. % laponite suspensions: (i) creep ($\sigma_c$), (ii) elastic stress method ($\sigma_{d,e}$) and stress vs strain from ASB ($\sigma_{l,s}$), (iii) stress vs strain rate from ASB ($\sigma_{l,e}$), (iv) the intersection of two power-law extrapolations at low- and high-stresses in $G'$ vs the stress amplitude ($\sigma_{d,p}$), the characteristic modulus where $G' = G''$ ($\sigma_{d,c}$), and the comparison between these above-determined yield stress and (b) 1.25 wt. % xanthan gum solutions. Frequency: 1 Hz.

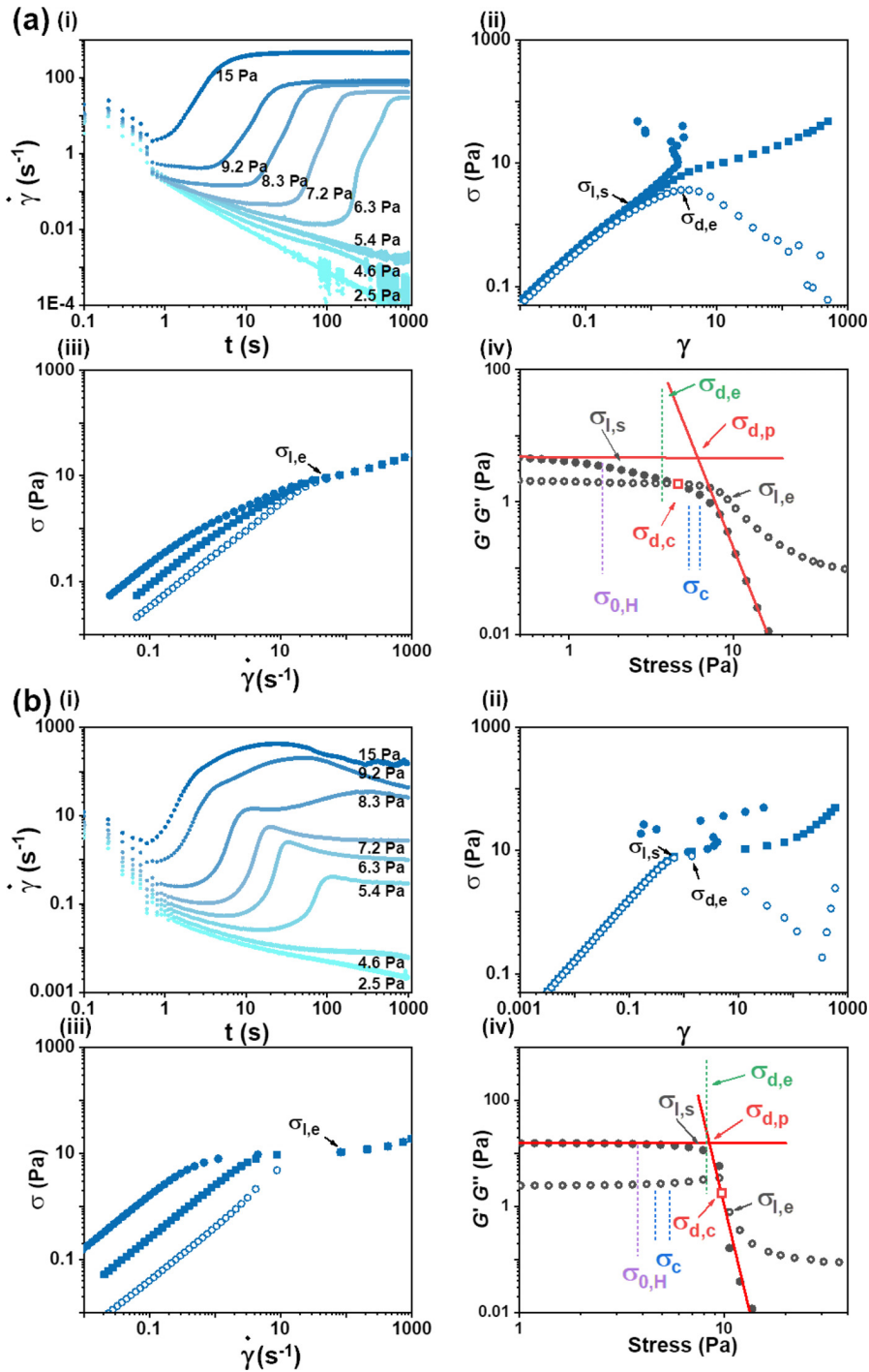

**FIG. 22.** Comparison of yield stresses obtained by using different methods: (a) 0.65 wt. % TEMPO-mediated oxidation cellulose nanofiber (CNF) suspensions and (b) 0.4 wt. % welan gum solutions. Frequency: 1 Hz.

To sum up, Fig. 24 shows that the yielding processes within the determined solid–liquid transition region of the two samples had insignificant relationships with the number of both the sampling points and the repeated cycles at each sampling point. The yield process is continuous and discontinuous for the 0.1 wt. % carbopol gel and 1.25 wt. % laponite suspension, respectively. This phenomenon indicates that the ASB method is not significantly influenced by the sampling points and the repeated cycles, which is important for the application of this method.

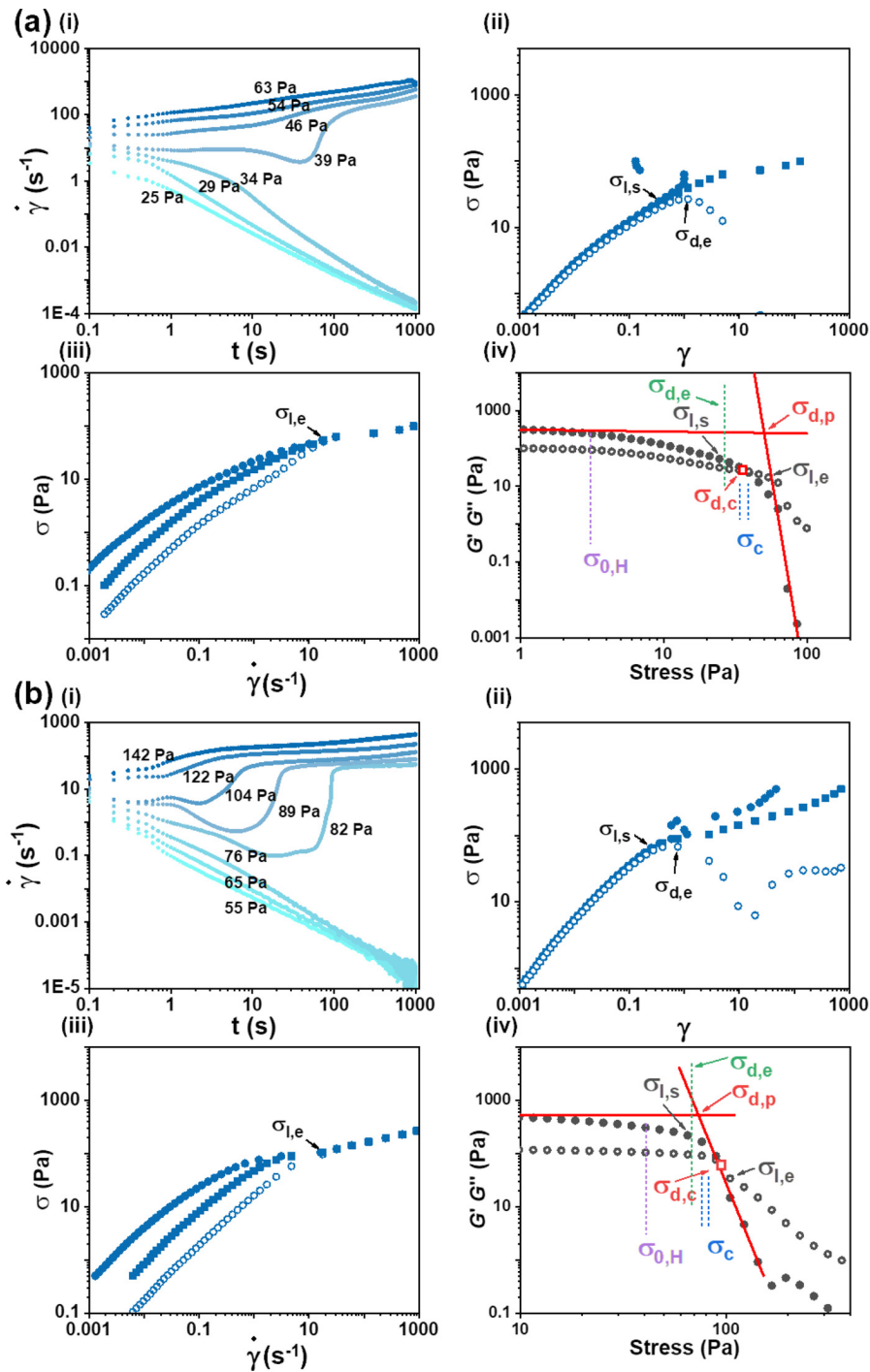

**FIG. 23.** Comparison of yield stresses obtained by using different methods: (a) 100 wt. % body lotion emulsions and (b) 100 wt. % ketchup. Frequency: 1 Hz.

## APPENDIX G: SUPPLEMENTARY RESULTS FROM STRESS DECOMPOSITION

Stress decomposition to divide Lissajous curves into elastic and viscous parts (sample: 3 wt. % hydrogel particle suspension) (Fig. 25). Stress decomposition results (Fig. 26).

## APPENDIX H: SUPPLEMENTARY RESULTS FROM SPP

First, SPP was used to investigate elastic Lissajous curves.[53] The solid–liquid transition regions of the three samples are denoted in the corresponding panels. Figure 27 displays the stress response to strain amplitudes (inner curves to outer) including pre-yielding (left), mid-yielding (middle), and post-yielding (right) regions. In the pre-yielding region, the curves behave like pure elastic materials. In the mid-yielding region, the shape of Lissajous curves can be interpreted by the cage model. Beginning at zero stress, a slightly increased strain caused linear elastic cage deformation. The sample began to flow when the applied strain was beyond the yield strain. The sample continued to flow until the instantaneous strain rate reached zero ($\dot{\gamma}_0$) when the cage instantaneously reformed.[53] Pure viscous responses appeared also in the post-yielding region. During the solid–liquid transition, the elastic Lissajous curves of carbopol gel and laponite suspension showed the most gentle and significant changes, respectively.

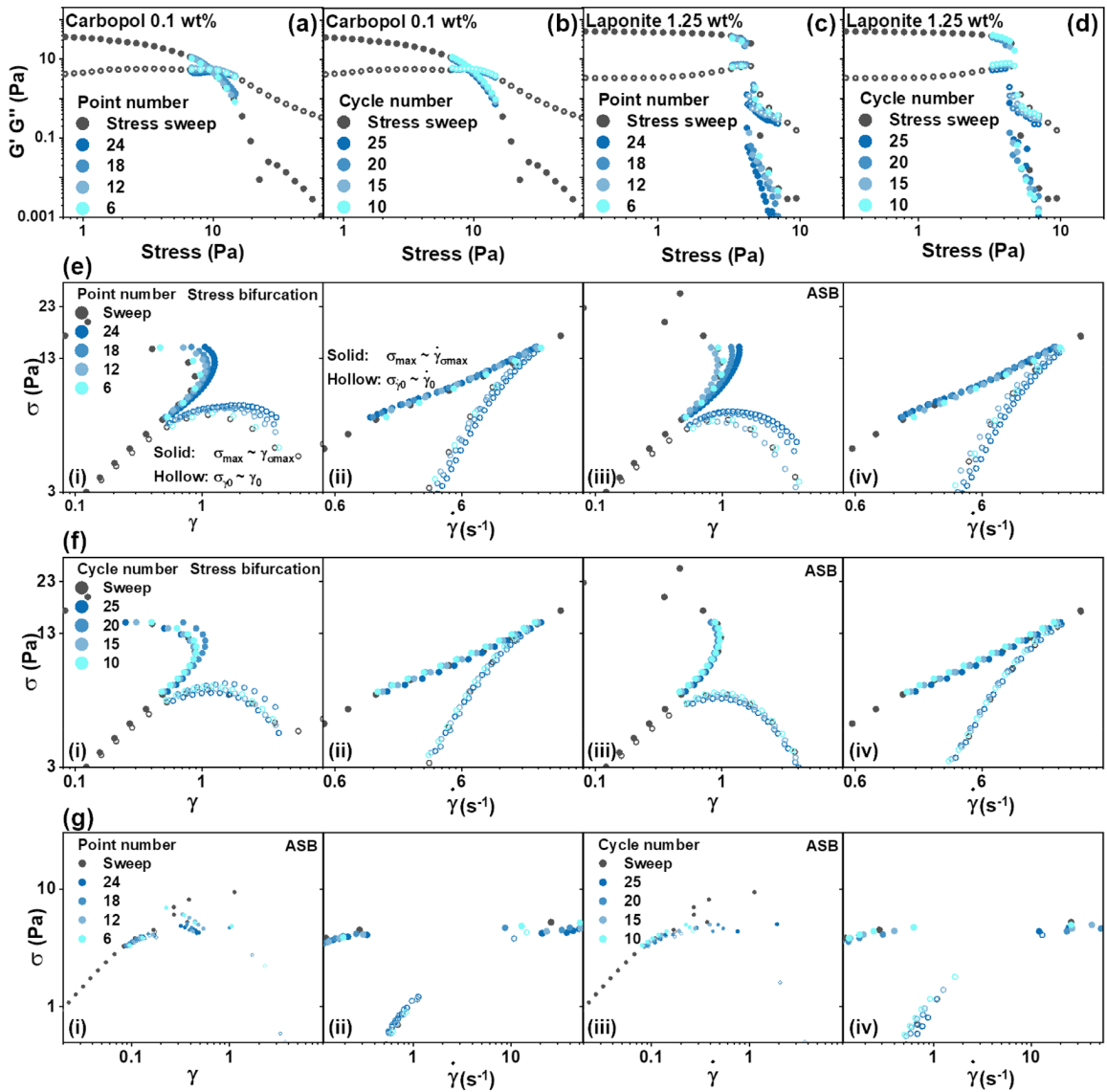

**FIG. 24.** The influence of the point number and cycle number on the yielding processes by testing 0.1 wt. % carbopol gels and 1.25 wt. % laponite suspensions: dynamic moduli from amplitude sweep tests at 1 Hz by changing (a) and (c) point number and (b) and (d) cycle number for (a) and (b) 0.1 wt. % carbopol gels and (c) and (d) 1.25 wt. % laponite suspensions; (e) 0.1 wt. % carbopol gels at different point numbers: (i) $\sigma_{max}$–$\dot{\gamma}_{\sigma max}$ curves and $\sigma_{\dot{\gamma}0}$–$\gamma_0$ curves, and (ii) $\sigma_{max}$–$\dot{\gamma}_{\sigma max}$ and $\sigma_{\dot{\gamma}0}$–$\dot{\gamma}_0$ curves from stress bifurcation, (iii) and (iv) the results by using ASB; (f) 0.1 wt. % carbopol gels at different cycle numbers; (g) The ASBs of 1.25 wt. % laponite suspensions at different (i) and (ii) point numbers and (iii) and (iv) cycle numbers.

Since the $G_{cage}$ is helpful to denote and quantify the slope of the stress at zero stress loads, which is specific for treating YSFs, the values of $G_{cage}$ were plotted with $G'$ vs stress [Fig. 28(a)]. As can be seen, the $G_{cage}$ and $G'$ correspond well at small stress levels. After reaching $\sigma_{1,s}$, the $G_{cage}$ decreased as the stress increased. At $\sigma_{1,e}$, big differences were detected between the $G_{cage}$ and $G'$ for the three samples. The values of $G_{cage}$ were higher than $G'$ at high stress levels, indicating that the elastic behaviors of the three samples can be partly reformed. Thus, elastic responses at large stress levels were identified as indicated by SPP, which is not shown by $G'$ from the rheometric tests. In addition, considering the changes in $G_{cage}$, the recovery abilities of the samples apparently followed the order of 0.1 wt. % carbopol gel > 1.25 wt. % xanthan gum solution > 1.25 wt. % laponite suspension. In addition, it is anticipated that when the frequency is set low enough, the values of $G_{cage}$ at high stress levels are comparable to the $G'$ value at low stress levels because all elastic structures can recover within an oscillatory cycle.

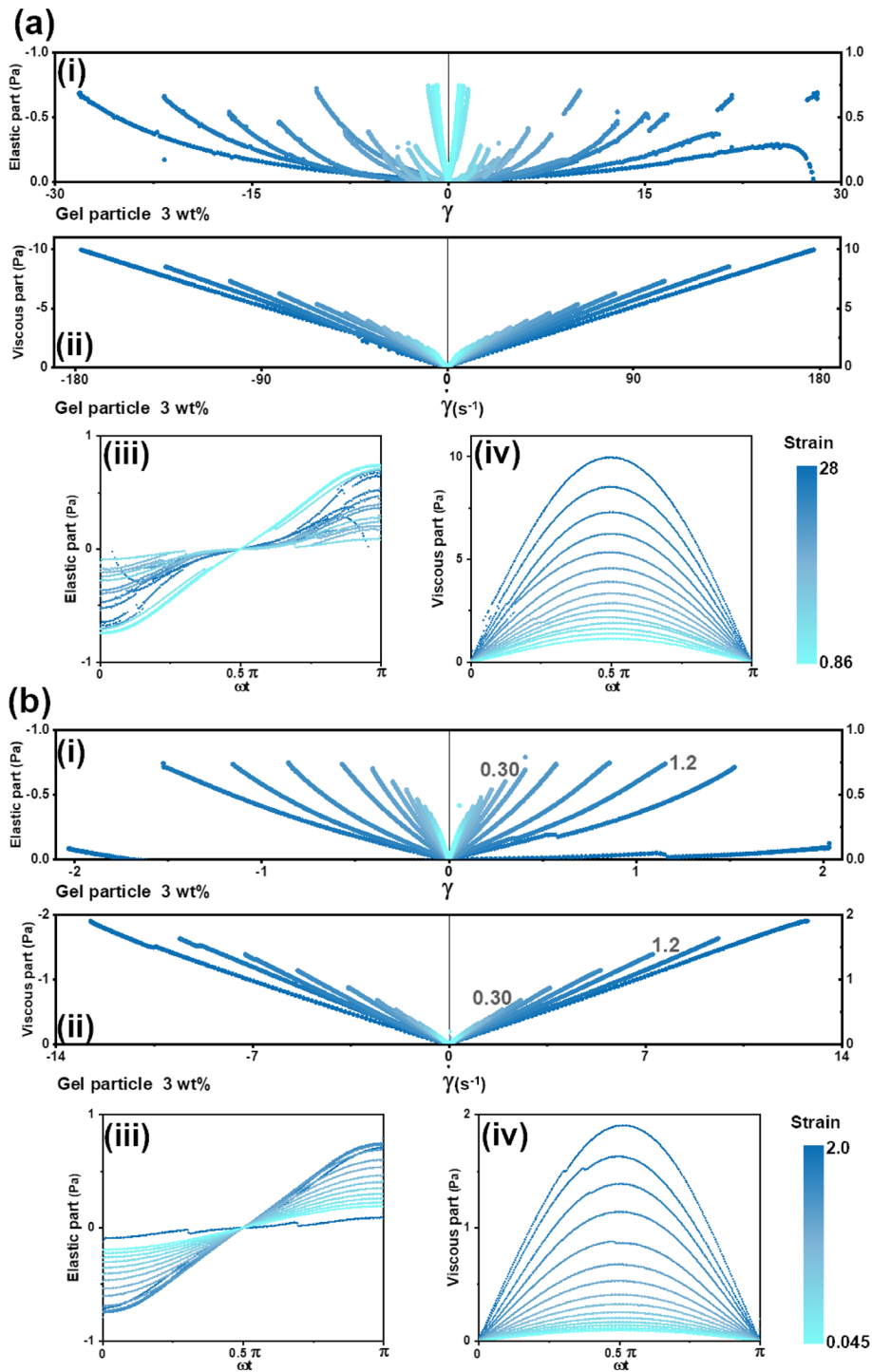

**FIG. 25.** Stress decomposition to divide Lissajous curves into elastic and viscous parts (sample: 3 wt. % hydrogel particle suspension). (a) The results in the strain amplitude range of 0.86–28: (i) elastic stress vs strain; (ii) viscous stress vs strain rate; (iii) elastic and (iv) viscous stress vs angle. (b) The results in the strain amplitude range of 0.045 ~2.0 [solid–liquid transition region: 0.30–1.15 (strain)].

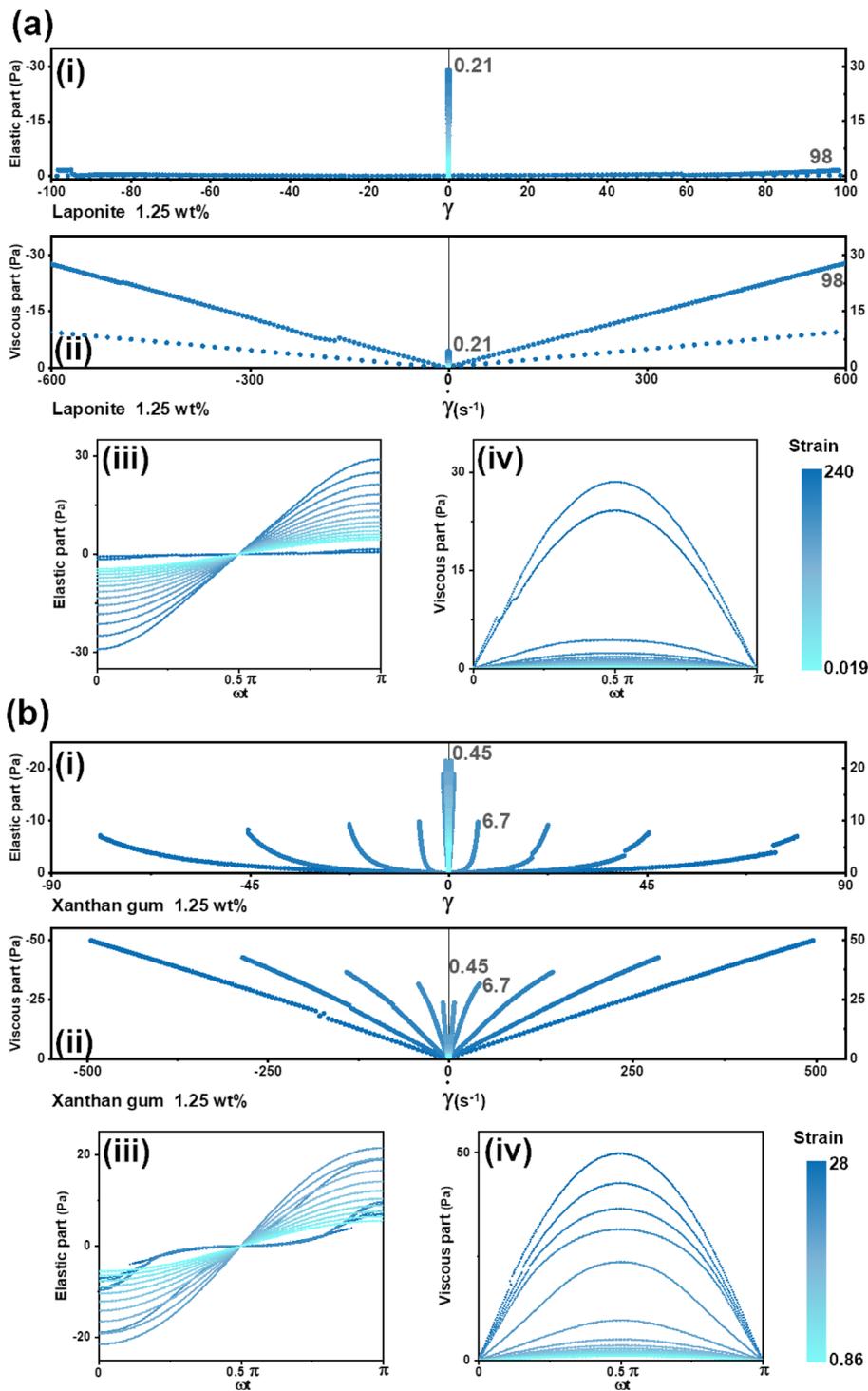

**FIG. 26.** Stress decomposition results. The results in the strain amplitude range of 0.019–240 for (a) 1.25 wt. % laponite suspension and in the strain amplitude range of 0.11–79 for 1.25 wt. % xanthan gum solution.

The amount of the strain acquired at the stress maximum ($\gamma$ at $\sigma_{max}$) as a function of the strain amplitude is plotted in Fig. 28(b). For a truly elastic material, $\gamma$ at $\sigma_{max}$ was equal to $2\gamma_0$. $\sigma_{max}$ occurs for a truly viscous material once the system acquires a strain of $\gamma_0$.

Therefore, both the elastic and viscous deformation regions can be defined in Fig. 28(b). The results of the xanthan gum solution and laponite suspension are in direct agreement with the corresponding solid–liquid transition regions. However, deviations arise for carbopol gel, which may be attributed to the gradual structure failure, as described for the difference between $\sigma_c$ and $\sigma_{1,s}$.

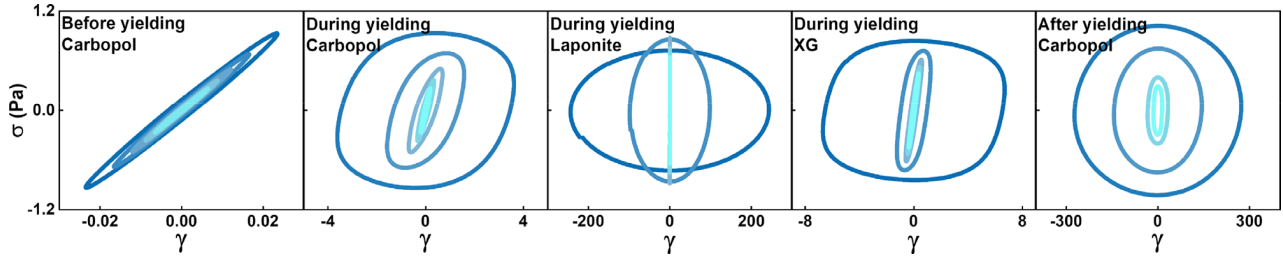

FIG. 27. Waveforms of the response to oscillatory shear at different regions (samples: 0.1 wt. % carbopol gel, 1.25 wt. % laponite suspension, and 1.25 wt. % xanthan gum solution). Frequency: 1 Hz.

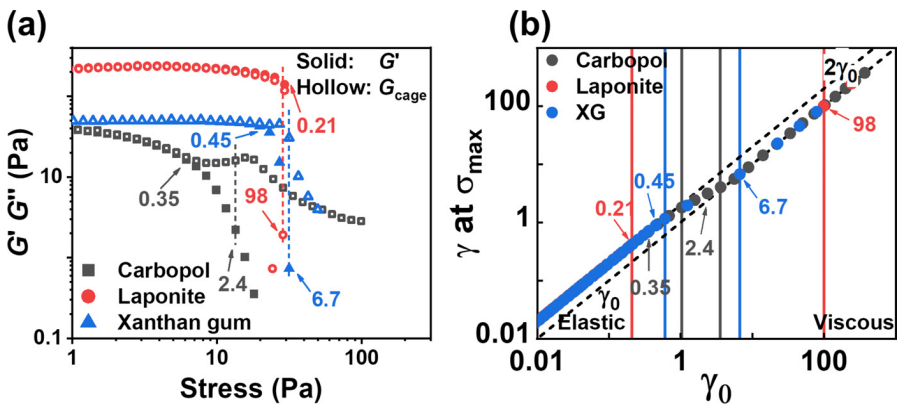

FIG. 28. (a) Apparent cage modulus and (b) strain acquired at the point of maximum stress of 0.1 wt. % carbopol gel, 1.25 wt. % laponite suspension, and 1.25 wt. % xanthan gum solution. Frequency: 1 Hz.

The behavior of the post-yielding inside an elastic Lissajous curve was demonstrated by assuming the occurrence of flow. The upper parts of elastic Lissajous curves within the range of $0-\gamma_0$ were used to generate the flow curves from oscillation tests since no obvious yield point appeared, which refers to Rogers *et al.*[53] with slight modification (more information of the LAOS-based flow curve was provided in Sec. II E). Figure 29 shows the flow curves from oscillation tests with the data of the steady shear and the viscous parts of the stress decomposition. The viscous parts of the stress decomposition were shown in Fig. 29 by simply plotting the time-dependent decomposed viscous stress vs the time-dependent strain rate. Deviations arose between the results of the two methods, which corresponds well with the conclusion in the literature.[84] It is foreseeable that the flow curves generated from LAOS may not reproduce the steady-state flow curve because LAOS signals (sinusoidal and distorted sinusoidal signals) obviously deviate from the steady-state requirement for measuring the steady-state flow curve. Meanwhile, it was found that a higher maximum stress/strain rate results in a lower minimum stress/strain rate, which may be attributed to the violent structural failure under a high stress/strain rate along with a short time for the structural recovery. The deviation may be because the time for structure recovery is too short, where the duration for the structural recovery may be less than 0.1 s when the frequency is 1 Hz. Furthermore, Fig. 29 also contains the viscous parts from stress decomposition. The bifurcation points between the flow curves from oscillation tests and the viscous parts from stress decomposition seem to fit well with the steady shear data. It is noticed that, for one flow curve from LAOS [e.g., the rightmost curve in Fig. 29(a)], the degree of deviation of this curve and the steady-state flow curve in the region, the strain rate region below the bifurcation point between this LAOS-based flow curve and the corresponding viscous parts, increases as the strain rate decreases. It may be attributed to the lack of the stress provided by the elastic structures that cannot recover within a short time during half of an oscillatory shear cycle. A situation can be assumed: (i) for one flow curve from LAOS, the stress values in the large strain rate region are dominantly provided by the dissipated structures in the sample; and (ii) when the strain rate decreases sharply, the stress provided by the dissipated structures declines significantly while the elastic structures in the sample cannot recover immediately. From such assumption, the deviations between the steady-state flow curve and the LAOS-based flow curves can be explained and the differences between the two kinds of results correspond to the unrecovered elastic structures. Each sample reached a flow at $\sigma_{1,s}$. At $\sigma_{1,e}$, the flow curve from LAOS was lower and higher than the steady shear curve for non-thixotropic fluid and thixotropic fluid, respectively.

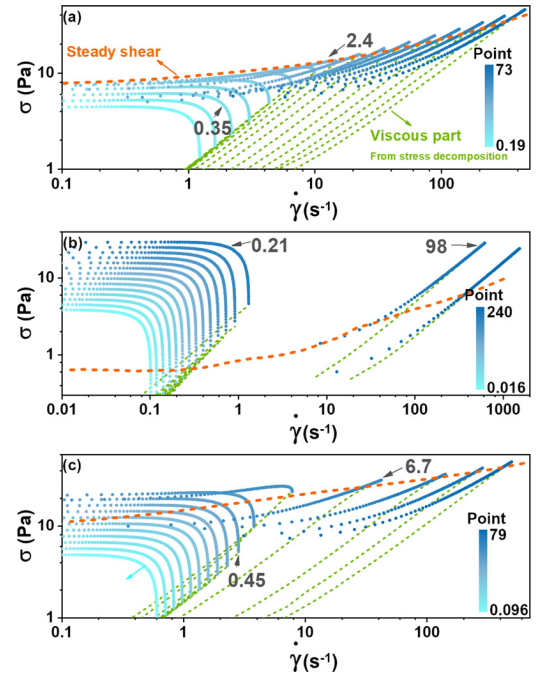

FIG. 29. LAOS-based flowing curves from Lissajous curves, viscous parts from stress decomposition, and steady shear as functions of the strain rate for (a) 0.1 wt. % carbopol gel, (b) 1.25 wt. % laponite suspension, and (c) 1.25 wt. % xanthan gum. Frequency: 1 Hz.

## APPENDIX I: PROCESSING DATA IN LITERATURE

It is known that $G''$ contains two contributions from the recoverable and unrecoverable strains, while the unrecoverable part corresponds to flow.[44] However, in ASB, the total $G''$ is adopted.

Therefore, the validity of ASB must be further tested. The following Figs. 31(a-i), 31(b-i) and 31(c-i) show the decomposed amplitude sweep data of the samples of carbopol, xanthan gum, and concentrated Ludox, which were extracted from the original Fig. 3 in the literature.[44] The ASB curves based on the extracted data in the literature,[44] are shown in Figs. 31(a-ii), 31(b-ii) and 31(c-ii) for determining the start yield points, and in Figs. 31(a-iii), 31(b-iii), and 31(c-iii) for determining the end yield points, respectively.

As shown in Fig. 31, based on our ASB treatment, using either the data of $G'$ and $G''$ or those of $G'_{solid}$ and $G''_{fluid}$ can offer the same start [Figs. 31(a-ii), 31(b-ii) and 31(c-ii)] and end [Figs. 31(a-iii), 31(b-iii), and 31(c-iii)] yield points. Furthermore, according to the original data in the literature [Figs. 31(a-i), 31(b-i), and 31(c-i)], the values of $G''_{fluid}$ and $G''$ are very close to each other in the immediate vicinity of our determined start and end yield points. Therefore, even if $G''$ is applied in our ASB, since $G''_{fluid} \gg G''_{solid}$, there is no significant influence on the determination of the yield stress. As a result, our ASB method should be reliable for the determination of yield stress. However, the concept of $G''_{fluid}$ should be helpful in the construction of a more solid and cogent theoretical foundation in ASB.

Based on Fig. 31, the physical meanings of the start and end yield points may be able to relate to the conclusion given by the recovery rheology. It is noticed that the contributions of $G''_{fluid}$ provided by the three representative samples are all non-negligible at the corresponding start yield points. In other words, when the stress/strain amplitude exceeds the start yield point, the $G''_{fluid}$ value is close to the $G'_{solid}$ value, and the contribution of $G''_{fluid}$ to the total stress/strain response can not be ignored. Therefore, the start yield point represents that the $G''_{fluid}$ contribution becomes significant. At the three end yield points in Fig. 31, it is also clearly shown that the responses of $G''_{fluid}$ are dominant, where the values of $G'_{solid}$ and $G''_{solid}$ are much lower than the $G''_{fluid}$ value.

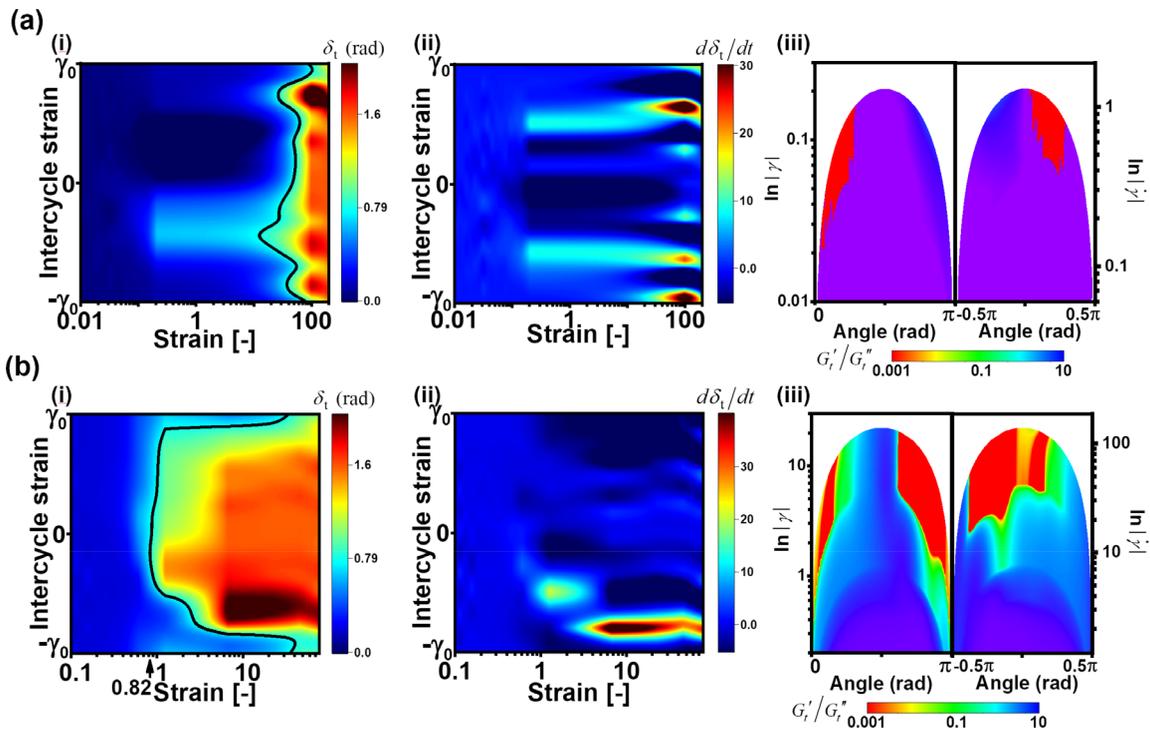

FIG. 30. Transient-modulus-related methods in SPP to treat (a) 1.25 wt. % laponite suspension and (b) 1.25 wt. % xanthan gum solution: (i) contour plot of phase angle $\delta_t$; (ii) contour plot of the phase angle velocity $d\delta_t/dt$. The denoted strain values represent the first appearance of $\delta_t = \pi/4$ ($G_t' = G_t''$); (iii) in $|\gamma|$ (intracycle strain) and in $|\dot{\gamma}|$ (intracycle strain rate) vs angle ($\theta$, $\gamma(t) = \gamma_0 \sin\theta$) to show the $G_t'/G_t''$ ratio with the color mapping.

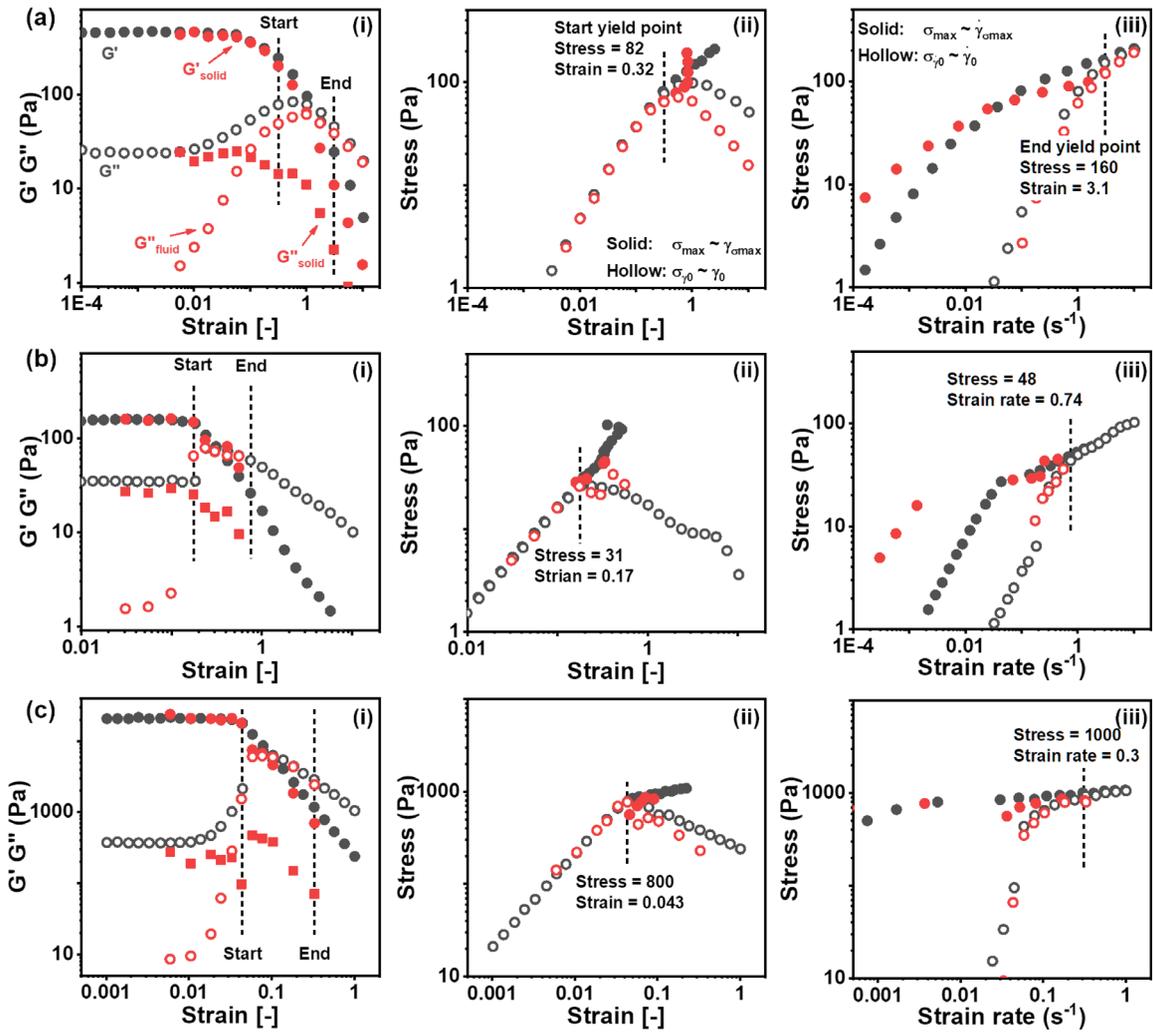

**FIG. 31.** The determination of start and end yield points in ASB by using the data in the literature:[44] Decomposed amplitude sweeps for the samples: (a-i) carbopol, (b-i) xanthan gum, and (c-i) concentrated Ludox, and their corresponding stress–strain curves (a-ii), (b-ii), and (c-ii) and the stress–strain rate curves (a-iii), (b-iii), and (c-iii) obtained by ASB treatment.

## APPENDIX J: GENERAL PROTOCOL FOR ASB

The frequency range can be set depending on the research requirements carrying out the ASB analysis based on the results demonstrated in Sec. IV D by using models at different frequencies. It should be mentioned that the end yield stress is generally positive to the frequency value because nearly complete structural destruction within a shorter timescale will occur at a larger stress/strain amplitude. The minimum stress/strain amplitude should be set in the SAOS region and the maximum stress/strain amplitude must be in the LAOS region (e.g., the strain amplitude of 10 or higher values is proper for the ASB method proved by the results demonstrated in this work). Eqs. (30) and (31) must be applied to carry out the ASB method, which requires the values of $G'$, $G''$, and the stress/strain amplitude. The start yield point is taken from the one before the obvious bifurcation point on the stress–strain plot, while the end yield point is based on the one after the final point showing obvious bifurcation on the stress–strain rate plot. The absence of a start yield stress is attributed to the fact that the $G'$ value is close to $G''$, which leads to the separation of the two stress–strain curves provided by the ASB method in the SAOS region while many YSFs present $G' \gg G''$. The lack of an end yield point is due to the fact that the sample has already behaved like a pure viscous fluid in the SAOS region, possessing $G' \ll G''$.

08 December 2023 13:34:06